\documentclass[10pt,a4paper]{article}
\pdfoutput=1
\usepackage{jheppub}
\usepackage{amsmath}
\usepackage{verbatim}
\usepackage{amssymb}
\usepackage{inputenc, array}
\usepackage{textcomp}
\usepackage{appendix}
\usepackage{amsfonts}
\usepackage{amsthm}
\usepackage{tikz}
\usetikzlibrary{decorations.markings}
\usepackage{mathtools}
\usepackage{bm}
\allowdisplaybreaks
\title{Intrinsic Approach to $1+1$D Carrollian Conformal Field Theory}
\author{Amartya Saha}
\affiliation{Indian Institute of Technology Kanpur, Kanpur 208016, INDIA} 
\emailAdd{amartyas@iitk.ac.in}
\preprint{}
\abstract{The 3D Bondi-Metzner-Sachs (BMS$_3$) algebra that is the asymptotic symmetry algebra at null infinity of the $1+2$D asymptotically flat space-time is isomorphic to the $1+1$D Carrollian conformal algebra. Building on this connection, various preexisting results in the BMS$_3$-invariant field theories are reconsidered in light of a purely Carrollian perspective in this paper. In direct analogy to the covariant transformation laws of the Lorentzian tensors, the flat Carrollian multiplets are defined and their conformal transformation properties are established. A first-principle derivation of the Ward identities in a $1+1$D Carrollian conformal field theory (CCFT) is presented. This derivation introduces the use of the complex contour-integrals (over the space-variable) that provide a strong analytic handle to CCFT. The temporal step-function factors appearing in these Ward identities enable the translation of the operator product expansions (OPEs) into the language of the operator commutation relations and vice versa, via a contour-integral prescription. Motivated by the properties of these step-functions, the $i\epsilon$-forms of the Ward identities and OPEs are proposed that permit for the hassle-free use of the algebraic properties of the latter. Finally, utilizing the computational techniques developed, it is shown that the modes of the quantum energy-momentum tensor operator generate the centrally extended version of the infinite-dimensional $1+1$D Carrollian conformal algebra.}
\begin{document}
\maketitle
\flushbottom
\section{Introduction}
In recent years, physical systems with an underlying Carrollian symmetry group \cite{LL,SG} have emerged as active research avenues. Some of the interesting research programs involving the Carrollian physics are: the geometry of the Carrollian space-time \cite{Henneaux:1979vn,Ciambelli:2019lap,Perez:2021abf}, cosmology \cite{deBoer:2021jej,Concha:2021jnn}, holography in asymptotically flat space-times \cite{Donnay:2022aba,Bagchi:2016bcd} through the connection with the Bondi-Metzner-Sachs (BMS) group \cite{Duval:2014uva,Bagchi:2010zz} and the tension-less limit of string theory \cite{Is, Bagchi:2019cay}. 

\medskip

In this paper, we will be interested in field theories invariant under the $1+1$D Carrollian conformal algebra (at level 2) that is isomorphic to the BMS$_3$ algebra \cite{Duval:2014lpa}. Explicit constructions of such field theories have been carried out e.g. in \cite{Bagchi:2019clu,Banerjee:2020qjj,Gupta:2020dtl}. Interestingly, only in $1+1$D, the Carrollian algebra is isomorphic to the Galilean algebra with this isomorphism extending to their conformal counterparts \cite{Duval:2014lpa, Duval:2014uoa}. 

\medskip

The $1+1$D (level 2) Carrollian conformal transformations are given by \cite{Duval:2014lpa}:
\begin{align*}
x\rightarrow x^\prime=F(x)\hspace{2mm},\hspace{2mm}t\rightarrow t^\prime=tF^\prime(x)+G(x)
\end{align*}
where $F(x)$ and $G(x)$ are arbitrary functions of $x$ and are called the super-rotations and the super-translations respectively.  The generators $\mathcal{L}_n$ and $\mathcal{M}_n$ (with $n\in\mathbb{Z}$) of these transformations in the space of ordinary scalar-valued functions give rise to the following infinite-dimensional algebra known as the $1+1$D Carrollian/Galilean \cite{Bagchi:2009my} conformal algebra:
\begin{align*}
\left[\mathcal{L}_n\hspace{1mm},\hspace{1mm}\mathcal{L}_m\right]=i(n-m)\mathcal{L}_{n+m}\hspace{2.5mm};\hspace{2.5mm}\left[\mathcal{L}_n\hspace{1mm},\hspace{1mm}\mathcal{M}_m\right]=i(n-m)\mathcal{M}_{n+m}\hspace{2.5mm};\hspace{2.5mm}\left[\mathcal{M}_n\hspace{1mm},\hspace{1mm}\mathcal{M}_m\right]=0
\end{align*}

\medskip

This $1+1$D (level 2) Carrollian conformal algebra (CCA$_{1+1}$) can be obtained by a Inonu-Wigner contraction of two copies of the relativistic Witt algebra \cite{Bagchi:2009pe}, thus paving the way to examining the Carrollian conformal field theories systematically in this singular limit. One of the major stumbling blocks in the study of these theories, as opposed to their 2D relativistic parents, is the lack of separation between the left and the right modes. This is because the contraction of the two copies of the Witt algebra to the CCA$_{1+1}$ involves a mixing between the chiral and anti-chiral sectors of the parent algebra. So far, this has been the reason why there has been very little use of complex analysis techniques for $1+1$D CCFTs. In this paper, by using an intrinsic\footnote{We shall not use any `limiting' argument in this work.} approach, we propose a solution to this problem by elevating the space coordinate to a complex variable.

\medskip

There is another aspect of $1+1$D CCFTs we reconsider in the wake of a simple observation: the action of the Carrollian boost generator on space-time is non-diagonalizable and has a Jordan block structure. In direct analogy to how tensors are defined using Lorentz covariance in special relativity, we define $1+1$D local Carrollian multiplets transforming under the reducible but indecomposible representations of this generator. Similar looking Jordan block structures have been noticed in the literature earlier to arise in the Hilbert space representation of the $1+1$D (quantum) Carrollian boost generator and have been thoroughly investigated in \cite{Chen:2020vvn}, in analogy with similar structures arising in Logarithmic CFTs. In this construction, an assumption of a state-operator correspondence is implicit. But, one should distinguish this from our case. It is important to stress that our construction is based on the observation that the local fields in the Carrollian theory need to arrange themselves in representations of the space-time (i.e. classical) Carrollian boost generator, like local fields arrange themselves in representations of the Lorentz group in all relativistic quantum field theories (QFTs). Thus, this does not at all require any reference to the aspects of the quantum algebra (e.g. Hilbert space, state-operator map). Moving on to conformal representations, we go on to define the quasi-primary and primary fields. 

\medskip

The word `local' is important here. In absence of the above mentioned motivation stemming from the space-time physics, the quantum Carrollian boost generator in \cite{Chen:2020vvn} was given non-zero diagonal entries in matrix representations in the Hilbert space. It was shown in \cite{Chen:2022cpx} that the correlation function, first derived in \cite{Bagchi:2009ca}, between two fields corresponding to states (by the assumption of state-operator correspondence) transforming under such a `singlet' representation (as defined in \cite{Chen:2020vvn}) can arise from a non-local action. The conclusion that this correlator can originate only from a non-local action can be reached by taking a Fourier transformation of the same: the momentum-space form of the Carrollian correlator makes it manifest that the equation of motion, of which this correlator is a Green's function, must contain spatial derivatives of negative order whenever $\xi\neq0$, irrespective of $\Delta$ (notations are as used in \cite{Chen:2020vvn} whose $\xi=0$ counterpart is \cite{Chen:2022jhx}). This conclusion also applies for any higher rank multiplets \cite{Chen:2020vvn} with $\xi\neq0$. Moreover, the `limiting' perspective in \cite{Bagchi:2009pe} has given rise only to such `singlet' fields. Instead, in this paper, we will be concerned with exploring the properties of CCA$_{1+1}$ invariant field theories describable by local actions.

\medskip

Next, we focus on the Energy-Momentum (EM) tensor of the CCA$_{1+1}$ invariant classical local field theories in $1+1$D flat (Carrollian) background. We begin by considering invariance of the action only under the six parameter global subalgebra of CCA$_{1+1}$. Concentrating first on the Carrollian boost and the temporal special conformal transformation (TSCT, the quadratic super-translation), we show (under an assumption on the Lagrangian) that:
\begin{itemize}
\item Contrary to the claim made in \cite{deBoer:2021jej}, invariance only under the Carrollian boost (and space-time translation) does not always permit a Belinfante-improvement to a vanishing $T^x_{\hspace{1.5mm}t}$ component. This result is in agreement with the conclusion made in \cite{Petkou:2022bmz}.
\item Rather, it is the TSCT symmetry that guarantees $T^x_{\hspace{1.5mm}t}=0$.
\item Moreover, the conditions for the TSCT invariance include the requirement of the Carrollian boost symmetry. Thus, TSCT invariance implies Carrollian boost invarinace but the converse is not true.
\end{itemize}

\medskip

In \cite{Baiguera:2022lsw}, the consequence of the local (i.e. space-dependent) Carrollian boost invariance of the action of a classical theory in a general curved Carrollian background of arbitrary dimensions was investigated. It was shown that the EM tensor off-shell satisfies such a condition that, for flat Carrollian background, simplifies into $T^i_{\hspace{1.5mm}t}=0$. Our conclusion agrees with that since the $1+1$D TSCT is already a local flat-Carrollian boost.

\medskip

We repeat the analysis involving the dilation and the spatial special conformal transformation (SSCT) and show that:
\begin{itemize}
\item Dilation invariance does not guarantee the existence of a trace-less EM tensor. The relativistic counterpart of this result is detailed in \cite{Polchinski:1987dy}.
\item SSCT invariance implies the trace-less condition $T^\mu_{\hspace{1.5mm}\mu}=0$.
\item In addition to that, SSCT symmetry `predicts' the Carrollian boost and the dilation invariance but the converse is not true.
\end{itemize}  
We have explicitly found an action that is invariant under the other global CC transformations but the SSCT. Construction of this example is rather easy in comparison to finding an example \cite{Riva:2005gd} of an action invariant under Euclidean transformations and dilation but not under the special conformal transformations (SCTs).  

\medskip

Finally, moving on to the full CCA$_{1+1}$ invarince, we show that the conserved Noether currents corresponding to the $1+1$D Carrollian conformal (CC) transformations can all be expressed in a very simple schematic form:
\begin{align*}
j^\mu_{\hspace{1.5mm}a}={T}^\mu_{\hspace{1.5mm}\nu}f^\nu_{\hspace{1.5mm}a}
\end{align*}
where $f^\mu_{\hspace{1.5mm}a}$ is the vector field of a CC transformation.

\medskip

We then investigate the $1+1$D CC Ward identities where the above form of the currents plays a crucial role. To derive the Ward identities, we introduce the notion of complexification of the space coordinate, thus paving the way to make use of the complex contour integral techniques in $1+1$D CCFTs. It also marks the beginning of manifestly different treatments for the time and space coordinates. We first derive the general forms of the $1+1$D super-translation and super-rotation Ward identities. Surprisingly, a notion of time-ordering inside the correlation functions involved automatically arises through a temporal $\theta$-function in these Ward identities. We then specialize to the specific cases involving only primary and quasi-primary fields, respectively. Except the explicit representation of the fields under the Carrollian boost and the $\theta$-functions, these results agree with those derived in \cite{Hijano:2018nhq} and mentioned in \cite{Hao:2021urq} in the context of BMS$_3$ field theory.

\medskip

Building on the hint of the time-ordering through the temporal $\theta$-functions in the Ward identities, we formulate the operator formalism for $1+1$D CCFTs. We show that there is no need to perform a `radial quantization' by assuming a `plane-to-cylinder' map \cite{Bagchi:2013qva}. Rather it is the magic of the temporal $\theta$-function that executes a vital role in relating the operator product expansions (OPEs) with the operator commutation relations via a contour integral over the complex variable $x$. This relation has been implicitly used in the literature, e.g. in \cite{Hao:2021urq}, as it is what is expected from the `limiting' procedure. But, to the best of our knowledge, an intrinsic derivation using only the principles of $1+1$D Carrollian physics has not appeared before.

\medskip

However useful the temporal $\theta$-functions may be, they make it difficult to use the algebraic properties of the OPEs. To rectify this issue, we derive a novel $i\epsilon$-form of the Ward identities and the OPEs. This form, while retaining the signatures of time-ordering, also makes it easier to analytically continue those quantities back to the real $x$. With this prescription at our disposal, we are finally able to fix the contours in the definition of the quantum conserved charge operators. Note that the problem of determining this contour is a non-trivial one, since in the quantum theory $x$ is being treated as a complex variable whereas in the classical theory, $x$ is a real variable. It is worth emphasizing that, though similar definitions have already been anticipated in the literature in the BMS$_3$ context \cite{Hijano:2018nhq, Hao:2021urq}, we provide a mathematically consistent understanding of the same purely from the perspective of Carrollian physics.

\medskip

We now turn to the explicit calculation of the 2-point and 3-point (time-ordered) correlation functions of the quasi-primary fields. Like the relativistic 2D CFT, the generic coordinate dependence of these correlators are completely determined by symmetry arguments. Our results here are similar to those obtained in \cite{Chen:2020vvn} with $\xi=0$. We then re-express the same in the $i\epsilon$-form.

\medskip

We want to point out that recently, in the context of flat holography, the (modified-)Mellin transformations of the $1+3$D space-time scattering amplitudes of mass-less particles have been shown in \cite{Bagchi:2022emh} to be equal to the $1+2$D CCFT correlators that have spatial Dirac-delta function factors. This kind of CCFT correlators was earlier observed in \cite{deBoer:2021jej,Chen:2021xkw}. But, in the BMS$_3$ free scalar model \cite{Hao:2021urq}, if the following basic correlator \cite{Chen:2021xkw}:
\begin{align*}
\left\langle\phi(t_1,x_1)\phi(t_2,x_2)\right\rangle=t_{12}\delta(x_{12})
\end{align*}
is used to derive the correlators between other fields, e.g. the primaries $\partial_t\phi$ and $\partial_x\phi$ or the EM tensor components, the results do not agree with those obtained from the Ward identities. So, we do not consider such correlators in this work. It will be very interesting to resolve this contradiction in future.

\medskip

Finally, we further explore the operator formalism. Motivated by the structure of the 2-point quasi-primary correlators, we write down the mode-expansions for arbitrary quasi-primary multiplet fields. After that, we look into the properties of the quantum EM tensor. We begin by deriving the mode-expansions of the EM tensors, taking cues from the classical and quantum conservation laws. Our results agree with those obtained in \cite{Bagchi:2010vw} in the Galilean context. That the EM tensor modes are the generators of the $1+1$D CC transformations in the space of quantum fields is shown next. We now derive the OPEs between the EM tensor components starting from the general form of the super-translation and super-rotation Ward identities and using nothing but the expected bosonic exchange property between two EM tensor components and the assumption that no field in the theory possesses a negative scaling dimension. The resulting $TT$ OPEs that resembles the form earlier appeared in \cite{Bagchi:2021gai} reveals that:
\begin{itemize}
\item In $1+1$D CCFTs, the EM tensor components $T^t_{\hspace{1.5mm}x}$ and $T^t_{\hspace{1.5mm}t}$ form a rank-$\frac{1}{2}$ quasi-primary multiplet with Carollian boost charge\footnote{Our definition of the boost charge differs from the one in \cite{Chen:2020vvn}.} $\xi=2$ and scaling dimension $\Delta=2$.
\end{itemize}
From the $TT$ OPEs, we then show that the EM tensor modes indeed generate the following centrally extended (quantum) version of the CCA$_{1+1}$ that is isomorphic to the centrally extended quantum BMS$_3$ algebra:
\begin{align*}
&i\left[M_n\hspace{1mm},\hspace{1mm}M_m\right]=0\nonumber\\
&i\left[L_n\hspace{1mm},\hspace{1mm}M_m\right]=(n-m)M_{n+m}+\frac{C_2}{12}(n^3-n)\delta_{n+m,0}\\
&i\left[L_n\hspace{1mm},\hspace{1mm}L_m\right]=(n-m)L_{n+m}-i\frac{C_1}{12}(n^3-n)\delta_{n+m,0}\nonumber
\end{align*}
where the constants $C_1$ and $C_2$ are called the central charges of this algebra. First appearing in \cite{Barnich:2006av}, this central extension, with $C_1=0$, arose as the Poisson algebra of (classical) BMS$_3$ charges in the context of Einstein gravity.

\medskip

We conclude this work by defining a hermitian conjugation relation for the quasi-primary multiplet fields, in the same spirit as the BPZ conjugation \cite{Belavin:1984vu} in 2D relativistic CFTs. The $1+1$D Carrollian conformal inversion transformation is at the core of this construction. By extending the `stereographic projection of the circle $(\theta)$ onto the line $(x)$' to the following Carrollian conformal transformation:
\begin{align*}
\theta\rightarrow x=-\cot\frac{\theta}{2}\hspace{2.5mm};\hspace{2.5mm}\tau\rightarrow t=\frac{\tau}{2}\csc^2\frac{\theta}{2}
\end{align*}
we interpret the inversion transformation in the $(t,x)$ space-time as the space-time reflection in the $(\tau,\theta)$ coordinates. 

\medskip

It is then seen that the vacuum-expectation-value (VEV) of a product of a hermitian conjugated multiplet $\left[\bm{\Phi}(t,x)\right]^\dagger$ with an ordinary multiplet $\bm{\Phi^\prime}(t,x)$ does not depend on $t$ or $\tau$. This fact, combined with the above mentioned reflection property, may have an important consequence on the $1+2$D flat holography. Finally, from the conjugation relation for the EM tensor, we extract the following hermitian conjugation properties of the EM tensor modes:
\begin{align*}
L_n^\dagger=(-)^{n+1}L_{-n}\hspace{4.5mm}\text{ and }\hspace{4.5mm}M_n^\dagger=(-)^{n}M_{-n}
\end{align*}  
which are a little different from the standard BPZ mode-conjugation relations \cite{Belavin:1984vu} arising in the radial-quantization of 2D relativistic CFTs. 

\medskip

The rest of the paper is organized as follows. Section \ref{97} contains the study on the Carrollian conformal transformation properties of the fields and the transformations themselves. In section \ref{98}, we introduce the notion of the Carrollian multiplets transforming under the indecomposible representations of the classical Carrollian boost generator. We study the infinitesimal transformation properties of the Carrollian multiplets, under the $1+1$D Carrollian conformal group, in section \ref{99}. This helps us deduce the corresponding finite transformation rules in section \ref{100} where we also define the Carrollian quasi-primary and primary multiplet fields according to their $1+1$D CC transformation properties. In section \ref{101}, we are concerned with classical properties of the EM tensor of a CCA$_{1+1}$ invariant local field theory in a $1+1$D flat Carrollian background. The conditions and consequences of the Carrollian boost and TSCT invariance are explored in section \ref{102} and of the dilatation and the SSCT invariance in section \ref{103}. The consequences of the full CCA$_{1+1}$ symmetry is then discussed in section \ref{104}. We move on to the quantum aspects of this symmetry in section \ref{105}, by deriving the general form of the super-translation and super-rotation Ward identities in $1+1$D. In sections \ref{106} and \ref{107}, we specialize in the Ward identities involving only primary and quasi-primary fields respectively. Section \ref{84} is devoted towards the establishment of the relation between OPEs and operator commutation relations in $1+1$D CCFTs, thus opening the door to the operator formalism. In section \ref{108}, we derive the more convenient $i\epsilon$-form of the OPEs and the Ward identities and finalize the definition of the quantum charge operator. Next, we extract, using only symmetry arguments, the general structure of the 2-point and 3-point quasi-primary correlators in sections \ref{110} and \ref{111} respectively. Finally, in section \ref{112}, we further explore the operator formalism. Having proposed a mode-expansion of the quasi-primary fields in section \ref{113}, we establish various properties of the quantum EM tensor from the first principles in section \ref{114}, e.g. mode-expansion, $TT$ OPEs etc and show that the algebra of the EM tensor modes is the centrally extended CCA$_{1+1}$ or the BMS$_3$ algebra. We conclude in section \ref{131} by defining the hermitian conjugation relation for the quasi-primary multiplets and applying that to find the conjugation properties of the EM tensor modes.

\medskip

\section{Transformation Properties of Carrollian Fields}\label{97}
We first review some of the properties of the $1+1$D (level 2) Carrollian conformal (CC) transformations.

\medskip

In \cite{Hijano:2017eii}, it was shown by finding an explicit map between the generators that the $1+2$D Poincare algebra is isomorphic to the global subalgebra of the $1+1$D Galilean conformal algebra. As we have stated earlier, in $1+1$D the Galilean and the Carrollian conformal algebras are also isomorphic. This isomorphism should extend to the respective six-parameter groups.

\medskip

In $1+1$D, the Carroll group \cite{LL,SG} is formed by the space-time translations and the Carrollian boost. These transformations augmented by dilation, TSCT and SSCT generate the $1+1$D Carrollian conformal group. All of these transformations are collectively expressed as:
\begin{align}
x\rightarrow x^\prime=\frac{ax+b}{cx+d}\text{ { },{ } }t\rightarrow t^\prime=\frac{t}{{(cx+d)}^2}+\lambda x^2+\mu x+\nu\label{56}
\end{align}
with $a,b,c,d,\lambda,\mu,\nu\in \mathbb{R}$ and $ad-bc=1$.

\medskip

Strictly speaking, to form a group, the group transformations must be invertible. Barring the SSCT, the remaining five are fine in this regard. To include the SSCT, defined as:
\begin{align}
x\rightarrow x^\prime=\frac{x}{1-ax}\text{ { },{ } }t\rightarrow t^\prime=\frac{t}{{(1-ax)}^2}\label{126}
\end{align}
as a group element by making it invertible, we must compactify \cite{Oblak:2015qia} the domain of the space coordinate $x$ from the real line $\mathbb{R}$ to the `Riemann circle' $S^1\simeq\mathbb{R}\cup\{\infty\}$. Conformally compactifying the time coordinate is unnecessary (and does not bring in new advantages) since the `compactification arithmetic' on $S^1$ does not apply on $t$. Thus, the transformations \eqref{56} form the global Carrollian conformal group of $\mathbb{R}_t\times S^1_x$ .

\medskip 

As is usual in physics, one is more interested in local aspects of transformations. As noted earlier, the $1+1$D CC transformations, not necessarily globally defined\footnote{For a transformation to be globally defined, we shall also demand globally non-singular behavior of the corresponding generators \cite{Blu:2009zz}. In this sense, the other super-translations not included in \eqref{56} fail to be globally defined.} on $\mathbb{R}\times S^1$, have the following finite form:
\begin{align}
x\rightarrow x^\prime=F(x)\hspace{2mm},\hspace{2mm}t\rightarrow t^\prime=tF^\prime(x)+G(x)\label{55}
\end{align}
with the arbitrary functions $F(x)$ and $G(x)$ being known as the super-rotations and the super-translations respectively. The infinitesimal versions of these transformations, compactly expressed as $x^{\mu}\rightarrow x^{\mu}+\epsilon^af^{\mu}_{\hspace{1.5mm}a}(\mathbf{x})$, `minimally' is (with $\epsilon^af^{\mu}_{\hspace{1.5mm}a}(\mathbf{x})$ being `infinitesimal'):
\begin{align}
&x\rightarrow x+\epsilon^xf(x)\text{  }\text{  }\text{ and }\text{  }\text{  }t\rightarrow t+\epsilon^xtf^\prime(x)+\epsilon^tg(x)\label{eq:4}\\
\Rightarrow\hspace{1.5mm}\hspace{1.5mm} & f^x_{\hspace{1.5mm}x}=f(x)\hspace{1.5mm},\hspace{1.5mm}f^x_{\hspace{1.5mm}t}=0 \text{\hspace{1.5mm}(`minimal')}\hspace{1.5mm},\hspace{1.5mm}f^t_{\hspace{1.5mm}x}=tf^\prime(x)\hspace{1.5mm},\hspace{1.5mm}f^t_{\hspace{1.5mm}t}=g(x)\nonumber
\end{align}
From \eqref{56}, it is clear that the global infinitesimal $1+1$D CC transformations are given by such $f(x)$ and $g(x)$ that are at most quadratic polynomials in $x$. Since, now $x\in\mathbb{R}\cup\{\infty\}$, we can have the following power series expansions for general $f(x)$ and $g(x)$ around e.g. $x=0$ :
\begin{align}
f(x)=\sum_{n\in\mathbb{Z}}a_nx^{n+1}\hspace{2.5mm};\hspace{2.5mm}g(x)=\sum_{n\in\mathbb{Z}}b_nx^{n+1}
\end{align}
with $\{a_n\}$ and $\{b_n\}$ being real numbers.

\medskip

Next, we consider a multi-component field transforming under \eqref{eq:4}, as a finite matrix representation, schematically as ($i$ denote collection of suitable indices):
\begin{align}
\Phi^i(\mathbf{x})\rightarrow{\tilde{\Phi}}^i(\mathbf{x^\prime})=\Phi^i(\mathbf{x})+\epsilon^a{(\mathcal{F}_a\cdot\Phi)}^i(\mathbf{x})\label{eq:5}
\end{align}
The generator of the above transformations is defined as \cite{DiFrancesco:1997nk}:
\begin{align}
\delta_{\bm{\epsilon}}\Phi^i(\mathbf{x})\equiv{\tilde{\Phi}}^i(\mathbf{x})-{{\Phi}}^i(\mathbf{x}):=-i\epsilon^aG_a(\mathbf{x})\Phi^i(\mathbf{x})\equiv -i\epsilon^aG_a\Phi^i(\mathbf{x})\label{73}
\end{align}
so that the generators are explicitly given by:
\begin{align}
-iG_a(\mathbf{x})\Phi^i(\mathbf{x})={(\mathcal{F}_a\cdot\Phi)}^i(\mathbf{x})-f^{\mu}_{\hspace{1.5mm}a}(\mathbf{x})\partial_\mu\Phi^i(\mathbf{x})\label{119}
\end{align}

\medskip

Thus, the generator of an infinitesimal space-time transformation $x^{\mu}\rightarrow x^{\mu}+\epsilon^af^{\mu}_{\hspace{1.5mm}a}(\mathbf{x})$ in the space of ordinary functions $\phi(\mathbf{x})$ (i.e. having ${(\mathcal{F}_a\cdot\phi)}(\mathbf{x})=0$) is obtained as:
\begin{align}
-i\epsilon^a\mathcal{G}_a\phi(\mathbf{x})=\phi(\mathbf{x}-\epsilon^a\mathbf{f}_{a}(\mathbf{x}))-\phi(\mathbf{x})\hspace{2.5mm}\Longrightarrow\hspace{2.5mm}\mathcal{G}_a(\mathbf{x})=-if^{\mu}_{\hspace{1.5mm}a}(\mathbf{x})\partial_\mu
\end{align}
Thus we have the following generators of the $1+1$D CC transformations (with $n\in\mathbb{Z}$) in the space of functions:
\begin{align}
&\text{the $x^{n+1}$ super-rotation is generated by: $\mathcal{L}_n=-ix^{n+1}\partial_x-i(n+1)x^nt\partial_t$}\label{109}\\
&\text{the $x^{n+1}$ super-translation is generated by: $\mathcal{M}_n=-ix^{n+1}\partial_t$}\label{115}
\end{align}
These generators were obtained in \cite{Bagchi:2009my} in the context of 2D Galilean conformal transformations.

\medskip

It is now easy to verify that these differential generators satisfy the CCA$_{1+1}$, as stated before:
\begin{align}
\left[\mathcal{L}_n\hspace{1mm},\hspace{1mm}\mathcal{L}_m\right]=i(n-m)\mathcal{L}_{n+m}\hspace{2.5mm};\hspace{2.5mm}\left[\mathcal{L}_n\hspace{1mm},\hspace{1mm}\mathcal{M}_m\right]=i(n-m)\mathcal{M}_{n+m}\hspace{2.5mm};\hspace{2.5mm}\left[\mathcal{M}_n\hspace{1mm},\hspace{1mm}\mathcal{M}_m\right]=0\label{116}
\end{align}

\medskip

From the explicit forms \eqref{109} and \eqref{115}, it is evident that the super-rotation and super-translation generators are singular at $x=0$ for $n<-1$. By performing the following global $1+1$D CC transformation:
\begin{align*}
x\rightarrow x^\prime=-\frac{1}{x}\text{ { },{ } }t\rightarrow t^\prime=\frac{t}{x^2}
\end{align*}
one concludes, on the other hand, that these generators are singular at $x=\infty$ for $n>1$. Thus, we again recover the fact that $\mathcal{L}_n$ and $\mathcal{M}_n$ with $n\in\{0,\pm1\}$ generate the $1+1$D global CC transformations of functions defined on $\mathbb{R}\times S^1$.

\medskip

We now turn our attention to the transformation properties of the Carrollian fields.

\medskip 

\subsection{Carrollian multiplets}\label{98}
The Carroll algebra in $1+1$D is given by \cite{LL}:
\begin{align}
\left[\mathbf{P}\hspace{1mm},\hspace{1mm}\mathbf{H}\right]=0\hspace{2.5mm};\hspace{2.5mm}\left[\mathbf{H}\hspace{1mm},\hspace{1mm}\mathbf{B}\right]=0\hspace{2.5mm};\hspace{2.5mm}\left[\mathbf{B}\hspace{1mm},\hspace{1mm}\mathbf{P}\right]=i\mathbf{H}
\end{align}
where $\mathbf{P}$, $\mathbf{H}$ and $\mathbf{B}$ respectively are the generators of the space-translation, time-translation and Carrollian boost. Out of these three transformations, only the Carrollian boost leaves the origin $(t,x)=(0,0)$ invariant. We note that there is no spatial rotation in $1+1$D.

\medskip

So, we investigate in detail the structure of the Carrollian boost which will form the basis of the construction of the Carrollian tensors, in the same way that one uses the Lorentz generators to define the transformation laws of the tensors in the relativistic setting. The following construction will be relevant even for a (non-conformal) Carrollian field theory.

\medskip

Before starting, we quickly note that under space-time translation, $x\rightarrow x^\prime=x+a,t\rightarrow t^\prime=t+b$, a (possibly multi-component) field transforms as:
\begin{align}
\mathbf{\Phi}(t,x)\longrightarrow\mathbf{\tilde{\Phi}}(t^\prime,x^\prime)=\mathbf{\Phi}(t,x)\label{123}
\end{align}

\medskip

In 1+1 space-time dimension, the `plane' Carrollian boost transformation (pCB) is defined as: $x\rightarrow x^\prime=x\text{ , }t\rightarrow t^\prime=t+vx${ }; or equivalently, as:
\begin{align}
\begin{pmatrix}
x\\
t
\end{pmatrix}\longrightarrow \begin{pmatrix}
x^\prime\\
t^\prime
\end{pmatrix}= \left[\exp{\begin{pmatrix}
0 & 0\\
v & 0
\end{pmatrix}}\right]\begin{pmatrix}
x\\
t
\end{pmatrix}\text{ }\Longleftrightarrow\text{ } x^\mu\rightarrow {x^{\prime}}^\mu={\left[e^{v\mathbf{B_{(2)}}}\right]}^{\mu}_{\hspace{2mm}\nu}\text{ }x^\nu
\end{align}
where 
\begin{align}
\mathbf{B_{(2)}}:=\begin{pmatrix}
0 & 0\\
1 & 0
\end{pmatrix}
\end{align}
is the 2D representation of the pCB generator $\mathbf{B}$ which is clearly not diagonalizable as its only generalized eigenvalue 0 has geometric multiplicity 1. Moreover, it will be evident below that indecomposable (but reducible) representations of $\mathbf{B}$ of dimension $\geq2$ all have 0 as their only generalized eigenvalue and it has geometric multiplicity 1. Since no matrix representation of $\mathbf{B}$ is thus diagonalizable, it can not have any non-trivial 1D (hence, irreducible) `plane' representation relevant for local field theory in $1+1$-dimensional space-time (i.e. $\mathbf{B_{(1)}}\equiv0$).

\medskip

Under the $1+1$ dimensional pCB, a rank-$n$ Carrollian Cartesian tensor field $\Phi$ with `boost-charge' $\xi$ transforms, similarly as the Lorentz covariance of Lorentz tensors, as:
\begin{align}
&{\Phi}^{\mu_1...\mu_n}(t,x)\longrightarrow{\tilde{\Phi}}^{\mu_1...\mu_n}(t^\prime,x^\prime)={\left[e^{-\xi v\mathbf{B_{(2)}}}\right]}^{\mu_1}_{\hspace{3mm}\nu_1}...{\left[e^{-\xi v\mathbf{B_{(2)}}}\right]}^{\mu_n}_{\hspace{3mm}\nu_n}{\Phi}^{\nu_1...\nu_n}(t,x)\nonumber\\
\Longleftrightarrow\qquad &{\mathbf{\Phi}}(t,x)\longrightarrow{{\mathbf{\tilde{\Phi}}}}(t^\prime,x^\prime)=\left[\bigotimes_{i=1}^n e^{-\xi v\mathbf{B_{(2)}}}\right]{\mathbf{\Phi}}(t,x)=e^{-\xi v\bigoplus\limits_{i=1}^n\mathbf{B_{(2)}}}{\mathbf{\Phi}}(t,x)\label{eq:1}
\end{align}
where $\mu_i, \nu_i$ are Carrollian space-time indices and for matrices, the left index denotes row while the right one denotes column; repeated indices are summed over and, in \eqref{eq:1}, indices are suppressed. We point out that the up/down appearance of a tensor-index is unimportant; only the left/right ordering matters. Clearly, the Carrollian scalar fields which, by definition, are invariant under pCB, must have $\xi=0$ (but formally, they can transform under any dimensional indecomposable representation of the pCB).

\medskip

Since the Carrollian Cartesian tensors defined above are decomposable, we now construct indecomposible Carrollian multiplets from these tensors. We begin by recognizing that:
\begin{align}
\mathbf{B_{(2)}}=\mathbf{J}_{(l=\frac{1}{2})}^\mathbf{-}
\end{align}
which is the lowering ladder operator in the $\text{su}(2)$ spin-$\frac{1}{2}$ representation. Thus, $\bigoplus\limits_{i=1}^n\mathbf{B_{(2)}}$ in \eqref{eq:1} can be decomposed into indecomposable representations of $\mathbf{J}^\mathbf{-} $ using the technique of `addition of $n$ spin-$\frac{1}{2}$ angular momenta' in quantum mechanics, such that:
\begin{align}
\mathbf{B_{(d)}}\equiv\mathbf{J}_{(l=\frac{d-1}{2})}^\mathbf{-}\label{118}
\end{align} 
A multi-component field transforming under the $d$-dimensional representation of pCB, $\mathbf{B_{(d)}}$, will be called a Carrollian multiplet of rank $\frac{d-1}{2}$ with $d$ components, denoted by 
\begin{align*}
\Phi_{(l=\frac{d-1}{2})}^m \text{\hspace{2.5mm} with \hspace{2.5mm}} m=\frac{1-d}{2},\frac{3-d}{2},...,\frac{d-1}{2}
\end{align*}
By treating the $\mu=t$ index as spin-$\frac{1}{2}$ up-state and the $\mu=x$ index as spin-$\frac{1}{2}$ down-state, components $\Phi_{(l)}^m$ of a Carrollian multiplet arise precisely as such linear combinations (with proper Clebsch-Gordon coefficients) of the components of a Cartesian tensor of an allowed rank $n$ that would appear while expanding the $|l,m\rangle$ states in an allowed $|s_1,s_2,...,s_n\rangle$ basis (where $|s_i\rangle$ are $\mathbf{J}_{(\frac{1}{2})}^\mathbf{z}$ eigenstates). So, as a linear combination of the components of a rank-$n$ Cartesian tensor, one can obtain multipltes of ranks: $0,1,2,...,\frac{n}{2}$ for even $n$ and $\frac{1}{2},\frac{3}{2},\frac{5}{2},...,\frac{n}{2}$ for odd $n$. As an example, we see how Carrollian multiplets of ranks $\frac{1}{2}$ and $\frac{3}{2}$ are constructed from a rank-3 Cartesian tensor:
\begin{align}
&\Phi_{(\frac{3}{2})}^{\frac{3}{2}}(t,x):=\Phi^{ttt}(t,x)\text{ }\text{ }\qquad\text{ }\text{ }\Phi_{(\frac{3}{2})}^{\frac{1}{2}}(t,x):=\frac{1}{\sqrt{3}}\left[\Phi^{ttx}+\Phi^{txt}+\Phi^{xtt}\right](t,x)\nonumber\\
&\Phi_{(\frac{3}{2})}^{-\frac{1}{2}}(t,x):=\frac{1}{\sqrt{3}}\left[\Phi^{txx}+\Phi^{xtx}+\Phi^{xxt}\right](t,x)\text{ }\text{ }\qquad\text{ }\text{ }\Phi_{(\frac{3}{2})}^{-\frac{3}{2}}(t,x):=\Phi^{xxx}(t,x)\label{54}\\
&\Phi_{(\frac{1}{2})}^{\frac{1}{2}}(t,x):=\frac{1}{\sqrt{a^2+b^2+c^2}}\left[a\Phi^{ttx}+b\Phi^{txt}+c\Phi^{xtt}\right](t,x) \text{ }\text{ }\qquad\text{with $a+b+c=0$}\nonumber\\
&\Phi_{(\frac{1}{2})}^{-\frac{1}{2}}(t,x):=\frac{1}{\sqrt{a^2+b^2+c^2}}\left[a\Phi^{xxt}+b\Phi^{xtx}+c\Phi^{txx}\right](t,x)\nonumber
\end{align}
(As the tuple $(a,b,c)$ in $\mathbb{R}^3$ lies on the plane $a+b+c=0$ which is spanned by two basis vectors, two linearly independent rank-$\frac{1}{2}$ multiplets arise.)

\medskip

A rank-$l$ Carrollian multiplet with boost-charge $\xi$ transforms under the $2l+1$ dimensional representation of the pCB as:
\begin{align}
\Phi_{(l)}^m(t,x)\longrightarrow\tilde{\Phi}_{(l)}^m(t^\prime,x^\prime)={\left[e^{-\xi v\mathbf{J}_{(l)}^{\mathbf{-}}}\right]}^m_{\hspace{2mm} m^\prime}{\Phi}_{(l)}^{m^\prime}(t,x)\label{eq:2}
\end{align}

\medskip

\textbf{Comment 1:} Since the finite dimensional indecomposable representations of $\mathbf{B}$ are not symmetric (or Hermittian), one can start with:
\begin{align*}
\begin{pmatrix}
t\\
x
\end{pmatrix}\longrightarrow \begin{pmatrix}
t^\prime\\
x^\prime
\end{pmatrix}= \left[\exp{\begin{pmatrix}
0 & v\\
0 & 0
\end{pmatrix}}\right]\begin{pmatrix}
t\\
x
\end{pmatrix}\text{ }\Longleftrightarrow\text{ } x^\mu\rightarrow {x^{\prime}}^\mu={\left[e^{v\mathbf{B^\prime_{(2)}}}\right]}^{\mu}_{\hspace{2mm}\nu}\text{ }x^\nu
\end{align*}
where 
\begin{align*}
\mathbf{B^\prime_{(2)}}:=\begin{pmatrix}
0 & 1\\
0 & 0
\end{pmatrix}
\end{align*}
and follow the preceding argument to conclude that:
\begin{align*}
\mathbf{B^\prime_{(d)}}\equiv\mathbf{J}_{(l=\frac{d-1}{2})}^\mathbf{+}
\end{align*}
which is the $\text{su}(2)$ raising ladder operator. But, as $\mathbf{J}_{(l)}^\mathbf{+}={(\mathbf{J}_{(l)}^\mathbf{-})}^{\textbf{T}}$, the raising and lowering operators' representation matrices are related to each other by the similarity transformation: 
\begin{align*}
S=\text{anti-diag }{(1,1,...,1)}_{2l+1}
\end{align*}
and consequently, $\mathbf{B}$ and $\mathbf{B^\prime}$ furnish two equivalent representations of the pCB generator.

\medskip

\textbf{Comment 2:} After constructing the Carrollian multiplets from the Cartesian tensors as demonstrated in \eqref{54}, one can always redefine the components of the multiplets by simply multiplying appropriate numbers with them such that:
\begin{align*}
\text{in \eqref{eq:2}, }\mathbf{J}_{(l)}^{\mathbf{-}}\text{ is replaced by }\mathbf{M}_{(l)}:=\text{sub-diag }{(1,1,...,1)}_{2l+1}\text{ .}
\end{align*}
Thus, instead of the actual $\mathbf{J}_{(l)}^{\mathbf{-}}$ matrix, only the indecomposable Jordan-block structure is important for defining the pCB transformation property of the Carrollian multiplets. However, in the present work, we choose not to do this. 

\medskip

\subsection{Infinitesimal transformations of fields}\label{99} 
From \eqref{56}, it is evident that the subgroup of the $1+1$D Carrollian conformal group that keeps the space-time origin invariant is generated by the Carrollian boost, dilation, TSCT and SSCT. The corresponding subalgebra is obtained by restricting $n$ to the set $\{0,1\}$ in \eqref{116} and is thus given by:
\begin{align}
&\left[\mathbf{D}\hspace{1mm},\hspace{1mm}\mathbf{B}\right]=0\hspace{2.5mm};\hspace{2.5mm}
\left[\mathbf{D}\hspace{1mm},\hspace{1mm}\mathbf{K}_t\right]=-i\mathbf{K}_t\hspace{2.5mm};\hspace{2.5mm}\left[\mathbf{D}\hspace{1mm},\hspace{1mm}\mathbf{K}_x\right]=-i\mathbf{K}_x\nonumber\\
&\left[\mathbf{B}\hspace{1mm},\hspace{1mm}\mathbf{K}_t\right]=0\hspace{2.5mm};\hspace{2.5mm}
\left[\mathbf{K}_x\hspace{1mm},\hspace{1mm}\mathbf{K}_t\right]=0\hspace{2.5mm};\hspace{2.5mm}
\left[\mathbf{K}_x\hspace{1mm},\hspace{1mm}\mathbf{B}\right]=i\mathbf{K}_t\label{117}
\end{align}
where $\mathbf{D}$, $\mathbf{K}_t$ and $\mathbf{K}_x$ are the generators of dilation, TSCT and SSCT respectively. This subalgebra is recognized to be identical to the $1+1$D Carrollian algebra augmented by dilation.

\medskip

Given that the pCB generator $\mathbf{B}$ has the finite-dimensional indecomposable matrix representation \eqref{118} in the Jordan block form, we would like to find the matrix representation of the algebra \eqref{117} corresponding to the `invariant subgroup' of the origin. From the commutation relations, one can easily conclude that while the matrices for $\mathbf{K}_t$ and $\mathbf{K}_x$ must vanish, the matrix representation of the dilation generator $\mathbf{D}$ is proportional to the identity. This last conclusion can also be reached by noticing the action of the dilation generator on space-time:
\begin{align*}
t\rightarrow t^\prime=\lambda t\hspace{1mm},\hspace{1mm}x\rightarrow x^\prime=\lambda x\hspace{2.5mm}\Longrightarrow\hspace{2.5mm} \begin{pmatrix}
t\\
x
\end{pmatrix}\longrightarrow \begin{pmatrix}
t^\prime\\
x^\prime
\end{pmatrix}=e^{(\log\lambda)\mathbf{I}}\begin{pmatrix}
t\\
x
\end{pmatrix}
\end{align*}

\medskip 
  
Thus, among the generators of the $1+1$D Carrollian conformal (CC) group, the pCB generator $\mathbf{B}$ and the space-time dilatation generator $\mathbf{D}$ form the maximal (and the only) set of the mutually commuting generators with finite dimensional matrix representations. This is closely related to the fact that the space-time action of only these two generators can be expressed as linear transformations. Moreover, $\mathbf{D}$ does not suffer from the non-diagonalizability problem unlike $\mathbf{B}$. Thus, it is convenient to perform field theory analysis in the diagonal (`eigenfield') basis of $\mathbf{D}$ because Carrollian multiplets can be simultaneously (and completely) labeled by scaling dimension $\Delta$, pCB rank $l$ and charge $\xi$ in this basis.

\medskip

A Carrollian multiplet or Cartesian tensor field with scaling dimension $\Delta$ transforms under the dilatation, as (suppressing tensor indices):
\begin{align}
\mathbf{\Phi}(t,x)\longrightarrow\mathbf{\tilde{\Phi}}(t^\prime,x^\prime)=\lambda^{-\Delta}\mathbf{\Phi}(t,x)\label{63}
\end{align}

\medskip

Thus, applying \eqref{119} on the infinitesimal versions of \eqref{eq:2} and \eqref{63}, we can deduce the following complete forms of $\mathbf{B}$ and $\mathbf{D}$ generating respectively the pCB and the dilation on the classical Carrollian multiplet fields:
\begin{align}
&\mathbf{B}(t,x)\Phi_{(l)}^m(t,x)=-i\left[\mathbf{I}x\partial_t+\xi\mathbf{J}_{(l)}^{\mathbf{-}}\right]^m_{\hspace{2mm}m^\prime}\Phi_{(l)}^{m^\prime}(t,x)\hspace{2.5mm}\Rightarrow\hspace{2.5mm}\mathbf{B}(\mathbf{0})=-i\xi\mathbf{J}^{\mathbf{-}}\hspace{5mm}\text{(on a multiplet)}\label{121}\\
&\mathbf{D}(t,x)\Phi_{(l)}^m(t,x)=-i\left[x\partial_x+t\partial_t+\Delta\right]\Phi_{(l)}^{m}(t,x)\hspace{2.5mm}\Rightarrow\hspace{2.5mm}\mathbf{D}(\mathbf{0})=-i\Delta\hspace{5mm}\text{(on a multiplet)}
\end{align} 

\medskip

We now proceed to find the actions of the generators $\mathbf{K}_t$ and $\mathbf{K}_x$ on the Carrollian multiplets. Recalling that these two generators  have no non-trivial finite dimensional matrix representation and that they keep the space-time origin invariant, we infer from their action on an arbitrary multiplet located at the origin that\footnote{The R.H.S. of the `$\Rightarrow$' signs in \eqref{121}-\eqref{120} are valid since $\mathbf{B}(\mathbf{0})$, $\mathbf{D}(\mathbf{0})$, $\mathbf{K}_t(\mathbf{0})$ and $\mathbf{K}_x(\mathbf{0})$ act as constant matrix multiplications and do not involve any differentiation.}
\begin{align}
&\mathbf{K}_t(\mathbf{0})\Phi_{(l)}^m(\mathbf{0})=0\hspace{2.5mm}\Rightarrow\hspace{2.5mm}\mathbf{K}_t(\mathbf{0})=0\hspace{5mm}\text{(while operting on multiplets)}\\
&\mathbf{K}_x(\mathbf{0})\Phi_{(l)}^m(\mathbf{0})=0\hspace{2.5mm}\Rightarrow\hspace{2.5mm}\mathbf{K}_x(\mathbf{0})=0\hspace{5mm}\text{(while operting on multiplets)}\label{120}
\end{align}
To find the corresponding actions on a multiplet situated at an arbitrary location in space-time, we make use of the finite actions of the space-time translation generators. Below comes an illustration (notations are as in \eqref{73}):
\begin{align}
&\mathbf{K}_t\Phi_{(l)}^m(t,x)=e^{i\mathbf{P}x+i\mathbf{H}t}\cdot\mathbf{K}_t\Phi_{(l)}^m(0,0)
=\left(e^{i\mathbf{P}x+i\mathbf{H}t}\text{ }\mathbf{K}_t(\mathbf{0})\text{ }e^{-i\mathbf{P}x-i\mathbf{H}t}\right)\Phi_{(l)}^m(t,x)\nonumber\\
\Rightarrow\text{ }&\mathbf{K}_t(t,x)=e^{i\mathbf{P}x+i\mathbf{H}t}\text{ }\mathbf{K}_t(\mathbf{0})\text{ }e^{-i\mathbf{P}x-i\mathbf{H}t}\label{122}
\end{align}
Obviously, this relation is valid for any arbitrary (differential/matrix) operator acting on a number-valued function. 

\medskip

The next step is to use the BCH lemma. For that we need the commutation relations of the generators of translations with those of the invariant subalgebra \eqref{117}; from \eqref{116}, these are found to be:
\begin{align}
&\left[\mathbf{P}\hspace{1mm},\hspace{1mm}\mathbf{B}\right]=-i\mathbf{H}\hspace{2.5mm};\hspace{2.5mm}\left[\mathbf{P}\hspace{1mm},\hspace{1mm}\mathbf{D}\right]=-i\mathbf{P}\hspace{2.5mm};\hspace{2.5mm}
\left[\mathbf{P}\hspace{1mm},\hspace{1mm}\mathbf{K}_t\right]=-2i\mathbf{B}\hspace{2.5mm};\hspace{2.5mm}\left[\mathbf{P}\hspace{1mm},\hspace{1mm}\mathbf{K}_x\right]=-2i\mathbf{D}\nonumber\\
&\left[\mathbf{H}\hspace{1mm},\hspace{1mm}\mathbf{B}\right]=0\hspace{2.5mm};\hspace{2.5mm}\left[\mathbf{H}\hspace{1mm},\hspace{1mm}\mathbf{D}\right]=-i\mathbf{H}\hspace{2.5mm};\hspace{2.5mm}
\left[\mathbf{H}\hspace{1mm},\hspace{1mm}\mathbf{K}_t\right]=0\hspace{2.5mm};\hspace{2.5mm}\left[\mathbf{H}\hspace{1mm},\hspace{1mm}\mathbf{K}_x\right]=-2i\mathbf{B}
\end{align}
Now, using the commutation relations involving $\mathbf{K}_t$ through the BCH lemma, from \eqref{122} we obtain:
\begin{align}
&\mathbf{K}_t(t,x)=\mathbf{K}_t(\mathbf{0})+ix\left[\mathbf{P}\hspace{1mm},\hspace{1mm}\mathbf{K}_t\right](\mathbf{0})-\frac{x^2}{2}\left[\mathbf{P}\hspace{1mm},\hspace{1mm}\left[\mathbf{P}\hspace{1mm},\hspace{1mm}\mathbf{K}_t\right]\right](\mathbf{0})=\mathbf{K}_t(\mathbf{0})+2x\mathbf{B}(\mathbf{0})+x^2\mathbf{H}\nonumber\\
\Rightarrow\text{ }&\mathbf{K}_t\Phi_{(l)}^m(t,x)=-i\left[\mathbf{I}x^2\partial_t+2x\xi\mathbf{J}_{(l)}^{\mathbf{-}}\right]^m_{\hspace{2mm}m^\prime}\Phi_{(l)}^{m^\prime}(t,x)\label{125}
\end{align}
Similarly, the action of the generator $\mathbf{K}_x$ is derived as:
\begin{align}
&\mathbf{K}_x(t,x)=\mathbf{K}_x(\mathbf{0})+2\left(x\mathbf{D}+t\mathbf{B}\right)(\mathbf{0})+x^2\mathbf{P}+2xt\mathbf{H}\nonumber\\
\Rightarrow\text{ }&\mathbf{K}_x\Phi_{(l)}^m(t,x)=-i\left[\mathbf{I}\left(x^2\partial_x+2xt\partial_t+2x\Delta\right)+2t\xi\mathbf{J}_{(l)}^{\mathbf{-}}\right]^m_{\hspace{2mm}m^\prime}\Phi_{(l)}^{m^\prime}(t,x)\label{127}
\end{align} 

\medskip

This concludes the discussion on the action of the generators of the $1+1$D CC group on the classical Carrollian conformal multiplets. Next, we lay out the finite transformation properties of those fields.

\medskip  

\subsection{Finite field transformations}\label{100}
To begin with, we note from \eqref{123}, \eqref{eq:2} and \eqref{63} that under the following finite Carrollian transformation augmented by dilation:
\begin{align}
x\rightarrow x^\prime=\lambda x+x_0\text{ { },{ } }t\rightarrow t^\prime=\lambda t+vx+t_0
\end{align}
a Carrollian multiplet field with pCB charge $\xi$, rank $l$ and scaling dimension $\Delta$ transforms as:
\begin{align}
\Phi_{(l)}^m(t,x)\longrightarrow\tilde{\Phi}_{(l)}^m(t^\prime,x^\prime)=\lambda^{-\Delta}{\left[e^{-\xi \frac{v}{\lambda}\mathbf{J}_{(l)}^{\mathbf{-}}}\right]}^m_{\hspace{2mm} m^\prime}{\Phi}_{(l)}^{m^\prime}(t,x)
\end{align}
This combined transformation rule can be re-expressed in the parameter-free form as:
\begin{align}
\tilde{\Phi}_{(l)}(t^\prime,x^\prime)={\left[\frac{x^\prime(x+h_x)-x^\prime(x)}{h_x}\right]}^{-\Delta}{\exp\left[{-\frac{\frac{t^\prime(t,x+h_t)-t^\prime(t,x)}{h_t}}{\frac{x^\prime(x+h_x)-x^\prime(x)}{h_x}}\text{ }\xi\mathbf{J}_{(l)}^{\mathbf{-}}}\right]}\cdot{\Phi}_{(l)}(t,x)\label{124}
\end{align}
where $h_x$ and $h_t$ are arbitrary non-zero real numbers. Since this finite transformation rule does not involve the transformation parameters themselves, it should be valid (with possibly some restrictions on $h_x$ and $h_t$) for all the transformations included in the $1+1$D CC group.

\medskip

Substituting the infinitesimal version (with small $a$) of TSCT defined as:
\begin{align}
x\rightarrow x^\prime=x\text{ { },{ } }t\rightarrow t^\prime=t+ax^2
\end{align}
into \eqref{124} and comparing with the infinitesimal transformation \eqref{125}, we discover that for consistency, the condition: $h_t\rightarrow0$ is required. Repeating this procedure for the infinitesimal version of SSCT \eqref{126} and comparing with \eqref{127}, one obtains another consistency condition: $h_x\rightarrow0$.

\medskip
 
Thus, under the global $1+1$D CC transformations \eqref{56}, a Carrollian conformal multiplet field transforms as following, as deduced from \eqref{124}:
\begin{align}
&\tilde{\Phi}_{(l)}(t^\prime,x^\prime)=\lim\limits_{h_x\rightarrow0}{\lim\limits_{h_t\rightarrow0}}\text{ }{\left[\frac{x^\prime(x+h_x)-x^\prime(x)}{h_x}\right]}^{-\Delta}{\exp\left[{-\frac{\frac{t^\prime(t,x+h_t)-t^\prime(t,x)}{h_t}}{\frac{x^\prime(x+h_x)-x^\prime(x)}{h_x}}\text{ }\xi\mathbf{J}_{(l)}^{\mathbf{-}}}\right]}\cdot{\Phi}_{(l)}(t,x)\nonumber\\
\Rightarrow\text{ }&\Phi_{(l)}^m(t,x)\longrightarrow\tilde{\Phi}_{(l)}^m(t^\prime,x^\prime)={(\partial_x x^\prime)}^{-\Delta}{\left[e^{-\xi \frac{\partial_x t^\prime}{\partial_x x^\prime}\mathbf{J}_{(l)}^{\mathbf{-}}}\right]}^m_{\hspace{2mm} m^\prime}{\Phi}_{(l)}^{m^\prime}(t,x)\label{eq:3}
\end{align}
The $1+3$D relativistic counterpart (applicable to tensor fields) of this transformation rule is discussed in \cite{Weinberg:2010fx}.  

\medskip

In $1+1$D, if a Carrollian multiplet field transforms as above under any arbitrary $1+1$D CC transformation \eqref{55}, it will be called a $1+1$D CC primary field with pCB rank $l$, boost-charge $\xi$ and scaling dimension $\Delta$. On the other hand, the multiplets that obey the above transformation rule only for the global transformations \eqref{56}, are known as the $1+1$D CC quasi-primary fields\footnote{To be precise, the quantum operators corresponding to these fields obey bosonic statistics; there is an extra sign-function factor \cite{Chen:2020vvn} in the transformation rule for the Fermionic operators (which will not be dealt with here).}.

\medskip

\section{Classical EM Tensor}\label{101}
We want to investigate on the consequences for a field theory (on $1+1$D space-time) possessing the global (subalgebra of) CCA$_{1+1}$ as the kinematical symmetry algebra of the action. Classically, by Noether's theorem, each of the continuous symmetries of the action must have an associated on-shell conserved current. The conserved charges constructed out of these currents give a dynamical realization of the continuous symmetries of the action. The conserved Noether currents corresponding to the kinematical symmetries are related to the EM tensor of the theory. Hence, we now take a look at the classical properties of the EM tensor of a global CCA$_{1+1}$ invariant theory.

\medskip

We consider a classical theory of fields in a $1+1$D flat Carrollian space-time. The `fundamental' fields that appear in the Lagrangian are postulated\label{85} to transform as multiplets (i.e. they have tensorial transformation property) under the full $1+1$D global CC group\footnote{For the relativistic case, similar assumption was made in \cite{Mack:1969rr}.}. In the language introduced in the previous section, these fields are assumed to be $1+1$D CC quasi-primary fields.

\medskip

Let the action describing the classical theory be, following \cite{Bagchi:2019clu}:
\begin{align}
S[\bm{\Phi}]=\int dt\int dx\text{ }\mathcal{L}(\bm{\Phi},\partial_\mu\bm{\Phi})\label{70}
\end{align}
This action transforms on-shell under any space-time transformation \eqref{eq:4} as \cite{DiFrancesco:1997nk}:
\begin{align}
S[\bm{\Phi}]\rightarrow S^\prime[\bm{\tilde{\Phi}}]=S[\bm{\Phi}]+\int dt\int dx\text{ }\epsilon^a\partial_\mu j^\mu_{\hspace{1.5mm}a}(\mathbf{x})\label{74}
\end{align} 
where $j^\mu_{\hspace{1.5mm}a}(\mathbf{x})$ is the corresponding Noether current (summation over the repeated indices $i$ are implicit):
\begin{align}
j^\mu_{\hspace{1.5mm}a}(\mathbf{x})= {T_{(c)}}^\mu_{\hspace{1.5mm}\nu} f^\nu_{\hspace{1.5mm}a}-{(\mathcal{F}_a\cdot\Phi)}^i\frac{\partial\mathcal{L}}{\partial(\partial_\mu\Phi^i)}\label{66}
\end{align}
Here, ${T_{(c)}}^\mu_{\hspace{1.5mm}\nu}$ is the canonical EM tensor which clearly is the Noether current associated to the global space-time translation. It is conserved if the action has global translation symmetry:
\begin{align}
\partial_\mu{T_{(c)}}^\mu_{\hspace{1.5mm}\nu}=0\label{57}
\end{align}

\medskip

Below we lay out the detailed ramifications of the invariance of the action under the remaining four generators of the global CCA$_{1+1}$.

\medskip

\subsection{Invariance under pCB and TSCT}\label{102}
Under an infinitesimal pCB, the fields appearing in the Lagrangian transform according to the infinitesimal form of \eqref{eq:2}:
\begin{align}
\Phi^i(t,x)\rightarrow{\tilde{\Phi}}^i(t+\epsilon^t x,x)=\Phi^i(t,x)-\epsilon^t{(\bm{\xi}\cdot\Phi)}^i(t,x)
\end{align}
where $\xi\mathbf{J}^{\mathbf{-}}$ of \eqref{eq:2} has been abbreviated as $\bm{\xi}$ . Thus, comparing with \eqref{eq:5}, the pCB current is given by:
\begin{align}
j^\mu_{(B)}= x{T_{(c)}}^\mu_{\hspace{1.5mm}t}+{(\bm{\xi}\cdot\Phi)}^i\frac{\partial\mathcal{L}}{\partial(\partial_\mu\Phi^i)}
\end{align}
pCB invariance of the action implies on-shell conservation of this current, leading to the following condition on the EM tensor:
\begin{align}
\partial_{\mu}j^\mu_{(B)}=0\text{ }\Longrightarrow\text{ } {T_{(c)}}^x_{\hspace{1.5mm}t}+\partial_{\mu}\left({(\bm{\xi}\cdot\Phi)}^i\frac{\partial\mathcal{L}}{\partial(\partial_\mu\Phi^i)}\right)=0\hspace{5mm}\text{(on-shell)}\label{58}
\end{align}
that, together with the conservation law \eqref{57}$_{\nu=t}$, leads to:
\begin{align}
\partial_t {T_{(c)}}^t_{\hspace{1.5mm}t}=\partial_x\partial_{t}\left({(\bm{\xi}\cdot\Phi)}^i\frac{\partial\mathcal{L}}{\partial(\partial_t\Phi^i)}\right)+\partial_{x^2}\left({(\bm{\xi}\cdot\Phi)}^i\frac{\partial\mathcal{L}}{\partial(\partial_x\Phi^i)}\right)\hspace{5mm}\text{(on-shell)}\label{60}
\end{align}

\medskip

In Poincare-invariant field theory, `Belinfante-improvement' is carried out by adding an identically (i.e. off-shell) divergence-less tensor to the canonical EM tensor, to render the `new' EM tensor on-shell symmetric and conserved. Similarly here, we want to `improve' the EM tensor to have off-shell ${T}^x_{\hspace{1.5mm}t}=0$ by adding an off-shell divergence-less quantity with components $\left(\partial_xV,-\partial_tV\right)$, as:
\begin{align*}
&{T}^x_{\hspace{1.5mm}t}={T_{(c)}}^x_{\hspace{1.5mm}t}-\partial_tV\hspace{5mm}\text{with ${T}^x_{\hspace{1.5mm}t}=0$ off-shell}\\
&{T}^t_{\hspace{1.5mm}t}={T_{(c)}}^t_{\hspace{1.5mm}t}+\partial_xV\hspace{5mm}\text{with $\partial_t{T}^t_{\hspace{1.5mm}t}=0$ on-shell}
\end{align*}
Now, the crucial point is that to be able to perform this `improvement', we need, as is evident from \eqref{58}, that:
\begin{align}
\partial_{x}\left({(\bm{\xi}\cdot\Phi)}^i\frac{\partial\mathcal{L}}{\partial(\partial_x\Phi^i)}\right)(\bm{\Phi},\partial_{\mu}\bm{\Phi},\partial_x\partial_{\mu}\bm{\Phi})=\partial_tW\hspace{5mm}\text{(off-shell/identically)}\label{59}
\end{align}
for some function $W$ depending explicitly only on fields and their (possibly higher) derivatives. Otherwise, improving ${T}^x_{\hspace{1.5mm}t}$ to 0 would simultaneously bring ${T}^t_{\hspace{1.5mm}t}$ to 0 that would be a disaster for conformal symmetry. 

\medskip

A priori, the condition \eqref{59} is not expected to hold for an arbitrary Lagrangian. But, curiously, each of the examples with pCB symmetry that we came across, identically satisfies:
\begin{align}
{(\bm{\xi}\cdot\Phi)}^i\frac{\partial\mathcal{L}}{\partial(\partial_x\Phi^i)}=0\label{62}
\end{align}
that trivially meets the condition \eqref{59}, hence allowing for an improved ${T}^x_{\hspace{1.5mm}t}=0$. Thus, we speculate if \eqref{62} is itself a condition for constructing a pCB invariant action. It will be very interesting to prove or to find a counter-example to this speculation.

\medskip

Thus, we see that, unlike the Lorentz boost invariance in $1+1$D that automatically implies the existence of an on-shell symmetric EM tensor, invariance under pCB alone does not guarantee the existence of an EM tensor with off-shell ${T}^x_{\hspace{1.5mm}t}=0$. But we know of no example where the EM tensor of a pCB invariant theory can not be improved to have vanishing ${T}^x_{\hspace{1.5mm}t}$.

\medskip

Next, we shall see that if there is a TSCT symmetry of the action, it will automatically lead to an off-shell vanishing ${T}^x_{\hspace{1.5mm}t}$ component. Since, the fields in the Lagrangian are postulated to be $1+1$D CC quasi-primary fields, their infinitesimal transformation property under TSCT are obtained from \eqref{eq:3} as:
\begin{align}
\Phi^i(t,x)\rightarrow{\tilde{\Phi}}^i(t+\epsilon^t x^2,x)=\Phi^i(t,x)-\epsilon^t{2x(\bm{\xi}\cdot\Phi)}^i(t,x)
\end{align}  
This gives rise to the following TSCT current:
\begin{align}
j^\mu_{(T)}= x^2{T_{(c)}}^\mu_{\hspace{1.5mm}t}+2x{(\bm{\xi}\cdot\Phi)}^i\frac{\partial\mathcal{L}}{\partial(\partial_\mu\Phi^i)}
\end{align}
It will be on-shell conserved if the action has a TSCT symmetry, leading to the following condition:
\begin{align}
\partial_{\mu}j^\mu_{(T)}=0\text{ }\Longrightarrow\text{ } 2x\left[{T_{(c)}}^x_{\hspace{1.5mm}t}+\partial_{\mu}\left({(\bm{\xi}\cdot\Phi)}^i\frac{\partial\mathcal{L}}{\partial(\partial_\mu\Phi^i)}\right)\right]+2{(\bm{\xi}\cdot\Phi)}^i\frac{\partial\mathcal{L}}{\partial(\partial_x\Phi^i)}=0\hspace{5mm}\text{(on-shell)}
\end{align}
This implies the pCB invariance condition \eqref{58} along with the following extra condition:
\begin{align}
{(\bm{\xi}\cdot\Phi)}^i\frac{\partial\mathcal{L}}{\partial(\partial_x\Phi^i)}=0\hspace{5mm}\text{(on-shell)}\label{61}
\end{align}
which trivially satisfies the condition \eqref{59}, hence allowing for a vanishing ${T}^x_{\hspace{1.5mm}t}$ component.

\medskip

Thus, the TSCT invariance of the action `predicts' the pCB symmetry and permits the following improved EM tensor components, as seen from \eqref{60}:
\begin{align}
{T}^x_{\hspace{1.5mm}t}=0\hspace{2.5mm},\hspace{2.5mm}{T}^t_{\hspace{1.5mm}t}={T_{(c)}}^t_{\hspace{1.5mm}t}-\partial_x\left({(\bm{\xi}\cdot\Phi)}^i\frac{\partial\mathcal{L}}{\partial(\partial_t\Phi^i)}\right)\hspace{5mm}\text{(off-shell)}\label{64}
\end{align}
The improvement ${T}^x_{\hspace{1.5mm}t}=0$ allows us to re-express the conserved currents $j^\mu_{(B)}$ and $j^\mu_{(T)}$ into the following simple form:
\begin{align}
j^\mu_{(B)}=x{T}^\mu_{\hspace{1.5mm}t}\hspace{2.5mm},\hspace{2.5mm}j^\mu_{(T)}=x^2{T}^\mu_{\hspace{1.5mm}t}
\end{align} 

\medskip

\subsection{Invariance under dilatation and SSCT}\label{103}
Under a global infinitesimal scale transformation, the fields in the Lagrangian transform as the infinitesimal version of \eqref{63}:
\begin{align}
\Phi^i(t,x)\rightarrow{\tilde{\Phi}}^i(t+\epsilon^x t,x+\epsilon^x x)=\Phi^i(t,x)-\epsilon^x(\Delta\Phi)^i(t,x)
\end{align}
Hence, the dilatation current is expressed as:
\begin{align}
j^\mu_{(D)}= x^\nu{T_{(c)}}^\mu_{\hspace{1.5mm}\nu}+{(\Delta\Phi)}^i\frac{\partial\mathcal{L}}{\partial(\partial_\mu\Phi^i)}
\end{align}
Global scale-invariance of the action implies the on-shell conservation of the dilatation current:
\begin{align}
\partial_{\mu}j^\mu_{(D)}=0\text{ }\Longrightarrow\text{ }{T_{(c)}}^\mu_{\hspace{1.5mm}\mu}+\partial_{\mu}\left({(\Delta\Phi)}^i\frac{\partial\mathcal{L}}{\partial(\partial_\mu\Phi^i)}\right)=0\hspace{5mm}\text{(on-shell)}\label{65}
\end{align}
from where, using the conservation law \eqref{57}$_{\nu=x}$, one reaches:
\begin{align}
\partial_t{T_{(c)}}^t_{\hspace{1.5mm}x}=\partial_x{T_{(c)}}^t_{\hspace{1.5mm}t}+\partial_x\partial_{\mu}\left({(\Delta\Phi)}^i\frac{\partial\mathcal{L}}{\partial(\partial_\mu\Phi^i)}\right)\hspace{5mm}\text{(on-shell)}
\end{align}

\medskip

We shall now see that if the action is invariant under TSCT and dilatation (but not SSCT), the EM tensor can not, in general, be improved to have on-shell vanishing trace ${T}^\mu_{\hspace{1.5mm}\mu}$. Again, to `Belinfante-improve' the EM tensor components, we must add an identically divergence-less quantity with components $(\partial_xU,-\partial_tU)$ such that:
\begin{align*}
{T}^t_{\hspace{1.5mm}x}={T_{(c)}}^t_{\hspace{1.5mm}x}+\partial_xU\hspace{3mm},\hspace{3mm}{T}^x_{\hspace{1.5mm}x}={T_{(c)}}^x_{\hspace{1.5mm}x}-\partial_tU\hspace{5mm}\text{with on-shell { } $\partial_\mu{T}^\mu_{\hspace{1.5mm}x}=0$}
\end{align*}
In a TSCT invariant theory, since the EM tensor can be improved to off-shell have the form \eqref{64}, the dilatation-invariance condition \eqref{65} implies that:
\begin{align}
{T}^\mu_{\hspace{1.5mm}\mu}=-\partial_tU-\partial_x\left({(\bm{\xi}\cdot\Phi)}^i\frac{\partial\mathcal{L}}{\partial(\partial_t\Phi^i)}\right)-\partial_{\mu}\left({(\Delta\Phi)}^i\frac{\partial\mathcal{L}}{\partial(\partial_\mu\Phi^i)}\right)\hspace{5mm}\text{(on-shell)}\label{68}
\end{align}
Thus, to execute the improvement to ${T}^\mu_{\hspace{1.5mm}\mu}=0$ on-shell, it is necessary to have:
\begin{align}
\partial_x\left({(\bm{\xi}\cdot\Phi)}^i\frac{\partial\mathcal{L}}{\partial(\partial_t\Phi^i)}+{(\Delta\Phi)}^i\frac{\partial\mathcal{L}}{\partial(\partial_x\Phi^i)}\right)(\bm{\Phi},\partial_{\mu}\bm{\Phi},\partial_x\partial_{\mu}\bm{\Phi})=\partial_tW^\prime\hspace{5mm}\text{(off-shell)}\label{67}
\end{align}
for some function $W^\prime$ depending explicitly only on fields and their (possibly higher) derivatives.

\medskip

Clearly, this condition is too stringent to be satisfied by an arbitrary Lagrangian. Indeed, the following action exemplifies this issue:
\begin{align}
&S[\bm{\Phi}]=\frac{1}{2}\int dt\int dx\text{ }\left[{\left(\partial_t\Phi^+-\xi\partial_x\Phi^-\right)}^2+{\left(\partial_t\Phi^-\right)}^2\right]\\
&\text{with { } }\Phi^-(t,x)\rightarrow{\tilde{\Phi}}^-(t^\prime,x^\prime)=\Phi^-(t,x)\hspace{2.5mm};\hspace{2.5mm}\Phi^+(t,x)\rightarrow{\tilde{\Phi}}^+(t^\prime,x^\prime)=\Phi^+(t,x)-\xi\frac{\partial_x t^\prime}{\partial_x x^\prime}\Phi^-(t,x)\nonumber\\
&\text{where $(t,x)\rightarrow(t^\prime,x^\prime)$ is a $1+1$D global CC transformation \eqref{56}.}\nonumber
\end{align}
It can be easily checked that this action is invariant under pCB, TSCT and dilatation but not under SSCT. The EM tensor of this theory can be improved to have a vanishing ${T}^x_{\hspace{1.5mm}t}$ component but can not be made trace-less since the condition \eqref{67} is not satisfied.

\medskip

Thus, we see that, like in the relativistic scenario, dilatation invariance does not lead to the (S)SCT symmetry. Also, relatedly, TSCT and dilatation invariance together are not strong enough to ensure the existence of even an on-shell trace-less EM tensor.

\medskip

In the following, we shall see that if the action is invariant under SSCT and TSCT, the EM tensor can definitely be improved to be trace-less off-shell. The fields in the Lagrangian, that are postulated to be $1+1$D CC quasi-primary fields, transform under an infinitesimal SSCT as, from \eqref{eq:3}:
\begin{align}
\Phi^i(t,x)\rightarrow{\tilde{\Phi}}^i(t+\epsilon^x2xt,x+\epsilon^xx^2)=\Phi^i(t,x)-\epsilon^x\left[{2t(\bm{\xi}\cdot\Phi)}^i(t,x)+{2x(\Delta\Phi)}^i(t,x)\right]
\end{align}  
Hence, we have the following SSCT current directly from \eqref{66}:
\begin{align}
j^\mu_{(S)}= x^2{T_{(c)}}^\mu_{\hspace{1.5mm}x}+2tx{T_{(c)}}^\mu_{\hspace{1.5mm}t}+\left[2t{(\bm{\xi}\cdot\Phi)}^i+2x(\Delta\Phi)^i\right]\frac{\partial\mathcal{L}}{\partial(\partial_\mu\Phi^i)}
\end{align}
The assumed SSCT symmetry of the action will lead to the on-shell conservation of this current:
\begin{align}
\partial_\mu j^\mu_{(S)}=0\text{ }\Longrightarrow\text{ }x&\left[{T_{(c)}}^\mu_{\hspace{1.5mm}\mu}+\partial_{\mu}\left({(\Delta\Phi)}^i\frac{\partial\mathcal{L}}{\partial(\partial_\mu\Phi^i)}\right)\right]+t\left[{T_{(c)}}^x_{\hspace{1.5mm}t}+\partial_{\mu}\left({(\bm{\xi}\cdot\Phi)}^i\frac{\partial\mathcal{L}}{\partial(\partial_\mu\Phi^i)}\right)\right]\nonumber\\
+&\left[{(\bm{\xi}\cdot\Phi)}^i\frac{\partial\mathcal{L}}{\partial(\partial_t\Phi^i)}+(\Delta\Phi)^i\frac{\partial\mathcal{L}}{\partial(\partial_x\Phi^i)}\right]=0\hspace{5mm}\text{(on-shell)}
\end{align}
This reproduces the pCB and dilatation invariance conditions \eqref{58} and \eqref{65} respectively in addition to revealing the following extra condition for SSCT invariance:
\begin{align}
{(\bm{\xi}\cdot\Phi)}^i\frac{\partial\mathcal{L}}{\partial(\partial_t\Phi^i)}+(\Delta\Phi)^i\frac{\partial\mathcal{L}}{\partial(\partial_x\Phi^i)}=0\hspace{5mm}\text{(on-shell)}\label{69}
\end{align}
that trivially satisfies the condition \eqref{67}, hence permitting the improvement to a trace-less EM tensor. We again emphasize that \eqref{67} is a necessary condition to have ${T}^\mu_{\hspace{1.5mm}\mu}=0$ when the action is invariant under TSCT so that we already certainly have ${T}^x_{\hspace{1.5mm}t}=0$.

\medskip

Thus, if the action possesses SSCT symmetry, it must also be invariant under pCB and dilatation. Moreover, from \eqref{68} using \eqref{69}, it is concluded that TSCT and SSCT invariance of the action together allow for the following on-shell conserved but off-shell trace-less EM tensor with ${T}^x_{\hspace{1.5mm}t}=0$:
\begin{align}
{T}^x_{\hspace{1.5mm}t}=0\hspace{2.5mm},\hspace{2.5mm}&{T}^t_{\hspace{1.5mm}t}={T_{(c)}}^t_{\hspace{1.5mm}t}-\partial_x\left({(\bm{\xi}\cdot\Phi)}^i\frac{\partial\mathcal{L}}{\partial(\partial_t\Phi^i)}\right)=-{T}^x_{\hspace{1.5mm}x}\nonumber\\
&{T}^t_{\hspace{1.5mm}x}={T_{(c)}}^t_{\hspace{1.5mm}x}-\partial_x\left({(\Delta\Phi)}^i\frac{\partial\mathcal{L}}{\partial(\partial_t\Phi^i)}\right)\hspace{5mm}\text{(off-shell)}
\end{align}
The improvement to ${T}^\mu_{\hspace{1.5mm}\mu}=0={T}^x_{\hspace{1.5mm}t}$ permits for the following simple expressions for the conserved currents $j^\mu_{(D)}$ and $j^\mu_{(S)}$ :
\begin{align}
j^\mu_{(D)}=x^\nu{T}^\mu_{\hspace{1.5mm}\nu}\hspace{2.5mm},\hspace{2.5mm}j^\mu_{(S)}=x^2{T}^\mu_{\hspace{1.5mm}x}+2tx{T}^\mu_{\hspace{1.5mm}t}
\end{align}

\medskip

\subsection{Invariance under CCA$_{1+1}$}\label{104}
If we insist that any arbitrary $1+1$D CC transformation \eqref{55} is a classical symmetry of the action \eqref{70}, following and extending the above arguments, one can show that the EM tensor of such a theory can always be `Belinfante-improved' to have:
\begin{align}
{T}^x_{\hspace{1.5mm}t}=0\hspace{3.5mm}\text{ and }\hspace{3.5mm}{T}^\mu_{\hspace{1.5mm}\mu}=0\hspace{5mm}\text{(off-shell)}\label{71}
\end{align}
This result follows even when the postulate that the fields appearing in the Lagrangian transforms as $1+1$D CC (quasi-)primary field is given up. We do not need to assume any specific transformation property under any $1+1$D CC transformation, of the `fundamental fields'; instead, a sufficient condition is that no field (`fundamental' or otherwise) in the theory has negative scaling dimension.

\medskip
  
As a consequence, the components of the improved EM tensor of such a theory classically (on-shell) satisfies:
\begin{align}
&{T}^x_{\hspace{1.5mm}t}=0 \text{ }\text{  and  }\text{ }\partial_\mu{T}^\mu_{\hspace{1.5mm}t}=0\text{ }\text{ }\Longrightarrow\text{ }\text{ }\partial_t{T}^t_{\hspace{1.5mm}t}=0\text{ }\text{ }\Longrightarrow\text{ }\text{ }{T}^t_{\hspace{1.5mm}t}(t,x)={T}^t_{\hspace{1.5mm}t}(x)\label{eq:34}\\
&{T}^\mu_{\hspace{1.5mm}\mu}=0 \text{ }\text{  and  }\text{ }\partial_\mu{T}^\mu_{\hspace{1.5mm}x}=0\text{ }\text{ }\Longrightarrow\text{ }\text{ }\partial_x{T}^t_{\hspace{1.5mm}t}=\partial_t{T}^t_{\hspace{1.5mm}x}\text{ }\text{ }\Longrightarrow\text{ }\text{ }{T}^t_{\hspace{1.5mm}x}(t,x)=t\partial_x{T}^t_{\hspace{1.5mm}t}(x)+p(x)\label{eq:35}
\end{align}
with $p(x)$ being an arbitrary function.

\medskip

Finally, if the theory classically boasts of the kinematical CCA$_{1+1}$ symmetry, it possesses an infinite number of conserved Noether currents correspondingly. The properties \eqref{71} of the EM tensor allow for the following simple form of the conserved current associated to an arbitrary infinitesimal symmetry transformation \eqref{eq:4} compactly written as $x^{\mu}\rightarrow x^{\mu}+\epsilon^af^{\mu}_{\hspace{1.5mm}a}(\mathbf{x})$:
\begin{align}
j^\mu_{\hspace{1.5mm}a}={T}^\mu_{\hspace{1.5mm}\nu}f^\nu_{\hspace{1.5mm}a}\hspace{5mm}\text{(off-shell)}\label{72}
\end{align}

\medskip

We stress that the properties \eqref{71} are the sufficient and necessary conditions to be able to express the conserved currents as \eqref{72}. This is so because to show that an arbitrary current given in this form is on-shell conserved, only both of those conditions are required. 

\medskip

Our derivation below of the Ward identities in the corresponding QFT will heavily rely on the form \eqref{72} of the Noether current(-operator)s.

\medskip

\section{Ward Identities}\label{105}
Having analyzed the implications of the classical CCA$_{1+1}$ symmetry in detail, we now turn to the quantum aspects of the same. The quantum analogues of the classical Noether's theorem are the Ward identities.

\medskip

In the path-integral formalism of QFT, correlators are the central objects of interest. A general $n$-point (time-ordered) correlator $\langle X\rangle$ is defined as (suppressing the field tensor indices):
\begin{align}
\langle X\rangle\equiv\langle{\mathcal{T}}\Phi_1(\mathbf{x_1})\Phi_2(\mathbf{x_2})...\Phi_n(\mathbf{x_n})\rangle:=\frac{\int[\mathcal{D}\bm{\Phi}]\text{ }\Phi_1(\mathbf{x_1})\Phi_2(\mathbf{x_2})...\Phi_n(\mathbf{x_n})\text{ }e^{iS[\bm{\Phi}]}}{\int[\mathcal{D}\bm{{\Phi}}]\text{ }e^{iS[\bm{{\Phi}}]}}\label{76}
\end{align}
A field transformation, e.g. \eqref{eq:5}, will be a symmetry of the QFT if any correlator $\langle X\rangle$ remains invariant under that transformation, i.e. if \cite{DiFrancesco:1997nk}:
\begin{align}
\langle X\rangle &={\langle X\rangle}^\prime\\
\text{where\hspace{5mm}}{\langle X\rangle}^\prime\equiv\langle{\mathcal{T}}\tilde{\Phi}_1(\mathbf{x_1})\tilde{\Phi}_2(\mathbf{x_2})...\tilde{\Phi}_n(\mathbf{x_n})\rangle &:=\frac{\int[\mathcal{D}\bm{\tilde{\Phi}}]\text{ }\tilde{\Phi}_1(\mathbf{x_1})\tilde{\Phi}_2(\mathbf{x_2})...\tilde{\Phi}_n(\mathbf{x_n})\text{ }e^{iS^\prime[\bm{\tilde{\Phi}}]}}{\int[\mathcal{D}\bm{\tilde{\Phi}}]\text{ }e^{iS^\prime[\bm{\tilde{\Phi}}]}}\nonumber
\end{align} 
With the assumption that the path-integral measure is invariant: $[\mathcal{D}\bm{\Phi}]=[\mathcal{D}\bm{\tilde{\Phi}}]$, this symmetry condition leads to the Ward identity (at the 1st order in $\bm\epsilon$), using \eqref{73} and \eqref{74}:
\begin{align}
i\sum_{i=1}^n\langle{\mathcal{T}}\Phi_1(\mathbf{x_1})...(\epsilon^aG_a\Phi_i(\mathbf{x_i}))...\Phi_n(\mathbf{x_n})\rangle=i\int dt\int dx\text{ }\epsilon^a\langle{\mathcal{T}}\partial_\mu j^\mu_{\hspace{1.5mm}a}(\mathbf{x})(X-I)\rangle\label{44}
\end{align}
For future references, we introduce the following notation denoting the change $\delta_{\bm{\epsilon}}\langle X\rangle$ of the correlator $\langle X\rangle$ :
\begin{align*}
\delta_{\bm{\epsilon}}\langle X\rangle\equiv i\sum_{i=1}^n\langle{\mathcal{T}}\Phi_1(\mathbf{x_1})...(\epsilon^aG_a\Phi_i(\mathbf{x_i}))...\Phi_n(\mathbf{x_n})\rangle
\end{align*}

\medskip

A comment on the nature of the composite operators in a QFT is in order. Let two fields $\Phi_1$ and $\Phi_2$ be inserted into the correlator at the same space-time point $\mathbf{x}$. Correlators typically diverge when the points of insertion of two or more fields coincide. Such singularities must be appropriately regularized and renormalized \cite{Polchinski:1998rq}. For free field theories, `normal-ordering' is a convenient prescription to achieve this goal. Under this prescription, the two fields $\Phi_1$ and $\Phi_2$ are treated as a single composite operator inserted at $\mathbf{x}$ into the correlator. The composite operator is denoted as $:\Phi_1\Phi_2:(\mathbf{x})$ where $::$ is the normal-ordering symbol.

\medskip

In a Poincare-invariant field theory, if two operators inside a correlator are inserted at two space-time points $\mathbf{x_1},\mathbf{x_2}$ that are mutually light-like separated, those two operators also need to be treated as a single composite operator\footnote{A direct way to verify this statement is to consider \eqref{75} for a Lorentz transformation}. So, the crucial point is that whenever the Poincare-invariant distance between the space-time points of insertion of two fields vanishes, they are to be treated as one composite operator.   

\medskip

Equivalent to the Poincare-invariant length of a space-time interval, the flat Carrollian counterpart is the Euclidean spatial distance $\vert \vec{x}_1-\vec{x}_2\vert$ between two Carrollian space-time points $(t_1,\vec{x}_1)$ and $(t_2,\vec{x}_2)$. In analogy to the Poincare-invariant field theory, then in a Carrollian field theory, two fields inside a correlator should be treated as one composite operator if they are inserted at the same spatial location (but at possibly different times) forcing the flat Carrollian invariant `norm' to vanish. This conclusion captures the essence of the spatial absoluteness property of Carrollian physics. Thus, when the correlator in a Carrollian field theory is written as \eqref{76}, it is implicit that $\vec{x}_i\neq \vec{x}_j$ for $i\neq j$. 

\medskip

Now we proceed to derive the Ward identities explicitly for a CCA$_{1+1}$ invariant theory. As we have discussed, the currents corresponding to the kinematical CCA$_{1+1}$ symmetry can all be expressed simply in the form \eqref{72}. So, to have the same as the symmetry algebra of the quantum theory, the Ward identity \eqref{44} must be valid for any vector field $f^{\mu}_{\hspace{1.5mm}a}(t,x)$ that may be unbounded and the generators of which may not be necessarily everywhere well-defined.  Moreover, since the integral in the R.H.S. of \eqref{44} is over the entire space-time, the integrand definitely contains singularities whenever the spatial integration variable $x$ coincides with any one of the spatial insertions of the fields in $X$, i.e. the integrand surely diverges for some finite values of $x$. 

\medskip

Keeping in mind the possible unbounded-ness property of $f^{\mu}_{\hspace{1.5mm}a}(t,x)$ in both $t$ and $x$ and the existence of infinities of the integrand at some finite values of $x$, it is not hard to convince oneself that the only way to make sure that $\delta_{\bm{\epsilon}}\langle X\rangle$ converges is to allow $\langle{\mathcal{T}}\partial_\mu T^\mu_{\hspace{1.5mm}\nu}(\mathbf{x})X\rangle$ and $\langle{\mathcal{T}} T^\mu_{\hspace{1.5mm}\nu}(\mathbf{x})X\rangle$ to have only delta-function (or derivatives thereof) singularities\footnote{In suitable cases, step-function discontinuities in some (but not all) of the variables are allowed, as we shall see shortly.}, as a function of the real multi-variable $\mathbf{x}$, at points of insertion of fields in $X$. Thus, $\langle{\mathcal{T}}\partial_\mu T^\mu_{\hspace{1.5mm}\nu}(\mathbf{x})X\rangle$ and $\langle{\mathcal{T}} T^\mu_{\hspace{1.5mm}\nu}(\mathbf{x})X\rangle$ both vanish away from the points of insertion; in particular:
\begin{align}
\lim\limits_{{x}\rightarrow\infty}\langle{\mathcal{T}}\partial_\mu T^\mu_{\hspace{1.5mm}\nu}(\mathbf{x})X\rangle=0\hspace{7.5mm}\text{and}\hspace{7.5mm}\lim\limits_{{x}\rightarrow\infty}\langle{\mathcal{T}}T^\mu_{\hspace{1.5mm}\nu}(\mathbf{x})X\rangle=0\text{\hspace{10mm}(at any $t$)}\label{eq:29}
\end{align}

\medskip

\textbf{Comment 3:} It is tempting to appeal to the divergence theorem in the R.H.S. of \eqref{44} to conclude that:
\begin{align}
\delta_{\bm{\epsilon}}\langle X\rangle\equiv i\sum_{i=1}^n\langle{\mathcal{T}}\Phi_1(\mathbf{x_1})...(\epsilon^aG_a\Phi_i(\mathbf{x_i}))...\Phi_n(\mathbf{x_n})\rangle=0\label{79}
\end{align}
for any field transformation $\eqref{eq:5}$. This is not generally true. This relation is valid only when the generators of such a transformation is well-defined everywhere on the space-time. The global conformal generators or global CC generators, for example, have this property (hence, they are called `global'). The relation \eqref{79} can be used to constrain the general form of the correlators in a QFT \cite{DiFrancesco:1997nk}.

\medskip

For CCA$_{1+1}$ invariant QFTs, the time-integral of the Ward identity \eqref{44} is inverted to obtain the following integral form (with time-ordering being implicit from now on):
\begin{align}
i\int dx\text{ }\partial_\mu\langle j^\mu_{\hspace{1.5mm}a}(\mathbf{x})(X-I)\rangle=\sum_{i=1}^n\text{ }\delta(t-t_i)\int dx \text{ }\delta(x-x_i)\text{ }\langle\Phi_1(\mathbf{x_1})...(iG_a\Phi_i(\mathbf{x_i}))...\Phi_n(\mathbf{x_n})\rangle\label{eq:6}
\end{align}

\medskip

Now comes the most important trick introduced in this paper\footnote{This is motivated from the presence of the power-law factors in spatial separations in the section \ref{132} correlators that can be derived without any knowledge of the Ward identities.}. To invert the space-integral, we analytically continue $x$ from the Riemann circle $\mathbb{R}\cup\{\infty\}$ to the Riemann sphere $\mathbb{C}\cup\{\infty\}$. Thus, we are treating $x$ and $t$ on different footing. As we shall see, this trick will open the door for the usage of the complex analytic method in $1+1$D CCFTs. Since the Riemann sphere is the one point (at $\infty$) compactification of $\mathbb{R}^2$, the boundary conditions \eqref{eq:29} are naturally extended to:
\begin{align}
\lim\limits_{|x|\rightarrow\infty}\langle{\mathcal{T}}\partial_\mu T^\mu_{\hspace{1.5mm}\nu}(t,x)X\rangle=0\hspace{7.5mm}\text{and}\hspace{7.5mm}\lim\limits_{|x|\rightarrow\infty}\langle{\mathcal{T}}T^\mu_{\hspace{1.5mm}\nu}(t,x)X\rangle=0\text{\hspace{10mm}(at any $t$)}\label{eq:28}
\end{align}
Thus, consider the following steps. The LHS of \eqref{eq:6} can be written as the following contour (complex) integral:
\begin{align}
&\int\limits^\infty_{-\infty}dx\text{ }\partial_\mu\langle j^\mu_{\hspace{1.5mm}a}(t,x)X\rangle\text{\hspace{15mm}(sum of real integrals of delta-function integrands)}\nonumber\\
=\text{ }&\frac{1}{2\pi i}\left(\int\limits_{-\infty+i0^-}^{\infty+i0^-}dx+\int\limits_{C_{|x|\rightarrow\infty}}dx\right)\text{ }\partial_\mu\langle j^\mu_{\hspace{1.5mm}a}(t,{x})X\rangle\text{\hspace{7.5mm}(complex integrand's form: $\sum\limits_{k\in\mathbb{Z}}\frac{a_k}{{(x-x_i)}^k}$ )}\nonumber\\
=\text{ }&\frac{1}{2\pi i}\oint\limits_{C} dx\text{ }\partial_\mu\langle j^\mu_{\hspace{1.5mm}a}(t,{x})X\rangle\text{ }=\text{ }\sum\limits_{p=1}^n\frac{1}{2\pi i}\oint\limits_{C_p} dx\text{ }\partial_\mu\langle j^\mu_{\hspace{1.5mm}a}(t,{x})X\rangle\label{eq:7}
\end{align}
where $C$ encloses all spatial insertions $\{x_i\}$ but $C_p$ encloses only the spatial insertion $x_p$. Thus, using the inverse of the residue formula, \eqref{eq:6} is inverted as (with $x$ now a complex variable):
\begin{align}
\partial_\mu\langle j^\mu_{\hspace{1.5mm}a}(t,{x})(X-I)\rangle\sim -i\sum\limits_{i=1}^n\text{ }\delta(t-{t_i})\left[\frac{i{(G_a)}_i\langle X\rangle}{x-x_i}+\sum_{k\geq2}\text{ }\frac{{\langle Y^{(k)}_a\rangle}_i(\mathbf{x_1},...\mathbf{x_n})}{{(x-x_i)}^{k}}\right]\label{eq:30}
\end{align}
where the correlators ${\langle Y^{(k)}_a\rangle}_i$ depend on the transformation properties of the fields in $X$ and the transformation itself and $\sim$ denotes `modulo terms regular (holomorphic) in $x-x_i$ inside $[...]$'. The correlators ${\langle Y^{(k)}_a\rangle}_i$ can not be inferred without knowing the transformation properties of the fields in $X$ since they do not contribute to the contour integral in \eqref{eq:7}. This is the general form of a Ward identity in a $1+1$D Carrollian QFT that is valid for any `internal' transformation also.
\begin{figure}[h]
\begin{center}
\begin{tikzpicture}[decoration={markings,
mark=at position 1.8cm with {\arrow[line width=1pt]{>}},
mark=at position 6.0cm with {\arrow[line width=1pt]{>}}}]
\draw[help lines,->] (-3.3,0) -- (3.3,0) coordinate (xaxis);
\draw[help lines,->] (0,-0.75) -- (0,2.75) coordinate (yaxis);
\path[draw,line width=0.8pt,postaction=decorate] (-2.5,-.25) -- (2.5,-.25) arc (0:180:2.5);
\node[above] at (xaxis) {$\text{Re }x$};
\node[left] at (yaxis) {$\text{Im }x$};
\draw (1.5,0) node{$\bullet$};
\draw (-2.0,0) node{$\bullet$};
\draw (-.3,0) node{$\bullet$};
\node at (1.5,0.3) {$x_3$};
\node at (-.3,0.3) {$x_2$};
\node at (-2.0,0.3) {$x_1$};
\node at (1.1,-0.55) {$C_{\text{Im }x=0^-}$};
\node at (-2.5,1.7) {$C_{|x|\rightarrow\infty}$};
\draw[help lines,->] (6.2,.9) -- (10.8,.9) coordinate (xaxis);
\draw[help lines,->] (8.5,-0.75) -- (8.5,2.75) coordinate (yaxis);
\path[draw,line width=0.8pt,postaction=decorate] (7.1,0.9) -- (7.1,0.9) arc (0:360:0.3);
\path[draw,line width=0.8pt,postaction=decorate] (8.5,0.9) -- (8.5,0.9) arc (0:360:0.3);
\path[draw,line width=0.8pt,postaction=decorate] (9.9,0.9) -- (9.9,0.9) arc (0:360:0.3);
\node[above] at (xaxis) {$\text{Re }x$};
\node[left] at (yaxis) {$\text{Im }x$};
\draw (6.7,0.9) node{$\bullet$};
\draw (9.5,0.9) node{$\bullet$};
\draw (8.1,0.9) node{$\bullet$};
\node at (6.5,1.4) {$C_1$};
\node at (9.3,1.4) {$C_3$};
\node at (7.9,1.4) {$C_2$};
\node at (4.9,0.9) {$\xrightarrow [\text{deformation}]{\text{continuous}}$};
\end{tikzpicture}
\caption{`theta' prescription in $1+1$D CCFT}
\end{center}
\end{figure}
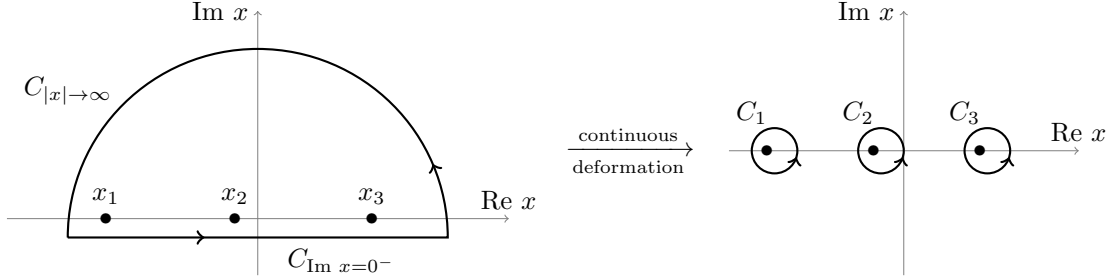

\medskip

Alternatively, this could have been formulated as the usual `$i\epsilon$' prescription by moving the poles into the upper half-plane instead of shifting the contour.

\medskip

For a $1+1$D CCFT on flat background, any space-time transformation current must satisfy: 
\begin{align*}
\partial_\mu\langle \hat{j}^\mu_{\hspace{1.5mm}a}(t,{x})\rangle=0
\end{align*}
as these currents are expressed as: $\hat{j}_{\hspace{1.5mm}a}^\mu=\hat{T}_{\hspace{1.5mm}\nu}^\mu f^\nu_{\hspace{1.5mm}a}$ and since, $\langle \hat{T}^\mu_{\hspace{1.5mm}\nu}\rangle=0$ on flat background. 

\medskip

The generator for global translation is: $iG_\nu\Phi^a(\mathbf{x})=\partial_\nu\Phi^a(\mathbf{x})$, giving the translation Ward identity:
\begin{align}
\partial_\mu\langle T^\mu_{\hspace{1.5mm}\nu}(\mathbf{x})X\rangle\sim -i\sum\limits_{p=1}^n\text{ }\delta({t}-{t_p})\text{ }\left[\frac{{\partial_{\nu_p}}\langle X\rangle}{x-x_p}+\sum_{k\geq2}\text{ }\frac{{\langle Y^{(k)}_\nu\rangle}_p(\mathbf{x_1},...\mathbf{x_n})}{{(x-x_p)}^k}\right]\label{eq:15}
\end{align}
From the infinitesimal version of \eqref{eq:2}, one obtains the Carrollian boost generator:
\begin{align}
iG_B\Phi^{m}_{(l)}(\mathbf{x})\text{ }=\text{ }x\partial_{t}\Phi^{m}_{(l)}(\mathbf{x})+{\xi}\sqrt{l(l+1)-m(m-1)}\text{ }\Phi^{m-1}_{(l)}(\mathbf{x})\text{ }:=\text{ }\left(x\partial_t+\bm{\xi}\right)\Phi^{m}_{(l)}(\mathbf{x})
\end{align}
that gives rise to the boost Ward identity:
\begin{align}
\langle T^x_{\hspace{1.5mm}t}(\mathbf{x})X\rangle\sim -i\sum\limits_{p=1}^n\text{ }\delta({t}-{t_p})\left[\frac{{ {\bm{\xi}}_p\langle X\rangle}-{\langle Y^{(2)}_t\rangle}_p}{x-x_p}+\sum_{k\geq2}\frac{{\langle Z^{(k)}_t\rangle}_p(\mathbf{x_1},...\mathbf{x_n})}{{(x-x_p)}^k}\right]\label{eq:16}
\end{align}
Finally, due to the dilation generator $iG_D\Phi^a(\mathbf{x})=(\Delta+x^\mu\partial_\mu)\Phi^a(\mathbf{x})$ where $\Delta$ is the scaling dimension of (all component of) the field $\Phi^a(\mathbf{x})$, the dilation Ward identity takes the following form:
\begin{align}
\langle T^\mu_{\hspace{1.5mm}\mu}(\mathbf{x})X\rangle\sim -i\sum\limits_{p=1}^n\text{ }\delta({t}-{t_p})\left[\frac{\Delta_p\langle X\rangle-{\langle Y^{(2)}_x\rangle}_p}{x-x_p}+\sum_{k\geq2}\text{ }\frac{{\langle Z_x^{(k)}\rangle}_p(\mathbf{x_1},...\mathbf{x_n})}{{(x-x_p)}^k}\right]\label{eq:17}
\end{align}
The translation, boost (classically assuming $T^i_{\hspace{1.5mm}t}(\mathbf{x})=0$) and dilation Ward identities are obtainable in the above forms in any space-time dimensions. But, further simplifications can be made in $1+1$D which we shall now achieve.

\medskip

\eqref{eq:15}$_{\nu=t}$ and \eqref{eq:16} give the following differential equation in $1+1$D:
\begin{align}
\partial_t\langle T^t_{\hspace{1.5mm}t}(\mathbf{x})X\rangle\sim -i\sum\limits_{p=1}^n\text{ }\delta(t-t_p)\left[\text{ }\sum_{k\geq3}\text{ }\frac{{\langle Y^{(k)}_t\rangle}_p+(k-1){\langle Z_t^{(k-1)}\rangle}_p}{{(x-x_p)}^k}+\frac{{{\bm{\xi}_p}}\langle X\rangle}{{(x-x_p)}^2}+\frac{{\partial_{t_p}}\langle X\rangle}{x-x_p}\right]\label{43}
\end{align}
To solve this, one needs an initial condition. We take it to be:
\begin{align}
\lim\limits_{t\rightarrow-\infty}\langle\mathcal{T} T^t_{\hspace{1.5mm}t}(t,x)X\rangle=0
\end{align}
The justification for this assumption will be given in the next section. Using this initial condition, \eqref{43} is solved as:
\begin{align}
\langle T^t_{\hspace{1.5mm}t}(t,x)X\rangle\sim -i\sum\limits_{p=1}^n\text{ }\theta(t-t_p)\left[\text{ }\sum_{k\geq3}\text{ }\frac{{\langle Y^{(k)}_t\rangle}_p+(k-1){\langle Z_t^{(k-1)}\rangle}_p}{{(x-x_p)}^k}+\frac{{{\bm{\xi}_p}}\langle X\rangle}{{(x-x_p)}^2}+\frac{{\partial_{t_p}}\langle X\rangle}{x-x_p}\right]\nonumber
\end{align}

\medskip

A crucial assumption in 2D relativistic CFT is that the EM tensor $T(z)$ is everywhere well-defined (finite) in the sense of correlation function \cite{Belavin:1984vu}. The extensive use of 2D CFT Ward identities rests on this assumption. Similarly, the finite-ness of $\langle T^t_{\hspace{1.5mm}t}(t,{x})X\rangle$ whenever $x\neq \{x_p\}$ is assumed in $1+1$D CCFT. In particular, $\langle T^t_{\hspace{1.5mm}t}(t,{x})X\rangle$ must be finite for $x\rightarrow\infty$ at any $t$. Thus the following function, holomorphic in $x$, is bounded:
\begin{align*}
H(x)\equiv\langle T^t_{\hspace{1.5mm}t}(t,x)X\rangle+i\sum\limits_{p=1}^n\text{ }\theta(t-t_p)\left[\text{ }\sum_{k\geq3}\text{ }\frac{{\langle Y^{(k)}_t\rangle}_p+(k-1){\langle Z_t^{(k-1)}\rangle}_p}{{(x-x_p)}^k}+\frac{{{\bm{\xi}_p}}\langle X\rangle}{{(x-x_p)}^2}+\frac{{\partial_{t_p}}\langle X\rangle}{x-x_p}\right]
\end{align*} 
Hence, by Liouville's theorem, this holomorphic function is a constant in $x$ and by the boundary condition \eqref{eq:28}, this constant is 0. This leads to the super-translation Ward identity:
\begin{align}
\langle T^t_{\hspace{1.5mm}t}(t,x)X\rangle=-i\sum\limits_{p=1}^n\text{ }\theta(t-t_p)\left[\text{ }\sum_{k\geq3}\text{ }\frac{{\langle Y^{(k)}_t\rangle}_p+(k-1){\langle Z_t^{(k-1)}\rangle}_p}{{(x-x_p)}^k}+\frac{{{\bm{\xi}_p}}\langle X\rangle}{{(x-x_p)}^2}+\frac{{\partial_{t_p}}\langle X\rangle}{x-x_p}\right]\label{eq:22}
\end{align}
The absence of holomorphic terms in \eqref{eq:22} is interpreted as the vanishing of the normal-ordered VEV $\langle{\mathcal{N}} T^t_{\hspace{1.5mm}t}(t,x)X\rangle$ and the terms singular at $x=\{x_p\}$ are the generalized `Wick contractions'.

\medskip
 
Finally, \eqref{eq:15}$_{\nu=x}$, \eqref{eq:16} and \eqref{eq:17} together lead to the differential equation:
\begin{align}
\partial_t\langle T^t_{\hspace{1.5mm}x}(\mathbf{x})X\rangle&\sim -i\sum\limits_{p=1}^n\text{ }\left[\delta(t-t_p)\left(\sum_{k\geq3}\text{ }\frac{{\langle Y^{(k)}_x\rangle}_p+(k-1){\langle Z_x^{(k-1)}\rangle}_p}{{(x-x_p)}^k}+\frac{{{\Delta_p}}\langle X\rangle}{{(x-x_p)}^2}+\frac{{\partial_{x_p}}\langle X\rangle}{x-x_p}\right)\right.\nonumber\\
&\left. \text{ }\text{ }\text{ }\text{ }\text{ }\text{ }\text{ }\text{ }-\theta(t-{t_p})\left(\text{ }\sum_{k\geq3}\text{ }k\text{ }\frac{{\langle Y^{(k)}_t\rangle}_p+(k-1){\langle Z_t^{(k-1)}\rangle}_p}{{(x-x_p)}^{k+1}}+\frac{{{2\bm{\xi}_p}}\langle X\rangle}{{(x-x_p)}^3}+\frac{{\partial_{t_p}}\langle X\rangle}{{(x-x_p)}^2}\right)\right]\nonumber
\end{align}
Again, the initial condition is assumed to be:
\begin{align}
\lim\limits_{t\rightarrow-\infty}\langle\mathcal{T} T^t_{\hspace{1.5mm}x}(t,x)X\rangle=0
\end{align}
Solving and further assuming the finite-ness property of $\langle T^t_{\hspace{1.5mm}x}(t,{x})X\rangle$ whenever $x\neq \{x_p\}$, we obtain the super-rotation Ward identity: 
\begin{align}
\langle T^t_{\hspace{1.5mm}x}({t,x})X\rangle=&-i\sum\limits_{p=1}^n\text{ }\theta(t-{t_p})\left[\sum_{k\geq3}\text{ }\frac{{\langle Y^{(k)}_x\rangle}_p+(k-1){\langle Z_x^{(k-1)}\rangle}_p}{{(x-x_p)}^k}+\frac{{{\Delta_p}}\langle X\rangle}{{(x-x_p)}^2}+\frac{{\partial_{x_p}}\langle X\rangle}{x-x_p}\right.\nonumber\\
&\left. -(t-{t_p})\left(\sum_{k\geq3}\text{ }k\text{ }\frac{{\langle Y^{(k)}_t\rangle}_p+(k-1){\langle Z_t^{(k-1)}\rangle}_p}{{(x-x_p)}^{k+1}}+\frac{{{2\bm{\xi}_p}}\langle X\rangle}{{(x-x_p)}^3}+\frac{{\partial_{t_p}}\langle X\rangle}{{(x-x_p)}^2}\right)\right]\label{eq:23}
\end{align}

\medskip

The reader may have already noticed that in the above Ward identity, the coefficient of a $(t-t_p)$ is the same as the $p$-th summand in $\partial_x\langle T^t_{\hspace{1.5mm}t}({t,x})X\rangle$. This corresponds to the following quantum version of the conservation equation \eqref{eq:35}:
\begin{align}
\partial_x\langle T^t_{\hspace{1.5mm}t}({t,x})X\rangle=\partial_t\langle T^t_{\hspace{1.5mm}x}({t,x})X\rangle\text{\hspace{5mm}(for $t\neq\{t_p\})$}
\end{align}
On the other hand, \eqref{43} is the quantum version of the classical conservation equation \eqref{eq:34}.

\medskip

So, the super-translation and super-rotation Ward identities capture the singular behavior of the $\langle T^t_{\hspace{1.5mm}t}({t,x})X\rangle$ and $\langle T^t_{\hspace{1.5mm}x}({t,x})X\rangle$ as the complex variable $x$ approaches the spatial insertions $\{x_p\}$ (real valued). The appearance of $\theta({t}-{t_p})$ discontinuity (in real variable $t$) in these Ward identities, on the other hand, is the manifestation of time-ordering inside the correlators. This step-function plays an important role in the operator formalism of $1+1$D CCFT.

\medskip

Thus, a general $1+1$D CC field with scaling dimension $\Delta$ and boost-charge $\xi$ and pCB rank $l$ satisfies the following OPEs, keeping in mind that OPEs make sense only if the L.H.S. are time-ordered: 
\begin{align}
&T^t_{\hspace{1.5mm}x}({t^\prime,x^\prime})\Phi_{(l)}^m(t,x)\sim -i{\theta(t^\prime-t)}\left[...+\frac{{{\Delta}}\Phi_{(l)}^m(t,x)}{{(x^\prime-x)}^2}+\frac{{\partial_{x}}\Phi_{(l)}^m(t,x)}{x^\prime-x}\right.\nonumber\\
&\left.\hspace{73mm}-({t^\prime}-{t})\left(...+\frac{{{2\bm{\xi}}}\Phi_{(l)}^m(t,x)}{{(x^\prime-x)}^3}+\frac{{\partial_{t}}\Phi_{(l)}^m(t,x)}{{(x^\prime-x)}^2}\right)\right]\nonumber\\
&T^t_{\hspace{1.5mm}t}(t^\prime,x^\prime)\Phi_{(l)}^m(t,x)\sim -i{\theta(t^\prime-t)}\left[...+\frac{\bm{\xi}\Phi_{(l)}^{m}(t,x)}{{(x^\prime-x)}^2}+\frac{{\partial_{t}} \Phi_{(l)}^m(t,x)}{x^\prime-x}\right]\label{eq:24}
\end{align}
where $...$ denotes higher-order poles in $ x^\prime-x$ and as before, $\sim$ denotes `modulo terms holomorphic in $x^\prime-x$ inside $[...]$ that have vanishing VEVs'. 

\medskip

From these Ward identities and OPEs, comparing the scaling behavior of both hand sides we readily conclude that:
\begin{align*}
\text{the EM tensor components $T^t_{\hspace{1.5mm}t}$ and $T^t_{\hspace{1.5mm}x}$ both have scaling dimension $\Delta=2$ .}
\end{align*} 

\medskip

If we insist that there is no field in the theory with negative scaling dimension, the above OPEs terminate by having finite order poles at $x^\prime=x$. For example, if the field $\Phi_{(l)}^m(t,x)$ has scaling dimension $\Delta$, the pole in the super-translation OPE is at most of the order $\lfloor \Delta\rfloor+2$, since each term in the OPE has a scaling dimension $\Delta+2$.

\medskip

\subsection{Ward identities for primary fields}\label{106}
Having derived the general (and somewhat schematic) form of the $1+1$D CCFT Ward identities, we now specialize to the case when all the fields appearing in $X$ are primaries. As we shall see, in this case the Ward identities are completely determined due to the `nice' transformation properties of the primary fields.

\medskip
 
As previously noted, a field transforming under any $1+1$D CC transformation \eqref{55} as:
\begin{align}
\Phi_{(l)}^m(t,x)\longrightarrow\tilde{\Phi}_{(l)}^m(t^\prime,x^\prime)={(\partial_x x^\prime)}^{-\Delta}{\left[e^{-\xi \frac{\partial_x t^\prime}{\partial_x x^\prime}\mathbf{J}_{(l)}^{\mathbf{-}}}\right]}^m_{\hspace{2mm} m^\prime}{\Phi}_{(l)}^{m^\prime}(t,x)\nonumber
\end{align}
is called a CC primary field with scaling dimension $\Delta$ and boost-charge $\xi$ and pCB rank $l$. From this, the generators of the infinitesimal CC transformation \eqref{eq:4} over the primary fields are readily obtained as:
\begin{align}
&iG_x\Phi_{(l)}^m(t,x)=[f(x)\partial_x+tf^\prime(x)\partial_t+\Delta f^\prime(x)]\Phi_{(l)}^m(t,x)+tf^{\prime\prime}(x)\xi\sqrt{l(l+1)-m(m-1)}\Phi_{(l)}^{m-1}(t,x) \nonumber\\
&iG_t\Phi_{(l)}^m(t,x)=g(x)\partial_t\Phi_{(l)}^m(t,x)+g^{\prime}(x)\xi\sqrt{l(l+1)-m(m-1)}\Phi_{(l)}^{m-1}(t,x)\label{51}
\end{align}
Thus, the transformation \eqref{eq:4} is a local quantum symmetry if a correlator $\langle X\rangle$ of only primary fields satisfies the following Ward identity; we reach there by starting from \eqref{44} and using \eqref{eq:7}:
\begin{align}
&\sum_{i=1}^n\langle{\mathcal{T}}\Phi_1(\mathbf{x_1}))...(i\epsilon^aG_a\Phi_i(\mathbf{x_i}))...\Phi_n(\mathbf{x_n})\rangle=\frac{1}{2\pi}\int dt\oint\limits_{C}dx\text{ }\epsilon^a\langle{\mathcal{T}}\partial_\mu j^\mu_{\hspace{1.5mm}a}(\mathbf{x})X\rangle\nonumber\\
\Rightarrow\text{ }&\epsilon^x\sum_{p=1}^n\text{ }[f(x_p)\partial_{x_p}+t_pf^\prime(x_p)\partial_{t_p}+\Delta_p f^\prime(x_p)+t_pf^{\prime\prime}(x_p)\bm{\xi}_p]\langle X\rangle+\epsilon^t\sum_{p=1}^n\text{ }[g(x_p)\partial_{t_p}+g^{\prime}(x_p)\bm{\xi}_p]\langle X\rangle\nonumber\\
=&\text{\hspace{2.5mm} }\frac{\epsilon^x}{2\pi}\int dt\oint\limits_{C} dx\text{ }[f(x)\partial_\mu\langle T^\mu_{\hspace{1.5mm}x} X\rangle+tf^\prime(x)\partial_\mu\langle T^\mu_{\hspace{1.5mm}t} X\rangle+f^\prime(x)\langle T^\mu_{\hspace{1.5mm}\mu} X\rangle+tf^{\prime\prime}(x)\langle T^x_{\hspace{1.5mm}t} X\rangle] \nonumber\\
&+\frac{\epsilon^t}{2\pi}\int dt\oint\limits_{C} dx\text{ }[g(x)\partial_\mu\langle T^\mu_{\hspace{1.5mm}t} X\rangle+g^\prime(x)\langle T^x_{\hspace{1.5mm}t} X\rangle]\label{80}
\end{align}
Since $f(x)$ and $g(x)$ are arbitrary but are assumed to be non-singular in the region enclosed by the contour $C$, it must hold from \eqref{eq:6} that:
\begin{align}
&\partial_\mu\langle T^\mu_{\hspace{1.5mm}\nu}(\mathbf{x})X\rangle\sim -i\sum_{p=1}^n\text{ }\delta(t-{t_p})\text{ }\frac{{\partial_{\nu_p}}\langle X\rangle}{x-x_p}\label{eq:8}\\
&\langle T^x_{\hspace{1.5mm}t}(\mathbf{x})X\rangle\sim -i\sum\limits_{p=1}^n\text{ }\delta({t}-{t_p})\text{ }\frac{{{\bm{\xi}_p}}\langle X\rangle}{x-x_p}\label{eq:9}\\
&\langle T^\mu_{\hspace{1.5mm}\mu}(\mathbf{x})X\rangle\sim -i\sum\limits_{p=1}^n\text{ }\delta({t}-{t_p})\text{ }\frac{\Delta_p\langle X\rangle}{x-x_p}\label{eq:10}
\end{align}
i.e. all the higher order poles must vanish in this case. These relations lead to the super-translation and super-rotation Ward identities respectively, for primary fields:
\begin{align}
&i\langle T^t_{\hspace{1.5mm}t}(t,x)X\rangle=\sum\limits_{p=1}^n\text{ }\theta({t}-{t_p})\text{ }\left[\frac{{{\bm{\xi}_p}}\langle X\rangle}{{(x-x_p)}^2}+\frac{{\partial_{t_p}}\langle X\rangle}{x-x_p}\right]\label{eq:11}\\
&i\langle T^t_{\hspace{1.5mm}x}({t,x})X\rangle=\sum\limits_{p=1}^n\text{ }{\theta({t}-{t_p})}\left[\frac{{{\Delta_p}}\langle X\rangle}{{(x-x_p)}^2}+\frac{{\partial_{x_p}}\langle X\rangle}{x-x_p}-({t}-{t_p})\left(\frac{{{2\bm{\xi}_p}}\langle X\rangle}{{(x-x_p)}^3}+\frac{{\partial_{t_p}}\langle X\rangle}{{(x-x_p)}^2}\right)\right]\label{eq:12}
\end{align}

\medskip

Thus, a $1+1$D CC primary field with scaling dimension $\Delta$ and boost-charge $\xi$ and pCB rank $l$ satisfies the following OPEs: 
\begin{align}
&T^t_{\hspace{1.5mm}x}({t^\prime,x^\prime})\Phi_{(l)}^m(t,x)\sim -i{\theta(t^\prime-t)}\left[\frac{{{\Delta}}\Phi_{(l)}^m(t,x)}{{(x^\prime-x)}^2}+\frac{{\partial_{x}}\Phi_{(l)}^m(t,x)}{x^\prime-x}\right.\nonumber\\
&\left.\hspace{73mm}-({t^\prime}-{t})\left(\frac{{{2\bm{\xi}}}\Phi_{(l)}^m(t,x)}{{(x^\prime-x)}^3}+\frac{{\partial_{t}}\Phi_{(l)}^m(t,x)}{{(x^\prime-x)}^2}\right)\right]\label{eq:13}\\
&T^t_{\hspace{1.5mm}t}(t^\prime,x^\prime)\Phi_{(l)}^m(t,x)\sim -i{\theta(t^\prime-t)}\left[\frac{\bm{\xi}\Phi_{(l)}^{m}(t,x)}{{(x^\prime-x)}^2}+\frac{{\partial_{t}} \Phi_{(l)}^m(t,x)}{x^\prime-x}\right]\label{eq:14}
\end{align}
The absence of pole-singularities of order $>2$ in OPE \eqref{eq:14} and in the part of OPE \eqref{eq:13} that is not the coefficient of $t^\prime-t$ is an alternative defining feature of the $1+1$D CC primary fields.

\medskip

\subsection{Ward identities for quasi-primary fields}\label{107}
The `minimal' infinitesimal version of the $1+1$D global CC transformation, in the form \eqref{eq:4}, has:
\begin{align*}
f(x)=\alpha x^2+\beta x+\gamma \text{\hspace{4.5mm}and\hspace{4.5mm}} g(x)=\lambda x^2+\mu x+\nu
\end{align*}
Thus, a correlator $\langle X\rangle$ of only quasi-primary fields satisfies \eqref{80} only with $f(x)$ and $g(x)$ both being at most quadratic polynomials. Thus, translation, boost and dilation Ward identities respectively take the following forms:
\begin{align}
&\partial_\mu\langle T^\mu_{\hspace{1.5mm}\nu}(\mathbf{x})X\rangle\sim -i\sum\limits_{p=1}^n\text{ }\delta({t}-{t_p})\text{ }\left[\frac{{\partial_{\nu_p}}\langle X\rangle}{x-x_p}+\sum_{k\geq4}\text{ }\frac{{\langle Y^{(k)}_\nu\rangle}_p(\mathbf{x_1},...\mathbf{x_n})}{{(x-x_p)}^k}\right]\\
&\langle T^x_{\hspace{1.5mm}t}(\mathbf{x})X\rangle\sim -i\sum\limits_{p=1}^n\text{ }\delta({t}-{t_p})\left[\frac{{ {\bm{\xi}}_p\langle X\rangle}}{x-x_p}+\sum_{k\geq3}\frac{{\langle Z^{(k)}_t\rangle}_p(\mathbf{x_1},...\mathbf{x_n})}{{(x-x_p)}^k}\right]\\
&\langle T^\mu_{\hspace{1.5mm}\mu}(\mathbf{x})X\rangle\sim -i\sum\limits_{p=1}^n\text{ }\delta({t}-{t_p})\left[\frac{\Delta_p\langle X\rangle}{x-x_p}+\sum_{k\geq3}\text{ }\frac{{\langle Z_x^{(k)}\rangle}_p(\mathbf{x_1},...\mathbf{x_n})}{{(x-x_p)}^k}\right]
\end{align}
Again, apart from the simple poles, only those other poles are retained in these Ward identities that do not identically contribute to the contour integral (by Cauchy integral formula) in \eqref{80}. These then lead to the super-tarnslation and super-rotation Ward identities for a quasi-primary correlator: 
\begin{align}
&i\langle T^t_{\hspace{1.5mm}t}(t,x)X\rangle=\sum\limits_{p=1}^n\theta(t-t_p)\left[\text{ }\sum_{k\geq4}\text{ }\frac{{\langle Y^{(k)}_t\rangle}_p+(k-1){\langle Z_t^{(k-1)}\rangle}_p}{{(x-x_p)}^k}+\frac{{{\bm{\xi}_p}}\langle X\rangle}{{(x-x_p)}^2}+\frac{{\partial_{t_p}}\langle X\rangle}{x-x_p}\right]\label{eq:18}\\
&i\langle T^t_{\hspace{1.5mm}x}({t,x})X\rangle=\sum\limits_{p=1}^n\theta({t}-{t_p})\left[\text{ }\sum_{k\geq4}\text{ }\frac{{\langle Y^{(k)}_x\rangle}_p+(k-1){\langle Z_x^{(k-1)}\rangle}_p}{{(x-x_p)}^k}+\frac{{{\Delta_p}}\langle X\rangle}{{(x-x_p)}^2}+\frac{{\partial_{x_p}}\langle X\rangle}{x-x_p}\right.\nonumber\\
&\left. \hspace{21mm}-({t}-{t_p})\left(\text{ }\sum_{k\geq4}\text{ }k\text{ }\frac{{\langle Y^{(k)}_t\rangle}_p+(k-1){\langle Z_t^{(k-1)}\rangle}_p}{{(x-x_p)}^{k+1}}+\frac{{{2\bm{\xi}_p}}\langle X\rangle}{{(x-x_p)}^3}+\frac{{\partial_{t_p}}\langle X\rangle}{{(x-x_p)}^2}\right)\right]\label{eq:19}
\end{align}

\medskip

Thus, a $1+1$D CC quasi-primary field with scaling dimension $\Delta$ and boost-charge $\xi$ and pCB rank $l$ satisfies the following OPEs: 
\begin{align}
&T^t_{\hspace{1.5mm}x}({t^\prime,x^\prime})\Phi_{(l)}^m(t,x)\sim -i{\theta(t^\prime-t)}\left[...+\frac{{{\Delta}}\Phi_{(l)}^m(t,x)}{{(x^\prime-x)}^2}+\frac{{\partial_{x}}\Phi_{(l)}^m(t,x)}{x^\prime-x}\right.\nonumber\\
&\left.\hspace{73mm}-({t^\prime}-{t})\left(...+\frac{{{2\bm{\xi}}}\Phi_{(l)}^m(t,x)}{{(x^\prime-x)}^3}+\frac{{\partial_{t}}\Phi_{(l)}^m(t,x)}{{(x^\prime-x)}^2}\right)\right]\nonumber\\
&T^t_{\hspace{1.5mm}t}(t^\prime,x^\prime)\Phi_{(l)}^m(t,x)\sim -i{\theta(t^\prime-t)}\left[...+\frac{\bm{\xi}\Phi_{(l)}^{m}(t,x)}{{(x^\prime-x)}^2}+\frac{{\partial_{t}} \Phi_{(l)}^m(t,x)}{x^\prime-x}\right]\label{81}
\end{align}
where $...$ denotes poles of order $\geq4$ in $x^\prime-x$. The absence of third-order poles in the part of the OPEs \eqref{81} that is not the coefficient of $t^\prime-t$ is an alternative defining feature of the $1+1$D CC quasi-primary fields. These OPEs will terminate at finite-order poles if we assume the non-existence of fields with negative scaling dimensions.

\medskip

\section{Time Ordering and OPE}\label{84}
The OPEs are meaningful only when the products of the two operators involved are time-ordered. The operator formalism works by distinguishing a time direction from the space directions. But so far, to discuss on the OPEs in $1+1$D CCFT, we have only used the path-integral formalism and got only a hint of time-ordering via the appearance of the temporal $\theta$-functions in the OPEs. To build further on this hint, we now begin the exploration of the operator formalism of $1+1$D CCFT.

\medskip

The distinction between space and time is evidently quite natural from the point of view of Carrollian transformations. Thus, as we proceed, we shall see that we do not need to assume any exotic `radial quantization' prescription here, contrary to 2D (Euclidean) CFTs. In 2D CFTs, radial ordering on the complex plane is manifestly related to time-ordering \cite{DiFrancesco:1997nk} in a theory living on a cylinder $\mathbb{R}\times {S}^1$. The 2D CFT plane-to-cylinder map very transparently expresses this fact. To the best of our knowledge, the corresponding $1+1$D CCFT `plane-to-cylinder' map existing in the literature \cite{Bagchi:2013qva} does not have any such physical interpretation without resorting to an analytic-continuation to Euclidean time \cite{Hao:2021urq}. Fortunately, as we shall see, we do not require the service of such a map to establish the operator formalism here. Rather, the temporal $\theta$-functions appearing in the OPEs, that can be naturally related to time-ordering, make a very significant contribution in relating the OPEs with the operator commutation relations via a complex contour integral (over $x$) prescription. Due to the richness of complex analytic methods, the operator formalism of $1+1$D CCFT will provide us with an extremely powerful computational tool.  

\medskip
  
In the operator formalism of QFT, the conserved charge $Q_a$ is the generator of an infinitesimal symmetry transformation on the space of the quantum fields:
\begin{align}
Q_a=\int\limits_{\Sigma^d} d^d\vec{x}\text{ }j^t_{\hspace{1.5mm}a}(t,\vec{x})\hspace{5mm};\hspace{5mm}\delta_{\bm{\epsilon}}{\Phi}(t,\vec{x})=-i\epsilon^a[Q_a\text{ },\text{ }{\Phi}(t,\vec{x})]\label{eq:21}
\end{align}
where $\Sigma^d$ is a space-like hypersurface. Since all field operators within a correlator must be time-ordered, so must be the L.H.S. of an OPE if it is to have an operator meaning. This statement has the following interpretation (in the limit $t^\pm=t+0^\pm$):
\begin{align}
[Q_a\text{ },{\Phi}(t,\vec{x})]=\int\limits_{\Sigma^d} d^d\vec{x}^\prime\text{ }j^t_{\hspace{1.5mm}a}(t^+,\vec{x}^\prime){\Phi}(t,\vec{x})-\int\limits_{\Sigma^d} d^d\vec{x}^\prime\text{ }{\Phi}(t,\vec{x})j^t_{\hspace{1.5mm}a}(t^-,\vec{x}^\prime)\text{\hspace{7.5mm}(as an OPE)}\label{78}
\end{align}
i.e. in the R.H.S., the OPE between the current $j^t_{\hspace{1.5mm}a}$ and the field ${\Phi}$ is to be used.

\medskip
 
In $1+1$D CCFT, the classical version of the conserved charges generating the transformation \eqref{eq:4} are given as:
\begin{align}
Q_x[f]=\int\limits_{\mathbb{R}\cup\{\infty\}} dx\text{ }\left[T^t_{\hspace{1.5mm}x}(t,x)f(x)+T^t_{\hspace{1.5mm}t}(t,x)f^\prime(x)t\right]\hspace{4mm},\hspace{4mm}Q_t[g]=\int\limits_{\mathbb{R}\cup\{\infty\}} dx\text{ }T^t_{\hspace{1.5mm}t}(t,x)g(x)\label{eq:25}
\end{align}
directly from the definition \eqref{eq:21} and using the form \eqref{72} for the currents. But since, to derive the Ward identities, $x$ was analytically continued from the Riemann circle to the Riemann sphere, we need to be careful to determine the integration region or contour for defining the quantum charges. For this purpose, the following calculation is considered. 

\medskip

Under \eqref{eq:4}, a general correlator $\langle X\rangle$ suffers the following infinitesimal change:
\begin{align*}
\delta_{\bm{\epsilon}}\langle X\rangle&=i\int dt\int\limits_{\mathbb{R}\cup\{\infty\}} dx\text{ }\epsilon^a\langle\partial_\mu j^\mu_{\hspace{1.5mm}a}(\mathbf{x})X\rangle\text{\hspace{15mm}from \eqref{44}}\\
&=\frac{\epsilon^a}{2\pi}\oint\limits_{C} dx\int dt\text{ }\partial_\mu\langle j^\mu_{\hspace{1.5mm}a}(t,{x})X\rangle\text{\hspace{15mm}using \eqref{eq:7}}\\
&=\frac{\epsilon^x}{2\pi}\oint\limits_{C} dx\int d{t}\text{ }\left[f(x)\partial_\mu\langle T^\mu_{\hspace{1.5mm}x} X\rangle+tf^\prime(x)\partial_\mu\langle T^\mu_{\hspace{1.5mm}t} X\rangle+f^\prime(x)\langle T^\mu_{\hspace{1.5mm}\mu} X\rangle+tf^{\prime\prime}(x)\langle T^x_{\hspace{1.5mm}t} X\rangle\right]\\
&+\frac{\epsilon^t}{2\pi}\oint\limits_{C} dx\int d{t}\text{ }\left[g(x)\partial_\mu\langle T^\mu_{\hspace{1.5mm}t} X\rangle+g^{\prime}(x)\langle T^x_{\hspace{1.5mm}t} X\rangle\right]\text{\hspace{15mm}using \eqref{72}}
\end{align*}
where the counter-clockwise contour $C$ encloses all the points of spatial insertion $\{x_p\}$ of the fields in $X$. Next, using the Ward identities \eqref{eq:15}-\eqref{eq:17} with `$\sim$' replaced by `$=$' in those equations, we obtain ($f^{(k)}$ and $g^{(k)}$ being $k$-th derivatives):
\begin{align*}
&\delta_{\bm{\epsilon}}\langle X\rangle\\
=&-\epsilon^x\sum_{p=1}^n\left[\left\{f(x_p)\partial_{x_p}+\Delta_pf^{(1)}(x_p)+f^{(1)}(x_p)t_p\partial_{t_p}+t_p\bm{\xi}_pf^{(2)}(x_p)\right\}\langle X\rangle\right.\nonumber\\
&\left.+\sum_{k\geq3}\frac{f^{(k-1)}(x_p){\langle Z^{(k-1)}_x\rangle}_p+t_pf^{(k)}(x_p){\langle Z^{(k-1)}_t\rangle}_p}{(k-2)!}+\sum_{k\geq3}\frac{f^{(k-1)}(x_p){\langle Y^{(k)}_x\rangle}_p+t_pf^{(k)}(x_p){\langle Y^{(k)}_t\rangle}_p}{(k-1)!}\right]\\
&-\epsilon^t\sum_{p=1}^n\left[\left\{g(x_p)\partial_{t_p}+\bm{\xi}_pg^{(1)}(x_p)\right\}\langle X\rangle+\sum_{k\geq3}\frac{g^{(k-1)}(x_p){\langle Y^{(k)}_t\rangle}_p}{(k-1)!}+\sum_{k\geq3}\frac{g^{(k-1)}(x_p){\langle Z^{(k-1)}_t\rangle}_p}{(k-2)!}\right]
\end{align*}
Clearly, to obtain the above, one needs to assume that $f(x)$ and $g(x)$ are non-singular inside the region enclosed by the contour $C$. We can now immediately re-express the above as the following:
\begin{align}
&\delta_{\bm{\epsilon}}\langle X\rangle\nonumber\\
=&\text{ }\frac{i\epsilon^x}{2\pi}\oint\limits_C dx\text{ } f(x)\sum\limits_{p=1}^n\text{ }\left[\text{ }\sum_{k\geq3}\text{ }\frac{{\langle Y^{(k)}_x\rangle}_p+(k-1){\langle Z_x^{(k-1)}\rangle}_p}{{(x-x_p)}^k}+\frac{{{\Delta_p}}\langle X\rangle}{{(x-x_p)}^2}+\frac{{\partial_{x_p}}\langle X\rangle}{x-x_p}\right.\nonumber\\
&\left. \text{ }\text{ }\text{ }\text{ }\text{ }\text{ }\text{ }\text{ }\text{ }\text{ }\text{ }\text{ }\text{ }\text{ }\text{ }\text{ }-(t-{t_p})\left(\text{ }\sum_{k\geq3}\text{ }k\text{ }\frac{{\langle Y^{(k)}_t\rangle}_p+(k-1){\langle Z_t^{(k-1)}\rangle}_p}{{(x-x_p)}^{k+1}}+\frac{{{2\bm{\xi}_p}}\langle X\rangle}{{(x-x_p)}^3}+\frac{{\partial_{t_p}}\langle X\rangle}{{(x-x_p)}^2}\right)\right]\nonumber\\
&+\frac{i\epsilon^x}{2\pi}\oint\limits_C dx\text{ } tf^\prime(x)\sum\limits_{p=1}^n\text{ }\left(\text{ }\sum_{k\geq3}\text{ }\text{ }\frac{{\langle Y^{(k)}_t\rangle}_p+(k-1){\langle Z_t^{(k-1)}\rangle}_p}{{(x-x_p)}^{k}}+\frac{{{\bm{\xi}_p}}\langle X\rangle}{{(x-x_p)}^2}+\frac{{\partial_{t_p}}\langle X\rangle}{{x-x_p}}\right)\nonumber\\
&+\frac{i\epsilon^t}{2\pi}\oint\limits_C dx\text{ } g(x)\sum\limits_{p=1}^n\text{ }\left[\text{ }\sum_{k\geq3}\text{ }\frac{{\langle Y^{(k)}_t\rangle}_p+(k-1){\langle Z_t^{(k-1)}\rangle}_p}{{(x-x_p)}^k}+\frac{{{\bm{\xi}_p}}\langle X\rangle}{{(x-x_p)}^2}+\frac{{\partial_{t_p}}\langle X\rangle}{x-x_p}\right]\label{77}
\end{align}
Recognizing the appearance of the super-rotation and super-translation Ward identities \eqref{eq:23} and \eqref{eq:22} in this expression, we are now led to the so-called $1+1$D CC Ward identity:
\begin{align}
\delta_{\bm{\epsilon}}\langle X\rangle=-\frac{\epsilon^t}{2\pi}\oint\limits_C dx\text{ } g(x)&\langle T^t_{\hspace{1.5mm}t}(t,x)X\rangle\bigg|_{t>\{t_p\}}\nonumber\\
&-\frac{\epsilon^x}{2\pi}\oint\limits_C dx\text{ }\left[f(x)\langle T^t_{\hspace{1.5mm}x}(t,x)X\rangle+tf^{\prime}(x)\langle T^t_{\hspace{1.5mm}t}(t,x)X\rangle\right]\bigg|_{t>\{t_p\}}
\label{eq:20}
\end{align}
The condition $t>\{t_p\}$ evidently arises from the existence of the temporal $\theta$-function in \eqref{eq:22} and \eqref{eq:23}.

\medskip

This expression can be further manipulated to be brought into another useful form. The importance of the $\theta$-function factor in the Ward identities will now become manifest. We proceed by using the property of the $\theta$-function to write:
\begin{align}
-\delta_{\bm{\epsilon}}\langle X\rangle&=\frac{\epsilon^t}{2\pi}\sum\limits_{p=1}^n\oint\limits_{C} dx\left[g(x)\langle T^t_{\hspace{1.5mm}t}(t_p^+,x)X\rangle-g(x)\langle T^t_{\hspace{1.5mm}t}(t_p^-,x)X\rangle\right]+\frac{\epsilon^x}{2\pi}\sum\limits_{p=1}^n\oint\limits_{C} dx\left[f(x)\langle T^t_{\hspace{1.5mm}x}(t_p^+,x)X\rangle\right.\nonumber\\
&\left.+t_p^+f^{\prime}(x)\langle T^t_{\hspace{1.5mm}t}(t_p^+,x)X\rangle-f(x)\langle T^t_{\hspace{1.5mm}x}(t_p^-,x)X\rangle-t_p^-f^{\prime}(x)\langle T^t_{\hspace{1.5mm}t}(t_p^-,x)X\rangle\right]\label{38}
\end{align}
In both of \eqref{eq:20} and \eqref{38}, we notice the appearances of the current-components $j^t_{\hspace{1.5mm}x}$ and $j^t_{\hspace{1.5mm}t}$ corresponding to the transformation \eqref{eq:4}, inside of the correlators that are the integrands of the contour integrals over $x$. A spatial integral of a $j^t_{\hspace{1.5mm}a}$ component of a current operator should give rise to the quantum conserved charge operator. Since, the conserved charge operator remains unvaried with time, we are free to evaluate the spatial integral of the charge density operator at any time. Using this trick and comparing \eqref{38} to the correlator version of \eqref{78}, we conclude that:
\begin{align}
-\delta_{\bm{\epsilon}}\langle X\rangle=&\frac{\epsilon^t}{2\pi}\sum\limits_{p=1}^n\text{ }\langle\Phi_1(t_1,x_1)...[\oint\limits_{C} dx\text{ }g(x) T^t_{\hspace{1.5mm}t}(t,x)\text{ },\Phi_p(t_p,x_p)]...\Phi_n(t_n,x_n)\rangle\nonumber\\
+&\frac{\epsilon^x}{2\pi}\sum\limits_{p=1}^n\text{ }\langle\Phi_1(t_1,x_1)...[\oint\limits_{C} dx\left\{f(x) T^t_{\hspace{1.5mm}x}(t,x)+tf^{\prime}(x) T^t_{\hspace{1.5mm}t}(t,x)\right\},\Phi_p(t_p,x_p)]...\Phi_n(t_n,x_n)\rangle\nonumber\\
=&\frac{\epsilon^t}{2\pi}\langle[\oint\limits_{C} dx\left\{g(x) T^t_{\hspace{1.5mm}t}(t,x)\right\},X]\rangle+\frac{\epsilon^x}{2\pi}\langle[\oint\limits_{C} dx\left\{f(x) T^t_{\hspace{1.5mm}x}(t,x)+tf^{\prime}(x) T^t_{\hspace{1.5mm}t}(t,x)\right\},X]\rangle
\end{align}
This is another version of the $1+1$D CC Ward identity. Comparing it to \eqref{eq:21}, we obtain:
\begin{align}
&\langle[Q_x[f],X]\rangle=\frac{1}{2\pi i}\langle[\oint\limits_{C} dx\left\{f(x) T^t_{\hspace{1.5mm}x}(t,x)+tf^{\prime}(x) T^t_{\hspace{1.5mm}t}(t,x)\right\},X]\rangle\nonumber\\
&\langle[Q_t[g],X]\rangle=\frac{1}{2\pi i}\langle[\oint\limits_{C} dx\left\{g(x) T^t_{\hspace{1.5mm}t}(t,x)\right\},X]\rangle\label{eq:31}
\end{align} 
where the contour $C$ is the same as defined earlier. 

\medskip

Thus, by simply removing the $\langle...\rangle$ symbol from \eqref{eq:31}, we reach the contour-integral prescription relating OPEs with commutation relations in $1+1$D CCFT without using any radial-quantization procedure. As has been demonstrated, it is the temporal $\theta$-function in the OPEs that makes this prescription possible.

\medskip

Since \eqref{eq:31} is valid for any $X$, for a field $\Phi(t,x)$, we have, e.g. with $Q_t[g]$:  
\begin{align}
[Q_t[g]\text{ },{\Phi}(t,x)]=\frac{1}{2\pi i}\oint\limits_{x} dx^\prime\text{ }g(x^\prime){\mathcal{T}}T^t_{\hspace{1.5mm}t}(t^+,x^\prime){\Phi}(t,x)-\frac{1}{2\pi i}\oint\limits_{x} dx^\prime\text{ }g(x^\prime){\mathcal{T}}T^t_{\hspace{1.5mm}t}(t^-,x^\prime){\Phi}(t,x)\label{eq:26}
\end{align}
where the integration contour encloses $x$ and inside the enclosed region, $g(x^\prime)$ is non-singular. Due to the presence of the step-function, it is evident from \eqref{eq:24} that the OPE $T^t_{\hspace{1.5mm}t}(t^-,x^\prime){\Phi}(t,x)$ has no singularity at $x^\prime=x$ but $T^t_{\hspace{1.5mm}t}(t^+,x^\prime){\Phi}(t,x)$ has a pole at $x^\prime=x$. Since, clearly only singular terms of the OPE contribute to this integral by Cauchy integral theorem, \eqref{eq:26} is simplified into: 
\begin{align}
[Q_t[g]\text{ },{\Phi}(t,x)]&=\frac{1}{2\pi i}\oint\limits_{x} dx^\prime\text{ }g(x^\prime)T^t_{\hspace{1.5mm}t}(t^+,x^\prime){\Phi}(t,x)
\end{align} 
Similar simplification occurs for $Q_x[f]$ also.

\medskip

Inspired by the above discussion, we propose the contour prescription to relate OPEs with commutation relations in $1+1$D CCFT for any conserved charge\footnote{The contour in its definition will be specified later.} operator $Q_a$:
\begin{align}
Q_a=\frac{1}{2\pi i}\oint dx\text{ }j^t_{\hspace{1.5mm}a}(t,x)\text{\hspace{5mm}generates\hspace{5mm}}[Q_a\text{ },{\Phi}(t,x)]=\frac{1}{2\pi i}\oint\limits_{x} dx^\prime\text{ }j^t_{\hspace{1.5mm}a}(t^+,x^\prime){\Phi}(t,x)\label{eq:27}
\end{align}
Here, $Q_a$ may also be a charge generating an `internal' field transformation. This will be applicable e.g. in the analysis of $1+1$D CCFT with affine gauge symmetry.

\medskip

\subsection{The `$i\epsilon$' prescription}\label{108}
The $\theta$-function factors in the OPEs and Ward identities look manifestly non-covariant with jump-discontinuities in the real variable $t$. This makes it difficult to analytically continue back to real variable $x$. The following `$i\epsilon$' prescription comes to the remedies of these problems.

\medskip
 
As noted earlier, the $1+1$D CC Ward identity can be recast into the alternative expression \eqref{38}:
\begin{align*}
-\delta_{\bm{\epsilon}}\langle X\rangle&=\frac{\epsilon^t}{2\pi}\sum\limits_{p=1}^n\oint\limits_{C} dx\left[g(x)\langle T^t_{\hspace{1.5mm}t}(t_p^+,x)X\rangle-g(x)\langle T^t_{\hspace{1.5mm}t}(t_p^-,x)X\rangle\right]+\frac{\epsilon^x}{2\pi}\sum\limits_{p=1}^n\oint\limits_{C} dx\left[f(x)\langle T^t_{\hspace{1.5mm}x}(t_p^+,x)X\rangle\right.\nonumber\\
&\left.+t_p^+f^{\prime}(x)\langle T^t_{\hspace{1.5mm}t}(t_p^+,x)X\rangle-f(x)\langle T^t_{\hspace{1.5mm}x}(t_p^-,x)X\rangle-t_p^-f^{\prime}(x)\langle T^t_{\hspace{1.5mm}t}(t_p^-,x)X\rangle\right]
\end{align*}
which motivates the following `$i\epsilon$' form of the super-translation and the super-rotation Ward identities respectively, with $\Delta \tilde{x}_p:=x-x_p-i\epsilon(t-t_p)$ :
\begin{align}
&\langle T^t_{\hspace{1.5mm}t}(t,x)X\rangle=\lim\limits_{\epsilon\rightarrow0^+}-i\sum\limits_{p=1}^n\left[\sum_{k\geq3}\text{ }\frac{{\langle Y^{(k)}_t\rangle}_p+(k-1){\langle Z_t^{(k-1)}\rangle}_p}{{(\Delta \tilde{x}_p)}^k}+\frac{{{\bm{\xi}_p}}\langle X\rangle}{{(\Delta \tilde{x}_p)}^2}+\frac{{\partial_{t_p}}\langle X\rangle}{\Delta \tilde{x}_p}\right]\label{40}\\
&\langle T^t_{\hspace{1.5mm}x}({t,x})X\rangle=\lim\limits_{\epsilon\rightarrow0^+}-i\sum\limits_{p=1}^n\left[\sum_{k\geq3}\text{ }\frac{{\langle Y^{(k)}_x\rangle}_p+(k-1){\langle Z_x^{(k-1)}\rangle}_p}{{(\Delta \tilde{x}_p)}^k}+\frac{{{\Delta_p}}\langle X\rangle}{{(\Delta \tilde{x}_p)}^2}+\frac{{\partial_{x_p}}\langle X\rangle}{\Delta \tilde{x}_p}\right.\nonumber\\
&\left.\hspace{19mm}-(t-t_p)\left(\sum_{k\geq3}\text{ }k\text{ }\frac{{\langle Y^{(k)}_t\rangle}_p+(k-1){\langle Z_t^{(k-1)}\rangle}_p}{{(\Delta \tilde{x}_p)}^{k+1}}+\frac{{{2\bm{\xi}_p}}\langle X\rangle}{{(\Delta \tilde{x}_p)}^3}+\frac{{\partial_{t_p}}\langle X\rangle}{{(\Delta \tilde{x}_p)}^2}\right)\right]\label{41}
\end{align}
i.e., since all the $\{x_p\}$ are real, the poles of these Laurent series (as functions of the complex variable $x$) are pushed into the upper-half plane for $t>t_p$ and into the lower-half plane for $t<t_p$. 

\medskip

Equivalently, in the OPE language, we have, with $\Delta \tilde{x}^\prime:=x^\prime-x-i\epsilon(t^\prime-t)$ :
\begin{align}
&T^t_{\hspace{1.5mm}x}({t^\prime,x^\prime})\Phi_{(l)}^m(t,x)\sim\lim\limits_{\epsilon\rightarrow0^+} -i\left[...+\frac{{{\Delta}}\Phi_{(l)}^m(t,x)}{{(\Delta\tilde{x}^\prime)}^2}+\frac{{\partial_{x}}\Phi_{(l)}^m(t,x)}{\Delta\tilde{x}^\prime}\right.\nonumber\\
&\left.\hspace{77mm}-({t^\prime}-{t})\left(...+\frac{{{2\bm{\xi}}}\Phi_{(l)}^m(t,x)}{{(\Delta\tilde{x}^\prime)}^3}+\frac{{\partial_{t}}\Phi_{(l)}^m(t,x)}{{(\Delta\tilde{x}^\prime)}^2}\right)\right]\nonumber\\
&T^t_{\hspace{1.5mm}t}(t^\prime,x^\prime)\Phi_{(l)}^m(t,x)\sim\lim\limits_{\epsilon\rightarrow0^+} -i\left[...+\frac{\bm{\xi}\Phi_{(l)}^{m}(t,x)}{{(\Delta\tilde{x}^\prime)}^2}+\frac{{\partial_{t}} \Phi_{(l)}^m(t,x)}{\Delta\tilde{x}^\prime}\right]\label{42}
\end{align} 
where $\sim$ denotes `modulo terms regular in $x^\prime$ '.

\medskip

Since in a QFT, the correlators should be treated as distributions, the equivalence of the `$i\epsilon$' form of the Ward identities and OPEs to their temporal $\theta$-function involving counterpart will be revealed when the above Ward identities are used to calculate $\delta_{\bm{\epsilon}}\langle X\rangle$. For this purpose, we begin by noticing the following equality in a sample contour integral calculation (with $\Delta x_p\equiv x-x_p$ and $\Delta t_p\equiv t-t_p$ and prescriptions are denoted as subscripts):
\begin{align}
&\frac{1}{2\pi i}\oint\limits_{C_u} dx\text{ } g(x)\langle T^t_{\hspace{1.5mm}t}(t,x)X\rangle_{i\epsilon}\nonumber\\
=&\lim\limits_{\epsilon\rightarrow0^+}\frac{-i}{2\pi i}\oint\limits_{C_u} dx\text{ } g(x)\sum\limits_{p=1}^n\left[\sum_{k\geq3}\text{ }\frac{{\langle Y^{(k)}_t\rangle}_p+(k-1){\langle Z_t^{(k-1)}\rangle}_p}{{(\Delta x_p-i\epsilon\Delta t_p)}^k}+\frac{{{\bm{\xi}_p}}\langle X\rangle}{{(\Delta x_p-i\epsilon\Delta t_p)}^2}+\frac{{\partial_{t_p}}\langle X\rangle}{\Delta x_p-i\epsilon\Delta t_p}\right]\nonumber\\
=&-i\sum_{\substack{p\\t>t_p}}\left[\left\{g(x_p)\partial_{t_p}+\bm{\xi}_pg^{(1)}(x_p)\right\}\langle X\rangle+\sum_{k\geq3}\text{ }g^{(k-1)}(x_p)\text{ }\frac{{\langle Y^{(k)}_t\rangle}_p+(k-1){\langle Z_t^{(k-1)}\rangle}_p}{(k-1)!}\right]\nonumber\\
=&\frac{1}{2\pi i}\oint\limits_{C} dx\text{ }g(x)\langle T^t_{\hspace{1.5mm}t}(t_p^+,x)X\rangle_{\theta}\label{39}
\end{align}
with $C_u$ being the contour depicted below, enclosing the upper-half plane and $C$ is the contour enclosing all the positions of insertion $\{x_p\}$ as before. 
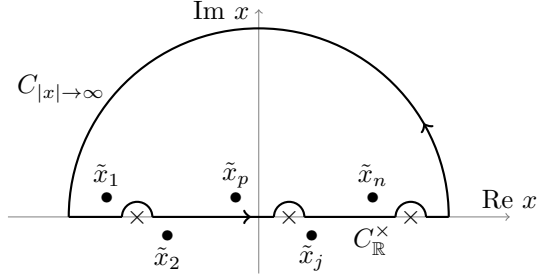
\begin{figure}[h]
\begin{center}
\begin{tikzpicture}[decoration={markings,
mark=at position 1.3cm with {\arrow[line width=1pt]{>}}}]
\draw[help lines,->] (-3.3,0) -- (3.3,0) coordinate (xaxis);
\draw[help lines,->] (0,-0.75) -- (0,2.75) coordinate (yaxis);
\path[draw,line width=0.8pt,postaction=decorate] (-2.5,0)  (2.5,0) arc (0:180:2.5);
\path[draw,line width=0.8pt,postaction=decorate] (0.2,0)  (.6,0) arc (0:180:.2);
\path[draw,line width=0.8pt,postaction=decorate] (1.8,0)  (2.2,0) arc (0:180:.2);
\path[draw,line width=0.8pt,postaction=decorate] (-1.8,0)  (-1.4,0) arc (0:180:.2);
\path[draw,line width=0.8pt,postaction=decorate] (-2.5,0)--(-1.8,0);
\path[draw,line width=0.8pt,postaction=decorate] (.6,0)--(1.8,0);
\path[draw,line width=0.8pt,postaction=decorate] (2.2,0)--(2.5,0);
\path[draw,line width=0.8pt,postaction=decorate] (-1.4,0)--(0.2,0);
\node[above] at (xaxis) {$\text{Re }x$};
\node[left] at (yaxis) {$\text{Im }x$};
\draw (.4,0) node{$\times$};
\draw (-1.6,0) node{$\times$};
\draw (2,0) node{$\times$};
\draw (1.5,0.25) node{$\bullet$};
\draw (-2.0,0.25) node{$\bullet$};
\draw (-.3,0.25) node{$\bullet$};
\draw (-1.2,-0.25) node{$\bullet$};
\draw (.7,-0.25) node{$\bullet$};
\node at (1.5,0.55) {$\tilde{x}_n$};
\node at (-.3,0.55) {$\tilde{x}_p$};
\node at (-2.0,0.55) {$\tilde{x}_1$};
\node at (-1.2,-0.55) {$\tilde{x}_2$};
\node at (0.7,-0.55) {$\tilde{x}_j$};
\node at (1.5,-0.3) {$C_{\mathbb{R}}^{\times}$};
\node at (-2.6,1.7) {$C_{|x|\rightarrow\infty}$};
\end{tikzpicture}
\caption{`$i\epsilon$' prescription for $1+1$D CCFT OPEs : $C_u=C_{\mathbb{R}}^\times+C_{|x|\rightarrow\infty}$ and $\tilde{x}_i\equiv x_i+i\epsilon(t-t_i)$ where $t$ is the `time of insertion' of the conserved charge density operator; $\times$ are the singularities of the vector field. The integration over the $C_{\mathbb{R}}^\times$ part of the contour is a real integration in the Cauchy principal value sense.} 
\end{center}
\end{figure}

\medskip

\eqref{39} shows the equivalence (in the sense of distribution) of the two prescriptions. Thus, within the `$i\epsilon$' prescription, the $1+1$D CC Ward identity \eqref{eq:20} is re-expressed as:
\begin{align}
&\delta_{\bm{\epsilon}}\langle X\rangle=-\frac{\epsilon^t}{2\pi}\oint\limits_{C_u} dx\text{ } g(x)\langle T^t_{\hspace{1.5mm}t}(t,x)X\rangle\bigg|_{t>\{t_p\}}\nonumber\\
&\hspace{32mm}-\frac{\epsilon^x}{2\pi}\oint\limits_{C_u} dx\text{ }\left[f(x)\langle T^t_{\hspace{1.5mm}x}(t,x)X\rangle+tf^{\prime}(x)\langle T^t_{\hspace{1.5mm}t}(t,x)X\rangle\right]\bigg|_{t>\{t_p\}}\\
\Rightarrow &\hspace{13mm}\langle[Q_x[f],X]\rangle=\frac{1}{2\pi i}\langle[\oint\limits_{C_u} dx\left\{f(x) T^t_{\hspace{1.5mm}x}(t,x)+tf^{\prime}(x) T^t_{\hspace{1.5mm}t}(t,x)\right\},X]\rangle\bigg|_{t>\{t_p\}}\nonumber\\
&\hspace{13mm}\langle[Q_t[g],X]\rangle=\frac{1}{2\pi i}\langle[\oint\limits_{C_u} dx\left\{g(x) T^t_{\hspace{1.5mm}t}(t,x)\right\},X]\rangle\bigg|_{t>\{t_p\}}
\end{align}
i.e. the integrands have all their poles in the upper half plane.

\medskip

Finally, we have gathered all the ingredients to fix the contour in the definition of the quantum conserved charge. Recall that this exercise turned into a non-trivial affair because $x$ was analytically continued from the Riemann circle to the Riemann sphere. We proceed to show that a conserved charge operator indeed has the contour integral representation as stated in \eqref{eq:27}, through the example \eqref{eq:26} using the OPE version of \eqref{39}:
\begin{align}
[Q_t[g]\text{ },{\Phi}(t,x)]&=\frac{1}{2\pi i}\oint\limits_{x} dx^\prime\text{ }g(x^\prime){\mathcal{T}}T^t_{\hspace{1.5mm}t}(t^+,x^\prime){\Phi}(t,x)-\frac{1}{2\pi i}\oint\limits_{x} dx^\prime\text{ }g(x^\prime){\mathcal{T}}T^t_{\hspace{1.5mm}t}(t^-,x^\prime){\Phi}(t,x)\nonumber\\
&=\frac{1}{2\pi i}\oint\limits_{C_u} dx^\prime\text{ }g(x^\prime){\mathcal{T}}T^t_{\hspace{1.5mm}t}(t^+,x^\prime){\Phi}(t,x)-\frac{1}{2\pi i}\oint\limits_{C_u} dx^\prime\text{ }g(x^\prime){\mathcal{T}}T^t_{\hspace{1.5mm}t}(t^-,x^\prime){\Phi}(t,x)\nonumber\\
&=\frac{1}{2\pi i}\oint\limits_{C^\prime_u} dx^\prime\text{ }g(x^\prime){\mathcal{T}}T^t_{\hspace{1.5mm}t}(t^\prime(>t),x^\prime){\Phi}(t,x)-\frac{1}{2\pi i}\oint\limits_{C^\prime_u} dx^\prime\text{ }g(x^\prime){\mathcal{T}}T^t_{\hspace{1.5mm}t}(t^\prime(<t),x^\prime){\Phi}(t,x)\nonumber\\
&=[\text{ }\frac{1}{2\pi i}\oint\limits_{C^\prime_u} dx^\prime\text{ }g(x^\prime)T^t_{\hspace{1.5mm}t}(t^\prime,x^\prime)\text{ },\text{ }{\Phi}(t,x)\text{ }]
\end{align}
where the contour $C^\prime_u$ encloses the upper half plane as well as the whole of the real line (hence, singularities, if any, of the vector field). The transition from the second to the third line in the above calculation is captured pictorially below: 
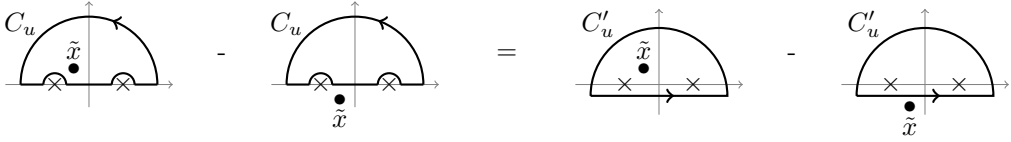
\begin{figure}[h]
\begin{center}
\begin{tikzpicture}[decoration={markings,
mark=at position 1.1cm with {\arrow[line width=1pt]{>}}}]
\draw[help lines,->] (-1.1,0) -- (1.1,0) coordinate (xaxis);
\draw[help lines,->] (0,-0.3) -- (0,1.1) coordinate (yaxis);
\path[draw,line width=0.8pt,postaction=decorate] (-.9,0)  (.9,0) arc (0:180:.9);
\path[draw,line width=0.8pt,postaction=decorate] (-0.6,0)  (-.3,0) arc (0:180:.15);
\path[draw,line width=0.8pt,postaction=decorate] (0.3,0)  (.6,0) arc (0:180:.15);
\path[draw,line width=0.8pt,postaction=decorate] (-.9,0)--(-.6,0);
\path[draw,line width=0.8pt,postaction=decorate] (-.3,0)--(.3,0);
\path[draw,line width=0.8pt,postaction=decorate] (.6,0)--(.9,0);
\draw (-.45,0) node{$\times$};
\draw (.45,0) node{$\times$};
\draw (-.2,0.2) node{$\bullet$};
\node at (-.2,0.45) {$\tilde{x}$};
\node at (-.9,.8) {$C_{u}$};
\draw[help lines,->] (2.4,0) -- (4.6,0) coordinate (xaxis);
\draw[help lines,->] (3.5,-0.3) -- (3.5,1.1) coordinate (yaxis);
\path[draw,line width=0.8pt,postaction=decorate] (2.6,0)  (4.4,0) arc (0:180:.9);
\path[draw,line width=0.8pt,postaction=decorate] (2.9,0)  (3.2,0) arc (0:180:.15);
\path[draw,line width=0.8pt,postaction=decorate] (3.8,0)  (4.1,0) arc (0:180:.15);
\path[draw,line width=0.8pt,postaction=decorate] (2.6,0)--(2.9,0);
\path[draw,line width=0.8pt,postaction=decorate] (3.2,0)--(3.8,0);
\path[draw,line width=0.8pt,postaction=decorate] (4.1,0)--(4.4,0);
\draw (3.05,0) node{$\times$};
\draw (3.95,0) node{$\times$};
\draw (3.3,-0.2) node{$\bullet$};
\node at (3.3,-0.45) {$\tilde{x}$};
\node at (2.6,.8) {$C_{u}$};
\node at (1.75,.4) {-};
\draw[help lines,->] (6.4,0) -- (8.6,0) coordinate (xaxis);
\draw[help lines,->] (7.5,-0.4) -- (7.5,1) coordinate (yaxis);
\path[draw,line width=0.8pt,postaction=decorate] (6.6,-0.15)--(8.4,-.15) arc (0:180:.9);
\draw (7.05,0) node{$\times$};
\draw (7.95,0) node{$\times$};
\draw (7.3,0.2) node{$\bullet$};
\node at (7.3,0.45) {$\tilde{x}$};
\node at (6.7,.8) {$C^\prime_{u}$};
\node at (5.5,.4) {=};
\draw[help lines,->] (9.9,0) -- (12.1,0) coordinate (xaxis);
\draw[help lines,->] (11,-0.4) -- (11,1) coordinate (yaxis);
\path[draw,line width=0.8pt,postaction=decorate] (10.1,-0.15)--(11.9,-.15) arc (0:180:.9);
\draw (10.55,0) node{$\times$};
\draw (11.45,0) node{$\times$};
\draw (10.8,-0.3) node{$\bullet$};
\node at (10.8,-0.55) {$\tilde{x}$};
\node at (10.2,.8) {$C^\prime_{u}$};
\node at (9.25,.4) {-};
\end{tikzpicture}
\caption{Equality of subtractions of contours in `$i\epsilon$' prescription} 
\end{center}
\end{figure}

\medskip

In the third and the fourth line we keep arbitrary $t^\prime$, since any calculation involving the conserved charge operator is expected ultimately to be independent of the time of insertion of the charge density operator, after performing the contour integral along the $C^\prime_u$ contour. This is a crucial step in the similar calculation involving the $Q_x[f]$ charge.
  
\medskip

Thus, we are led to the definition of the quantum conserved charge $Q_t[g]$:
\begin{align*}
Q_t[g]=\frac{1}{2\pi i}\oint\limits_{C^\prime_u} dx^\prime\text{ }g(x^\prime)T^t_{\hspace{1.5mm}t}(t^\prime,x^\prime)
\end{align*}
or, more generally, an arbitrary quantum conserved charge $\hat{Q}_a$ is given by:
\begin{align}
Q^{\text{classical}}_a=\int\limits_{\mathbb{R}\cup\{\infty\}} dx^\prime\text{ }j^t_{\hspace{1.5mm}a}(t^\prime,x^\prime)\text{ }\longrightarrow\text{ }\hat{Q}_a=\frac{1}{2\pi i}\oint\limits_{C^\prime_u} dx^\prime\text{ }j^t_{\hspace{1.5mm}a}(t^\prime,x^\prime)\label{50}
\end{align}
that gives rise to the following generator equation:
\begin{align}
[\hat{Q}_a\text{ },{\Phi}(t,x)]=\frac{1}{2\pi i}\oint\limits_{C_u} dx^\prime\text{ }j^t_{\hspace{1.5mm}a}(t^+,x^\prime){\Phi}(t,x)\label{52}
\end{align}

\medskip

\textbf{Comment 4}: We now clarify some of the assumptions by considering the equivalence between the two prescriptions:
\begin{enumerate}
\item All of the preceding calculations in this subsection seem to hold even if $i\epsilon\Delta t_i$ are replaced\footnote{Arbitrary odd powers of $\Delta t_i$ are not considered so as to give $\epsilon$ the dimension of speed.} by $i\lambda\Delta t_i$ with $\text{Re }\lambda>0$ in \eqref{40}-\eqref{42}. That it is not the case can be seen by comparing e.g. \eqref{43} to $\partial_t$\eqref{40}$\big|_\lambda$ at $t\neq\{t_p\}$: since the first one vanishes, so must the later that is possible only if $\text{Re }\lambda\rightarrow0^+$ and $\text{Im }\lambda=0$.
\item The two assumptions in the $\theta$-prescription:
\begin{align*}
\lim\limits_{t\rightarrow-\infty}\langle T^t_{\hspace{1.5mm}t}(t,x)X\rangle=0 \text{\hspace{4.5mm} and\hspace{4.5mm}} \lim\limits_{t\rightarrow-\infty}\langle T^t_{\hspace{1.5mm}x}(t,x)X\rangle=0
\end{align*}
have clearer meaning (in the sense of distribution) in the `$i\epsilon$'-prescription: in the $1+1$D CC Ward identity, their contributions are always 0 which is evident from the contour prescription \eqref{39}, since all the poles are pushed into the lower-half plane. 
\end{enumerate}
It is now trivial to analytically continue the OPEs and Ward identities \eqref{40}-\eqref{42} back to the real $x$.

\medskip

\section{Correlation Functions}\label{132}
Having discussed on the invariance of the correlation functions via the Ward identities and the OPEs, we shall now try to actually find the general structures of the same in $1+1$D quantum CCFT.

\medskip

The correlation functions are defined as the vacuum expectation values of the time-ordered products of fields. As stated before, in a $1+1$D CCFT, the fields inside the correlator symbol $\langle...\rangle$ must all be inserted at different spatial locations\footnote{This rules out the possibility of having the spatial Dirac delta-function as a factor of the correlation functions.}. Whenever the spatial location of two or more fields coincide, they must be treated as one composite operator. To avoid ambiguity in time-ordering, we also insert no two fields at a same time.    

\medskip

The vacuum is taken to be invariant under the global subalgebra of the quantum CCA$_{1+1}$ i.e. the quantum vacuum state is annihilated by the generators of the global quantum subalgebra. We also assume that the vacuum state is unique (i.e. non-degenerate).

\medskip

The readers willing to skip the details of this section may note that the main results are \eqref{45}, \eqref{49} and \eqref{82}. 

\medskip
   
\subsection{2-point quasi-primary correlator}\label{110}
\subsubsection*{2-point correlators of bosonic vector quasi-primaries:}
We first calculate the correlation function between two bosonic `vector' quasi-primary fields. $1+1$D Carrollian vector fields transform as rank-$\frac{1}{2}$ spherical multiplets with $\xi\neq0$ under pCB. As we shall see, an important example of a quasi-primary vector field is the EM tensor in $1+1$D CCFT. We denote the vector field components as $\Phi^\pm(t,x)$.

\medskip

The 2-point correlators $\langle\mathcal{T}\Phi^{s_1}_{\Delta_1,\xi_1}(t_1,x_1)\Phi^{s_2}_{\Delta_2,\xi_2}(t_2,x_2)\rangle$ (with $x_1\neq x_2$) are functions of $t_{12}:= t_1-t_2$ and $x_{12}:= x_1-x_2$ as a consequence of global translation-invariance, with $s_i\in\{-,+\}$; so, we denote:
\begin{align*}
\langle\mathcal{T}\Phi^{s_1}_{\Delta_1,\xi_1}(t_1,x_1)\Phi^{s_2}_{\Delta_2,\xi_2}(t_2,x_2)\rangle\equiv G^{s_1;s_2}_{\Delta_1,\xi_1;\Delta_2,\xi_2}(t_{12},x_{12})
\end{align*}
Due to invariance under the remaining $1+1$D global CC transformations, the 2-point correlators of the bosonic vector quasi-primaries satisfy the following coupled PDEs, due to \eqref{79}:
\begin{align}
&\left[x_{12}\partial_{t_{12}}+\bm{\xi}_1+\bm{\xi}_2\right]\text{ }G^{s_1;s_2}_{\Delta_1,\xi_1;\Delta_2,\xi_2}=0\nonumber\\
&\left[x_{12}\partial_{x_{12}}+t_{12}\partial_{t_{12}}+{\Delta}_1+\Delta_2\right]\text{ }G^{s_1;s_2}_{\Delta_1,\xi_1;\Delta_2,\xi_2}=0\label{46}\\
&\left[(x_1^2-x_{2}^2)\partial_{t_{12}}+2x_1\bm{\xi}_1+2x_2\bm{\xi}_2\right]\text{ }G^{s_1;s_2}_{\Delta_1,\xi_1;\Delta_2,\xi_2}=0\nonumber\\
&\left[(x_1^2-x_{2}^2)\partial_{x_{12}}+2(x_1t_1-x_2t_2)\partial_{t_{12}}+2(x_1\Delta_1+t_1\bm{\xi}_1)+2(x_2\Delta_2+t_2\bm{\xi}_2)\right]\text{ }G^{s_1;s_2}_{\Delta_1,\xi_1;\Delta_2,\xi_2}=0\nonumber
\end{align}
Since $x_{12}\neq0$, the solutions are obtained as (with real $x_{12}$):
\begin{align}
&G^{-;-}_{\Delta_1,\xi_1;\Delta_2,\xi_2}=0\nonumber\\
&G^{+;-}_{\Delta_1,\xi_1;\Delta_2,\xi_2}(t_{12},x_{12})=\frac{\lambda\xi_1\delta_{\Delta_1,\Delta_2}}{{|x_{12}|}^{2\Delta_1}}\hspace{5mm};\hspace{5mm}G^{-;+}_{\Delta_1,\xi_1;\Delta_2,\xi_2}(t_{12},x_{12})=\frac{\lambda\xi_2\delta_{\Delta_1,\Delta_2}}{{|x_{12}|}^{2\Delta_1}}\nonumber\\
&G^{+;+}_{\Delta_1,\xi_1;\Delta_2,\xi_2}(t_{12},x_{12})=\frac{\delta_{\Delta_1,\Delta_2}}{{|x_{12}|}^{2\Delta_1}}\left(\mu-2\lambda\xi_1\xi_2\frac{t_{12}}{x_{12}}\right)\label{32}
\end{align}
where $\mu$ and $\lambda$ are two independent 2-point constants. Thus, these correlators vanish if the scaling dimensions of the two fields involved are unequal. No such `selection rule' is imposed on the pCB charge.

\medskip

A desired property of the 2-point correlator is single-valuedness under the transformation $x\rightarrow xe^{2\pi i}$. To concretely investigate on this issue, we choose, without loss of generality, that: $x_{12}>0$. The 2-point correlators in \eqref{32} are evidently single-valued only if $2\Delta\in\mathbb{N}$. Negative integer values for $2\Delta$ are forbidden so as to prevent the following unphysical behavior of the 2-point correlators: had $\Delta$ been negative, the almost-equal-time correlator would increase infinitely with increase in spatial separation whereas at almost coincident spatial locations, the correlator would vanish. Thus, we assume that there is no field with negative scaling dimension in the field theory and the Identity field is the only primary field with $\Delta=0$.

\medskip

We notice from \eqref{32} that (for real $x_{12}\neq0$):
\begin{align}
\langle\mathcal{T}\Phi^{s_1}_{\Delta,\xi_1}(t_1,x_1)\Phi^{s_2}_{\Delta,\xi_2}(t_2,x_2)\rangle=\langle\mathcal{T}\Phi^{s_1}_{\Delta,\xi_1}(t_2,x_2)\Phi^{s_2}_{\Delta,\xi_2}(t_1,x_1)\rangle\label{47}
\end{align}
Moreover, due to the assumed bosonic property of the fields involved, it is expected that:
\begin{align}
&\langle\mathcal{T}\Phi^{s_1}_{\Delta_1,\xi_1}(t_1,x_1)\Phi^{s_2}_{\Delta_2,\xi_2}(t_2,x_2)\rangle=\langle\mathcal{T}\Phi^{s_2}_{\Delta_2,\xi_2}(t_2,x_2)\Phi^{s_1}_{\Delta_1,\xi_1}(t_1,x_1)\rangle\label{48}\\
\Longrightarrow\text{ }&\langle\mathcal{T}\Phi^{+}_{\Delta,\xi_1}(t_1,x_1)\Phi^{-}_{\Delta,\xi_2}(t_2,x_2)\rangle=\langle\mathcal{T}\Phi^{-}_{\Delta,\xi_2}(t_2,x_2)\Phi^{+}_{\Delta,\xi_1}(t_1,x_1)\rangle=\frac{\lambda\xi_1}{{|x_{12}|}^{2\Delta}}\nonumber\\
&\langle\mathcal{T}\Phi^{-}_{\Delta,\xi_1}(t_1,x_1)\Phi^{+}_{\Delta,\xi_2}(t_2,x_2)\rangle=\langle\mathcal{T}\Phi^{+}_{\Delta,\xi_2}(t_2,x_2)\Phi^{-}_{\Delta,\xi_1}(t_1,x_1)\rangle=\frac{\lambda\xi_2}{{|x_{12}|}^{2\Delta}}\hspace{5mm}\text{ (for real $x_{12}\neq0$)}\nonumber\\
&\langle\mathcal{T}\Phi^{+}_{\Delta,\xi_1}(t_1,x_1)\Phi^{+}_{\Delta,\xi_2}(t_2,x_2)\rangle=\langle\mathcal{T}\Phi^{+}_{\Delta,\xi_2}(t_2,x_2)\Phi^{+}_{\Delta,\xi_1}(t_1,x_1)\rangle=\frac{1}{{|x_{12}|}^{2\Delta}}\left(\mu-2\lambda\xi_1\xi_2\frac{t_{12}}{x_{12}}\right)\nonumber
\end{align}

\medskip

Since, the super-translation and super-rotation Ward identities, as functions of $\{\Delta x_p\}$ (real or complex) and $\{\Delta t_p\}$ take the form \eqref{40}-\eqref{41}, any two-point function should have a similar functional dependence on $t_{12}$ and complex $x_{12}$. For that, we need to analytically continue to complex $x_{12}$ on the Riemann sphere.  Thus, the analytically continued form of the bosonic 2-point correlators \eqref{32} are as follows (with $\tilde{x}_{12}:=x_{12}-i\epsilon t_{12}$):
\begin{align}
\langle\mathcal{T}\Phi^{-}_{\Delta_1,\xi_1}(t_1,x_1)\Phi^{-}_{\Delta_2,\xi_2}(t_2,x_2)\rangle&=\langle\mathcal{T}\Phi^{-}_{\Delta_2,\xi_2}(t_2,x_2)\Phi^{-}_{\Delta_1,\xi_1}(t_1,x_1)\rangle=0\nonumber\\
\langle\mathcal{T}\Phi^{+}_{\Delta_1,\xi_1}(t_1,x_1)\Phi^{-}_{\Delta_2,\xi_2}(t_2,x_2)\rangle&=\langle\mathcal{T}\Phi^{-}_{\Delta_2,\xi_2}(t_2,x_2)\Phi^{+}_{\Delta_1,\xi_1}(t_1,x_1)\rangle=\lim\limits_{\epsilon\rightarrow0^+}\frac{\lambda\xi_1\delta_{\Delta_1,\Delta_2}}{{\left(\tilde{x}_{12}\right)}^{2\Delta_1}}\nonumber\\
\langle\mathcal{T}\Phi^{-}_{\Delta_1,\xi_1}(t_1,x_1)\Phi^{+}_{\Delta_2,\xi_2}(t_2,x_2)\rangle&=\langle\mathcal{T}\Phi^{+}_{\Delta_2,\xi_2}(t_2,x_2)\Phi^{-}_{\Delta_1,\xi_1}(t_1,x_1)\rangle=\lim\limits_{\epsilon\rightarrow0^+}\frac{\lambda\xi_2\delta_{\Delta_1,\Delta_2}}{{\left(\tilde{x}_{12}\right)}^{2\Delta_1}}\nonumber\\
\langle\mathcal{T}\Phi^{+}_{\Delta_1,\xi_1}(t_1,x_1)\Phi^{+}_{\Delta_2,\xi_2}(t_2,x_2)\rangle&=\langle\mathcal{T}\Phi^{+}_{\Delta_2,\xi_2}(t_2,x_2)\Phi^{+}_{\Delta_1,\xi_1}(t_1,x_1)\rangle\nonumber\\
&=\lim\limits_{\epsilon\rightarrow0^+}\frac{\delta_{\Delta_1,\Delta_2}}{{\left(\tilde{x}_{12}\right)}^{2\Delta_1}}\left(\mu-2\lambda\xi_1\xi_2\frac{t_{12}}{\tilde{x}_{12}}\right)\label{45}
\end{align}   

\medskip

From \eqref{45}, we further note that, e.g.:
\begin{align}
\langle\mathcal{T}\Phi^{-}_{\Delta,\xi_2}(t_1,x_1)\Phi^{+}_{\Delta,\xi_1}(t_2,x_2)\rangle=\lim\limits_{\epsilon\rightarrow0^+}\frac{\lambda\xi_1}{{\left(\tilde{x}_{21}\right)}^{2\Delta}}={(-)}^{2\Delta}\langle\mathcal{T}\Phi^{-}_{\Delta,\xi_2}(t_2,x_2)\Phi^{+}_{\Delta,\xi_1}(t_1,x_1)\rangle
\end{align}
Thus, \eqref{47} holds for any complex $x_{12}$ if $\Delta\in\mathbb{N}$, i.e. the bosonic quasi-primary fields (of any rank-$l$) must have positive integer scaling dimensions.

\medskip

The above 2-point correlators depend on $x_{12}$ through the quantity $x_{12}-i\epsilon t_{12}$ where $\epsilon\rightarrow0^+$ has the dimension of speed. This may be thought of as the manifestation of the fact that the Carrollian limit is formally $c\rightarrow0^+$ (where $c$ is the speed of light).

\medskip

\subsubsection*{2-point correlators of arbitrary bosonic quasi-primaries:}
The method of solving the PDEs arising from \eqref{79} to find the 2-point correlators of two quasi-primary multiplets of arbitrary ranks is quite inefficient because of the large number of coupled PDEs involved. Fortunately, we have an alternative route leading to the 2-point correlators directly from symmetry considerations. 

\medskip

We begin by stating the finite version of \eqref{eq:5} : under a space-time transformation $\mathbf{x}\rightarrow \mathbf{x}^\prime$, the fields schematically transform as:
\begin{align}
{\Phi}(\mathbf{x})\longrightarrow{\tilde{\Phi}}(\mathbf{x}^\prime)={\mathcal{F}}({\Phi}(\mathbf{x}))
\end{align}
If this transformation is a symmetry of the action then, under the assumption of invariance of the path-integral measure, the correlators satisfy the following identity \cite{DiFrancesco:1997nk}:
\begin{align}
\langle\mathcal{T}{\Phi_1}(\mathbf{x_1^\prime}){\Phi_2}(\mathbf{x_2^\prime})\ldots{\Phi_n}(\mathbf{x_n^\prime})\rangle=\langle\mathcal{T}{\mathcal{F}}({\Phi_1}(\mathbf{x_1})){\mathcal{F}}({\Phi_2}(\mathbf{x_2}))\ldots{\mathcal{F}}({\Phi_n}(\mathbf{x_n}))\rangle\label{75}
\end{align}  
Translation-invariance implies that the correlators are functions of $\mathbf{x_i}-\mathbf{x_j}$. Scale-invariance is expressed as ($\lambda>0$):
\begin{align}
\langle\mathcal{T}{\Phi_1}(\lambda\mathbf{x_1}){\Phi_2}(\lambda\mathbf{x_2})\ldots{\Phi_n}(\lambda\mathbf{x_n})\rangle=\lambda^{-(\Delta_1+\Delta_2+\ldots+\Delta_n)}\langle\mathcal{T}{\Phi_1}(\mathbf{x_1}){\Phi_2}(\mathbf{x_2})\ldots{\Phi_n}(\mathbf{x_n})\rangle
\end{align}
while invariance under a time-ordering preserving $1+1$D `plane' Carrollian boost imposes the following constraint ($t_i^\prime=t_i+vx_i$):
\begin{align}
&\langle\mathcal{T}\Phi_{(l_1)}^{m_1}(t_1^\prime,x_1)\ldots\Phi_{(l_n)}^{m_n}(t_n^\prime,x_n)\rangle\nonumber\\
=\text{ }&{\left[e^{-\xi_1 v\mathbf{J}_{(l_1)}^{\mathbf{-}}}\right]}^{m_1}_{\hspace{3mm} m_1^\prime}\text{ }\ldots\text{ }{\left[e^{-\xi_n v\mathbf{J}_{(l_n)}^{\mathbf{-}}}\right]}^{m_n}_{\hspace{3mm} m_n^\prime}\text{ }\langle\mathcal{T}\Phi_{(l_1)}^{m_1^\prime}(t_1,x_1)\ldots\Phi_{(l_n)}^{m_n^\prime}(t_n,x_n)\rangle
\end{align}
For a correlator of only (bosonic) quasi-primary fields, additional constraints will come from transformation properties \eqref{eq:3} of (bosonic) quasi-primaries under special CCTs.

\medskip

All of the constraints imposed by the global CC symmetry together completely and uniquely determine the coordinate dependence of the two-point and three-point correlators of CC quasi-primaries. Due to relative simplicity, we shall calculate only the two-point functions in this paper.

\medskip

Using the facts that:
\begin{align*}
&{\left[{\left(\mathbf{J}_{(l)}^{\mathbf{-}}\right)}^n\right]}^{m}_{\hspace{3mm} m^\prime}={\left[\frac{(l+m)!(l-m^\prime)!}{(l-m)!(l+m^\prime)!}\right]}^{\frac{1}{2}}\delta_{n,m-m^\prime}\\\Longrightarrow\text{ }&{\left[e^{a\mathbf{J}_{(l)}^{\mathbf{-}}}\right]}^{m}_{\hspace{3mm} m^\prime}=\frac{a^{m-m^\prime}}{(m-m^\prime)!}{\left[\frac{(l+m)!(l-m^\prime)!}{(l-m)!(l+m^\prime)!}\right]}^{\frac{1}{2}}
\end{align*}
the (time-ordered) correlators of two non-scalar bosonic quasi-primary fields are obtained as (for real $x_{12}\neq0$):
\begin{align}
&\langle{\mathcal{T}}\Phi_{(l_1);\Delta_1,\xi_1}^{m_1}(t_1,x_1)\Phi_{(l_2);\Delta_2,\xi_2}^{m_2}(t_2,x_2)\rangle\nonumber\\
=\text{ }&\frac{\xi_1^{m_1-l_1}\xi_2^{m_2-l_2}\delta_{\Delta_1,\Delta_2}}{|x_{12}|^{2\Delta_1}}{\left[\frac{(l_1+m_1)!(l_2+m_2)!}{(l_1-m_1)!(l_2-m_2)!}\right]}^{\frac{1}{2}}\underset{m_1^\prime+m_2^\prime\geq l_1-l_2}{\sum\limits_{m_1^\prime=-l_1}^{m_1}\sum\limits_{m_2^\prime=-l_2}^{m_2}}\frac{{\left(-\frac{t_{12}}{x_{12}}\right)}^{m_1+m_2-m_1^\prime-m_2^\prime}}{\left(m_1-m_1^\prime\right)!\left(m_2-m_2^\prime\right)!}\nonumber\\
&\times C^{l_1;m_1^\prime+m_2^\prime-l_1}_{l_1,\Delta_1,\xi_1;l_2,\Delta_1,\xi_2}\text{ }\xi_2^{l_1+l_2-m_1^\prime-m_2^\prime}{\left[\frac{(l_1+l_2-m_1^\prime-m_2^\prime)!}{(2l_1)!(m_1^\prime+m_2^\prime-l_1+l_2)!}\right]}^{\frac{1}{2}}\qquad\hspace{10mm}\text{(for $l_1\geq l_2$)}\\
&\nonumber\\
&\text{Or}\nonumber\\
&\nonumber\\
=\text{ }&\frac{\xi_1^{m_1-l_1}\xi_2^{m_2-l_2}\delta_{\Delta_1,\Delta_2}}{|x_{12}|^{2\Delta_1}}{\left[\frac{(l_1+m_1)!(l_2+m_2)!}{(l_1-m_1)!(l_2-m_2)!}\right]}^{\frac{1}{2}}\underset{m_1^\prime+m_2^\prime\geq l_2-l_1}{\sum\limits_{m_1^\prime=-l_1}^{m_1}\sum\limits_{m_2^\prime=-l_2}^{m_2}}\frac{{\left(-\frac{t_{12}}{x_{12}}\right)}^{m_1+m_2-m_1^\prime-m_2^\prime}}{\left(m_1-m_1^\prime\right)!\left(m_2-m_2^\prime\right)!}\nonumber\\
&\times C^{m_1^\prime+m_2^\prime-l_2;l_2}_{l_1,\Delta_1,\xi_1;l_2,\Delta_1,\xi_2}\text{ }\xi_1^{l_1+l_2-m_1^\prime-m_2^\prime}{\left[\frac{(l_1+l_2-m_1^\prime-m_2^\prime)!}{(2l_2)!(m_1^\prime+m_2^\prime-l_2+l_1)!}\right]}^{\frac{1}{2}}\qquad\hspace{10mm}\text{(for $l_2\geq l_1$)}
\end{align}
where the $C^{\ldots}_{\ldots}$ are mutually independent 2-point coefficients, $2\min{\{l_1,l_2\}}+1$ in number. Thus, these correlators vanish if the scaling dimensions of the fields involved are unequal. When non-zero, the correlator is a polynomial in $\frac{t_{12}}{x_{12}}$ of degree $m_1+m_2-|l_1-l_2|$, multiplied by a power-law factor of ${|x_{12}|}^{-2\Delta}$. We see that no such conclusion is applicable for the pCB charge or for the ranks of the multiplets. 

\medskip

From the two above correlation functions, it appears that the 2-point coefficients between two fields are depending on the relative position of those fields inside the correlator. This must not be true. Indeed, the bosonic property gives rise to the following relation between the two sets of the 2-point coefficients:
\begin{align}
&\langle{\mathcal{T}}\Phi_{(l_1);\Delta,\xi_1}^{m_1}(t_1,x_1)\Phi_{(l_2);\Delta,\xi_2}^{m_2}(t_2,x_2)\rangle=\langle{\mathcal{T}}\Phi_{(l_2);\Delta,\xi_2}^{m_2}(t_2,x_2)\Phi_{(l_1);\Delta,\xi_1}^{m_1}(t_1,x_1)\rangle\nonumber\\
\Longrightarrow\hspace{5mm}&C^{l_1;m_1+m_2-l_1}_{l_1,\Delta,\xi_1;l_2,\Delta,\xi_2}=C^{m_1+m_2-l_1;l_1}_{l_2,\Delta,\xi_2;l_1,\Delta,\xi_1}\hspace{10mm}\text{(for $l_1\geq l_2$)}
\end{align}

\medskip

For $l_1=l_2=l$, the 2-point coefficients satisfy also the following consistency condition:
\begin{align}
{\left(\frac{\xi_1}{\xi_2}\right)}^{l-m_2}C^{m_1+m_2-l;l}_{l,\Delta,\xi_1;l,\Delta,\xi_2}={\left(\frac{\xi_2}{\xi_1}\right)}^{l-m_1}C^{l;m_1+m_2-l}_{l,\Delta,\xi_1;l,\Delta,\xi_2}
\end{align}

\medskip

We now note down the analytically continued version of the 2-point bosonic quasi-primary correlator as the following (as before, $\tilde{x}_{12}:=x_{12}-i\epsilon t_{12}$) :
\begin{align}
&\langle{\mathcal{T}}\Phi_{(l_1);\Delta_1,\xi_1}^{m_1}(t_1,x_1)\Phi_{(l_2);\Delta_2,\xi_2}^{m_2}(t_2,x_2)\rangle=\langle{\mathcal{T}}\Phi_{(l_2);\Delta_2,\xi_2}^{m_2}(t_2,x_2)\Phi_{(l_1);\Delta_1,\xi_1}^{m_1}(t_1,x_1)\rangle\nonumber\\
=\text{ }&\lim\limits_{\epsilon\rightarrow0^+}\text{ }\frac{\xi_1^{m_1-l_1}\xi_2^{m_2-l_2}\delta_{\Delta_1,\Delta_2}}{{(\tilde{x}_{12})}^{2\Delta_1}}{\left[\frac{(l_1+m_1)!(l_2+m_2)!}{(l_1-m_1)!(l_2-m_2)!}\right]}^{\frac{1}{2}}\underset{m_1^\prime+m_2^\prime\geq l_1-l_2}{\sum\limits_{m_1^\prime=-l_1}^{m_1}\sum\limits_{m_2^\prime=-l_2}^{m_2}}\frac{{\left(-\frac{t_{12}}{\tilde{x}_{12}}\right)}^{m_1+m_2-m_1^\prime-m_2^\prime}}{\left(m_1-m_1^\prime\right)!\left(m_2-m_2^\prime\right)!}\nonumber\\
&\times C^{l_1;m_1^\prime+m_2^\prime-l_1}_{l_1,\Delta_1,\xi_1;l_2,\Delta_1,\xi_2}\text{ }\xi_2^{l_1+l_2-m_1^\prime-m_2^\prime}{\left[\frac{(l_1+l_2-m_1^\prime-m_2^\prime)!}{(2l_1)!(m_1^\prime+m_2^\prime-l_1+l_2)!}\right]}^{\frac{1}{2}}\hspace{10mm}\text{(for $l_1\geq l_2$)}\label{49}
\end{align}

\medskip

From the above expressions, it is evident that, for $m_1+m_2<|l_1-l_2|$ :
\begin{align*}
\langle{\mathcal{T}}\Phi_{(l_1);\Delta_1,\xi_1}^{m_1}(t_1,x_1)\Phi_{(l_2);\Delta_2,\xi_2}^{m_2}(t_2,x_2)\rangle=\langle{\mathcal{T}}\Phi_{(l_2);\Delta_2,\xi_2}^{m_2}(t_2,x_2)\Phi_{(l_1);\Delta_1,\xi_1}^{m_1}(t_1,x_1)\rangle=0 
\end{align*}

\medskip

Also, the 2-point quasi-primary correlator involving exactly one component field of a scalar multiplet simply is (with $\xi_1\neq0$):
\begin{align}
\langle{\mathcal{T}}\Phi_{(l_1);\Delta_1,\xi_1}^{m_1}(t_1,x_1)\Phi_{(l_2);\Delta_2,0}^{m_2}(t_2,x_2)\rangle&=\langle{\mathcal{T}}\Phi_{(l_2);\Delta_2,0}^{m_2}(t_2,x_2)\Phi_{(l_1);\Delta_1,\xi_1}^{m_1}(t_1,x_1)\rangle\nonumber\\
&=\text{ }\lim\limits_{\epsilon\rightarrow0^+}\text{ }\frac{\delta_{l_1,m_1}\delta_{\Delta_1,\Delta_2}}{{(\tilde{x}_{12})}^{2\Delta_1}}\text{ }C^{l_1;m_2}_{l_1,\Delta_1,\xi_1;l_2,\Delta_1,0}
\end{align}
while the correlator of two quasi-primary scalar multiplets is:
\begin{align}
\langle{\mathcal{T}}\Phi_{(l_1);\Delta_1,0}^{m_1}(t_1,x_1)\Phi_{(l_2);\Delta_2,0}^{m_2}(t_2,x_2)\rangle&=\langle{\mathcal{T}}\Phi_{(l_2);\Delta_2,0}^{m_2}(t_2,x_2)\Phi_{(l_1);\Delta_1,0}^{m_1}(t_1,x_1)\rangle\nonumber\\
&=\text{ }\lim\limits_{\epsilon\rightarrow0^+}\text{ }\frac{\delta_{\Delta_1,\Delta_2}}{{(\tilde{x}_{12})}^{2\Delta_1}}\text{ }C^{m_1;m_2}_{l_1,\Delta_1,0;l_2,\Delta_1,0}
\end{align}
which resembles a CFT$_1$ bosonic quasi-primary 2-point correlator.

\medskip

\subsection{3-point quasi-primary correlator}\label{111}
We now find the general structure of 3-point correlation functions of bosonic vector quasi-primary fields. The 3-point vector quasi-primary correlators are denoted as, with $s_p\in\{+,-\}$ :
\begin{align*}
\langle\mathcal{T}\Phi^{s_1}_{\Delta_1,\xi_1}(t_1,x_1)\Phi^{s_2}_{\Delta_2,\xi_2}(t_2,x_2)\Phi^{s_3}_{\Delta_3,\xi_3}(t_3,x_3)\rangle\equiv{G}^{s_1s_2s_3}_{\{(\Delta_i,\xi_i)\}}
\end{align*}
with all different spatial insertions. The bosonic property (like in the 2-point case) ensures that the ordering of the quantum numbers in ${G}^{s_1s_2s_3}_{\{(\Delta_i,\xi_i)\}}$ is not important.

\medskip
 
Due to the $1+1$D global CC invariance, the 3-point correlators satisfy the following coupled PDEs (by treating $t_{12},t_{23},x_{12},x_{23}$ as the independent variables):
\begin{align}
&\left[x_{12}\partial_{t_{12}}+x_{23}\partial_{t_{23}}+\bm{\xi}_1+\bm{\xi}_2+\bm{\xi}_3\right]\text{ }G^{s_1s_2s_3}_{\{(\Delta_i,\xi_i)\}}=0\nonumber\\
&\left[x_{12}\partial_{x_{12}}+t_{12}\partial_{t_{12}}+x_{23}\partial_{x_{23}}+t_{23}\partial_{t_{23}}+{\Delta}_1+\Delta_2+\Delta_3\right]\text{ }G^{s_1s_2s_3}_{\{(\Delta_i,\xi_i)\}}=0\\
&\left[(x_1^2-x_{2}^2)\partial_{t_{12}}+(x_2^2-x_3^2)\partial_{t_{23}}+2x_1\bm{\xi}_1+2x_2\bm{\xi}_2+2x_3\bm{\xi}_3\right]\text{ }G^{s_1s_2s_3}_{\{(\Delta_i,\xi_i)\}}=0\nonumber\\
&\left[(x_1^2-x_{2}^2)\partial_{x_{12}}+(x_2^2-x_3^2)\partial_{x_{23}}+2(x_1t_1-x_2t_2)\partial_{t_{12}}+2(x_2t_2-x_3t_3)\partial_{t_{23}}\right.\nonumber\\
&\left.\hspace{32mm}+2(x_1\Delta_1+t_1\bm{\xi}_1)+2(x_2\Delta_2+t_2\bm{\xi}_2)+2(x_3\Delta_3+t_3\bm{\xi}_3)\right]\text{ }G^{s_1s_2s_3}_{\{(\Delta_i,\xi_i)\}}=0\nonumber
\end{align}
The solutions of these coupled PDEs, after analytic continuation to complex $x_{ij}$, are compactly expressed as (with $\Delta_{ijk}\equiv\Delta_i+\Delta_j-\Delta_k$ and $\tilde{x}_{ij}\equiv x_{ij}-i\epsilon t_{ij}$):
\begin{align*}
{G}^{s_1s_2s_3}_{\{(\Delta_i,\xi_i)\}}(t_{12},{x}_{12},t_{23},{x}_{23})=\lim\limits_{\epsilon\rightarrow0^+}{{(\tilde{x}_{12})}^{-\Delta_{123}}{(\tilde{x}_{23})}^{-\Delta_{231}}{(\tilde{x}_{13})}^{-\Delta_{312}}}\text{ }{\tilde{G}^{s_1s_2s_3}_{\{(\Delta_i,\xi_i)\}}(t_{12},\tilde{x}_{12},t_{23},\tilde{x}_{23})}
\end{align*}
where the reduced 3-point correlators $\tilde{G}^{s_1s_2s_3}_{\{(\Delta_i,\xi_i)\}}$ explicitly are:
\begin{align}
&\tilde{G}^{---}_{\{(\Delta_i,\xi_i)\}}=0\nonumber\\
&\tilde{G}^{+--}_{\{(\Delta_i,\xi_i)\}}=\lambda\xi_1\hspace{5mm};\hspace{5mm}\tilde{G}^{-+-}_{\{(\Delta_i,\xi_i)\}}=\lambda\xi_2\hspace{5mm;\hspace{5mm}}\tilde{G}^{--+}_{\{(\Delta_i,\xi_i)\}}=\lambda\xi_3\nonumber\\
&\tilde{G}^{++-}_{\{(\Delta_i,\xi_i)\}}=C^{++-}-2\lambda\xi_1\xi_2\frac{t_{12}}{\tilde{x}_{12}}\hspace{7.5mm};\hspace{7.5mm}\tilde{G}^{+-+}_{\{(\Delta_i,\xi_i)\}}=C^{+-+}-2\lambda\xi_1\xi_3\frac{t_{13}}{\tilde{x}_{13}}\nonumber\\
&\tilde{G}^{-++}_{\{(\Delta_i,\xi_i)\}}=C^{-++}-2\lambda\xi_2\xi_3\frac{t_{23}}{\tilde{x}_{23}}\label{82}\\
&\tilde{G}^{+++}_{\{(\Delta_i,\xi_i)\}}=C^{+++}-\xi_1C^{-++}\left(\frac{t_{12}}{\tilde{x}_{12}}+\frac{t_{13}}{\tilde{x}_{13}}-\frac{t_{23}}{\tilde{x}_{23}}\right)-\xi_2C^{+-+}\left(\frac{t_{12}}{\tilde{x}_{12}}+\frac{t_{23}}{\tilde{x}_{23}}-\frac{t_{13}}{\tilde{x}_{13}}\right)\nonumber\\
&\hspace{15mm}-\xi_3C^{++-}\left(\frac{t_{23}}{\tilde{x}_{23}}+\frac{t_{13}}{\tilde{x}_{13}}-\frac{t_{12}}{\tilde{x}_{12}}\right)-\lambda\xi_1\xi_2\xi_3\left[{\left(\frac{t_{12}}{\tilde{x}_{12}}+\frac{t_{23}}{\tilde{x}_{23}}-\frac{t_{13}}{\tilde{x}_{13}}\right)}^2-4\frac{t_{12}}{\tilde{x}_{12}}\frac{t_{23}}{\tilde{x}_{23}}\right]\nonumber
\end{align}
where $\lambda$ and $C^{\ldots}$ are 5 independent 3-point constants.

\medskip

As an example, we now calculate the following 3-point functions involving the EM tensor $\mathbf{T}(t,x)$:
\begin{align*}
\langle\mathcal{T}\mathbf{T}(t_1,x_1)\mathbf{\Phi}_{\Delta,\xi_i}(t_2,x_2)\mathbf{\Phi}_{\Delta,\xi_j}(t_3,x_3)\rangle
\end{align*}
where $\mathbf{\Phi}_{\Delta,\xi_i}(t_1,x_1)$ and $\mathbf{\Phi}_{\Delta,\xi_j}(t_2,x_2)$ are two bosonic vector primary fields. Let the 2-point correlator between these primary fields be given by \eqref{45}. Using the analytically continued Ward identities \eqref{40}-\eqref{41} applicable for primary fields, the 3-point functions are then obtained as below:
\begin{align}
&\langle\mathcal{T}T^t_{\hspace{1.5mm}t}{\Phi}^-_{\Delta,\xi_i}{\Phi}^-_{\Delta,\xi_j}\rangle=\langle\mathcal{T}T^t_{\hspace{1.5mm}t}{\Phi}^+_{\Delta,\xi_i}{\Phi}^-_{\Delta,\xi_j}\rangle=\langle\mathcal{T}T^t_{\hspace{1.5mm}t}{\Phi}^-_{\Delta,\xi_i}{\Phi}^+_{\Delta,\xi_j}\rangle=\langle\mathcal{T}T^t_{\hspace{1.5mm}x}{\Phi}^-_{\Delta,\xi_i}{\Phi}^-_{\Delta,\xi_j}\rangle=0\nonumber\\
&\langle\mathcal{T}T^t_{\hspace{1.5mm}x}{\Phi}^+_{\Delta,\xi_i}{\Phi}^-_{\Delta,\xi_j}\rangle=\lim\limits_{\epsilon\rightarrow0^+}-i\frac{\lambda\Delta\xi_i}{\tilde{x}_{23}^{2\Delta}}{\left(\frac{\tilde{x}_{23}}{\tilde{x}_{12}\tilde{x}_{13}}\right)}^2\hspace{3mm};\hspace{3mm}\langle\mathcal{T}T^t_{\hspace{1.5mm}x}{\Phi}^-_{\Delta,\xi_i}{\Phi}^+_{\Delta,\xi_j}\rangle=\lim\limits_{\epsilon\rightarrow0^+}-i\frac{\lambda\Delta\xi_j}{\tilde{x}_{23}^{2\Delta}}{\left(\frac{\tilde{x}_{23}}{\tilde{x}_{12}\tilde{x}_{13}}\right)}^2\nonumber\\
&\langle\mathcal{T}T^t_{\hspace{1.5mm}t}{\Phi}^+_{\Delta,\xi_i}{\Phi}^+_{\Delta,\xi_j}\rangle=\lim\limits_{\epsilon\rightarrow0^+}-i\frac{\lambda\xi_i\xi_j}{\tilde{x}_{23}^{2\Delta}}{\left(\frac{\tilde{x}_{23}}{\tilde{x}_{12}\tilde{x}_{13}}\right)}^2\\
&\langle\mathcal{T}T^t_{\hspace{1.5mm}x}{\Phi}^+_{\Delta,\xi_i}{\Phi}^+_{\Delta,\xi_j}\rangle=\lim\limits_{\epsilon\rightarrow0^+}-\frac{i}{\tilde{x}_{23}^{2\Delta}}\left[\Delta\mu-2\lambda\Delta\xi_i\xi_j\frac{t_{23}}{\tilde{x}_{23}}-2\lambda\xi_i\xi_j\left(\frac{t_{12}}{\tilde{x}_{12}}+\frac{t_{13}}{\tilde{x}_{13}}-\frac{t_{23}}{\tilde{x}_{23}}\right)\right]{\left(\frac{\tilde{x}_{23}}{\tilde{x}_{12}\tilde{x}_{13}}\right)}^2\nonumber
\end{align}
This result is consistent with the above derived general form of the 3-point bosonic vector quasi-primary correlators with the identifications: $T^t_{\hspace{1.5mm}x}\equiv T^+$ and $T^t_{\hspace{1.5mm}t}\equiv T^-$. 

\medskip

Comparison of this example to the general form \eqref{82} immediately reveals that:
\begin{align*}
\text{the EM tensor (multiplet) possesses pCB charge $\xi=2$ .}
\end{align*}

\medskip

General 3-point correlators involving quasi-primary multiplets of arbitrary rank can be calculated using the direct approach involving \eqref{75}. We shall not report the cumbersome and not-illuminating results here. 

\medskip

Sadly, the impressive run of extracting the general functional form of $n$-point correlation functions just by using general symmetry arguments meets an abrupt end at $n=3$. Due to the existence of the Carrollian conformal equivalents \cite{Bagchi:2016geg} of the conformal `invariant ratios', the space-time dependence of an $n(\geq4)$-point correlator can not be fixed by symmetry alone, without inputs from any particular dynamics.

\medskip

\section{Operator Formalism}\label{112}
Until now, we were directly concerned with various properties of the correlation functions at the expense of being indifferent to the fields themselves. We only worked in the path-integral formalism where the fields inside the correlators are even (mis)treated as if they are ordinary functions (distributions). It is the operator formalism of QFT where the true identity of the fields as operator valued function (distribution) come into play.

\medskip

We introduced the operator formalism of the $1+1$D CCFT in section \ref{84} to meet a demand concerning the correlation functions! In this section, we further develop this formalism and study the operator aspects of the CC quantum fields.

\medskip

\subsection{Mode expansion}\label{113}
We start by defining the mode-expansion of these fields since the modes (independent of space-time) are the actual operator parts of the quantum fields.

\medskip
 
From the general two-point quasi-primary correlators \eqref{49}, we observe that for a quasi-primary field $\Phi_{(l)}^m(t,x)$, the 2-point correlator of $\partial_t^{l+m+1}\Phi_{(l)}^m(t,x)$ with any quasi-primary field vanishes. If we assume that all the basis fields in a CCFT are either quasi-primaries themselves or global descendants (i.e. various degree space and time derivatives) thereof\footnote{This assumption is in accordance with the postulate stated in page \pageref{85}.}, the above implies that the 2-point correlator of $\partial_t^{l+m+1}\Phi_{(l)}^m(t,x)$ with any field in the theory must vanish. That is possible only if $\partial_t^{l+m+k}\Phi_{(l)}^m(t,x)$ for any $k\in\mathbb{N}$ is identically the `0-operator'. 

\medskip

This motivates the following mode expansion for an arbitrary quasi-primary (basis) field with scaling dimension $\Delta$, boost-charge $\xi$ and pCB rank $l$:
\begin{align}
&\Phi_{(l)}^m(t,x)=\sum_{p\in\mathbb{Z}}x^{-\Delta-p}\sum_{q=0}^{l+m}{\left(\frac{t}{x}\right)}^q\Phi_{(l);p,q}^m\label{eq:33}\\
\Rightarrow\text{ }&\Phi_{(l);p,q}^m=\oint\limits_0\frac{dt}{2\pi i}\text{ }\frac{1}{t^{q+1}}\oint\limits_0\frac{dx}{2\pi i}\text{ }x^{\Delta+p+q-1}\text{ }\Phi_{(l)}^m(t,x)\nonumber
\end{align}
From \eqref{eq:33}, we note that the $\Phi_{(l)}^{-l}$ component of a quasi-primary multiplet does not depend on $t$.

\medskip

There is nothing mysterious about the form of the quasi-primary mode expansion \eqref{eq:33}: we get a polynomial in real $t$ due to $\partial_t^{l+m+k}\Phi_{(l)}^m(t,x)$ for all $k\in\mathbb{N}$ being 0. After that, it is just the most general Laurent expansion (around the origin) of a distribution in the complex variable $x$. In the classical theory, the expansion in $x$ should be considered as a two-sided Taylor-expansion since $x\in\mathbb{R}\cup\{\infty\}$ now i.e. the point at $\infty$ is identified.

\medskip

\subsection{Quantum EM tensor}\label{114}
Finally, we shall now study the quantum aspects of the EM tensor field.

\medskip

\subsubsection*{EM tensor modes}
Earlier, we have seen that the EM tensor components have scaling dimension $\Delta=2$. Also, in a classical field theory invariant under CCA$_{1+1}$, their space-time dependence is given by \eqref{eq:34}-\eqref{eq:35}. 

\medskip

Guided by these properties, the EM tensor components' mode-expansion is defined as:
\begin{align}
&{T}^t_{\hspace{1.5mm}t}(t,x)=\sum_{n\in\mathbb{Z}}x^{-n-2}M_n\hspace{7mm};\hspace{7mm}{T}^t_{\hspace{1.5mm}x}(t,x)=\sum_{n\in\mathbb{Z}}x^{-n-2}\text{ }[L_n-(n+2)\frac{t}{x}M_n]\label{eq:37}\\
\Longrightarrow\hspace{2.5mm}&M_n=\oint\limits_0\frac{dx}{2\pi i}\text{ }x^{n+1}\text{ }{T}^t_{\hspace{1.5mm}t}(t,x)\hspace{4mm};\hspace{4mm}L_n=\oint\limits_0\frac{dx}{2\pi i}\left[x^{n+1}{T}^t_{\hspace{1.5mm}x}(t,x)+(n+1)x^nt\text{ }{T}^t_{\hspace{1.5mm}t}(t,x)\right]\label{eq:36}
\end{align}
This result was first obtained through a `limiting' perspective in \cite{Bagchi:2010vw} in the context of $1+1$D Galilean CFTs.

\medskip

Classically, as shown in \eqref{eq:34}, the component ${T}^t_{\hspace{1.5mm}t}$ depends only on $x$. Thus, the quantum mode expansion of ${T}^t_{\hspace{1.5mm}t}$ is simply a Laurent series around the origin in the complex variable $x$ (because in the quantum theory, we have analytically continued $x$ into the Riemann sphere from the Riemann circle in the classical theory). Now, obeying \eqref{eq:35} and writing another Laurent series in $x$ corresponding to the $p(x)$ there, we complete the mode-expansion of ${T}^t_{\hspace{1.5mm}x}$.

\medskip

\subsubsection*{The $1+1$D CC generators}
We shall now show that the EM tensor modes generate the $1+1$D CC transformations \eqref{55} in the space of the quantum fields.

\medskip

We begin by noting that the contour enclosing the spatial origin in \eqref{eq:36} can be continuously deformed into the contour $C_u^\prime$ enclosing the upper half plane along with the real line. Then comparing to the classical charges \eqref{eq:25} and appealing to \eqref{50}, one immediately concludes that $M_n$ and $L_n$ for $n\in\mathbb{Z}$ are such quantum conserved charge operators in the space of quantum fields that:
\begin{align*}
&M_n=Q_t[x^{n+1}] \hspace{2mm}\text{ generates }\hspace{2mm}x\rightarrow x\hspace{2mm},\hspace{2mm}t\rightarrow t+\epsilon^tx^{n+1}\\
&L_n=Q_x[x^{n+1}] \hspace{2mm}\text{ generates }\hspace{2mm}x\rightarrow x+\epsilon^xx^{n+1}\hspace{2mm},\hspace{2mm}t\rightarrow t+\epsilon^x(n+1)x^nt
\end{align*}

\medskip

This directly gives the infinitesimal transformation rule (first derived in \cite{Bagchi:2009ca} for `scalar' $\xi$) in the operator formalism for a primary field $\Phi_{(l)}^m(t,x)$, from \eqref{eq:21} and \eqref{51}:
\begin{align}
&[L_n\text{ }, \Phi_{(l)}^m(t,x)]=-i\left[x^{n+1}\partial_x+t(n+1)x^n\partial_t+\Delta(n+1)x^n+\bm{\xi}n(n+1)x^{n-1}t\right]\Phi_{(l)}^m(t,x)\nonumber\\
&[M_n\text{ }, \Phi_{(l)}^m(t,x)]=-i\left[x^{n+1}\partial_t+(n+1)x^n\bm{\xi}\right]\Phi_{(l)}^m(t,x)\label{53}
\end{align}
This can also be easily verified using the OPEs \eqref{42} in the form appropriate for a primary field and the prescription \eqref{52}.

\medskip

\subsubsection*{The $TT$ OPEs} 
We want to find the algebra of the EM tensor modes. For that, we first need to find the $TT$ OPEs between the EM tensor components.

\medskip
 
We can fix the $TT$ OPEs by general symmetry arguments along with the assumption that no field in the theory possesses negative scaling dimension. The bosonic nature of the EM tensor field will play a crucial role in this endeavour. Since the $i\epsilon$-form of the OPEs is the convenient one for applying the bosonic exchange properties of the two fields involved, we shall obtain the $TT$ OPEs in this form. 

\medskip

Though we expect the EM tensor to be a quasi-primary multiplet, we have not explicitly shown this fact anywhere. From our construction of the $TT$ OPEs, the quasi-primary transformation property of the EM tensor will be manifest.

\medskip

We start with the $T^t_{\hspace{1.5mm}t}({t^\prime,x^\prime})T^t_{\hspace{1.5mm}t}({t,x})$ OPE. The classical property \eqref{eq:34} tells us that $T^t_{\hspace{1.5mm}t}$ does not depend on $t$. Consequently, this component must be boost invariant, i.e.:
\begin{align}
\bm{\xi}\cdot T^t_{\hspace{1.5mm}t}=0
\end{align}
Thus, obeying the assumption of non-existence of fields with negative scaling dimensions, we write the following schematic OPE from the general form \eqref{42} of the Ward identities, with $\Delta\tilde{x}^\prime:=x^\prime-x-i\epsilon(t^\prime-t)$ as before: 
\begin{align}
&T^t_{\hspace{1.5mm}t}({t^\prime,x^\prime})T^t_{\hspace{1.5mm}t}({t,x})\sim\lim\limits_{\epsilon\rightarrow0^+}-i\left[\frac{A(t,x)}{{(\Delta\tilde{x}^\prime)}^4}+\frac{B(t,x)}{{(\Delta\tilde{x}^\prime)}^3}\right]\label{87}\\
\Longrightarrow\hspace{2.5mm}&T^t_{\hspace{1.5mm}t}({t,x})T^t_{\hspace{1.5mm}t}({t^\prime,x^\prime})\sim\lim\limits_{\epsilon\rightarrow0^+}-i\left[\frac{A(t^\prime,x^\prime)}{{(\Delta\tilde{x}^\prime)}^4}-\frac{B(t^\prime,x^\prime)}{{(\Delta\tilde{x}^\prime)}^3}\right]\label{86}
\end{align}
where $A$ and $B$ are two unidentified fields with scaling dimensions 0 and 1 respectively and are regular at $(t^\prime,x^\prime)=(t,x)$.

\medskip

Since the EM tensor is a bosonic field and the L.H.S. of the OPEs are time-ordered, we must have:
\begin{align*}
\mathcal{T}T^t_{\hspace{1.5mm}t}({t^\prime,x^\prime})T^t_{\hspace{1.5mm}t}({t,x})=\mathcal{T}T^t_{\hspace{1.5mm}t}({t,x})T^t_{\hspace{1.5mm}t}({t^\prime,x^\prime})
\end{align*}
To compare \eqref{87} with \eqref{86} in light of this bosonic property, we Taylor-expand $A$ and $B$ in the latter, around $(t,x)$ respectively upto the third and the second order. Comparing the terms order-by-order, we find that to satisfy the demand of the scaling dimensions specified for $A$ and $B$, the only possibility is to have the following OPE:
\begin{align}
T^t_{\hspace{1.5mm}t}({t^\prime,x^\prime})T^t_{\hspace{1.5mm}t}({t,x})\sim\lim\limits_{\epsilon\rightarrow0^+}-i\frac{C_3}{{(\Delta\tilde{x}^\prime)}^4}
\end{align} 
where $C_3$ is a constant, i.e. the field $A$ is proportional to the Identity field while no field-candidate is found to play the role of the $B$ field.

\medskip

Since, the scaling dimension of the EM tensor components is $\Delta=2$, from the general form \eqref{42}, we directly write the following schematic OPE:
\begin{align}
&T^t_{\hspace{1.5mm}x}({t^\prime,x^\prime})T^t_{\hspace{1.5mm}t}(t,x)\nonumber\\
\sim&\lim\limits_{\epsilon\rightarrow0^+} -i\left[\frac{D(t,x)}{{(\Delta\tilde{x}^\prime)}^4}+\frac{E(t,x)}{{(\Delta\tilde{x}^\prime)}^3}+\frac{{{2}}T^t_{\hspace{1.5mm}t}(t,x)}{{(\Delta\tilde{x}^\prime)}^2}+\frac{{\partial_{x}}T^t_{\hspace{1.5mm}t}(t,x)}{\Delta\tilde{x}^\prime}-({t^\prime}-{t})\frac{4C_3}{{(\Delta\tilde{x}^\prime)}^5}\right]\label{88}
\end{align} 
where the unspecified fields $D$ and $E$, regular at $(t^\prime,x^\prime)=(t,x)$, must have scaling dimensions 0 and 1 respectively.

\medskip

At the same time, we must have the schematic OPE below, according to the general form \eqref{42}:
\begin{align}
T^t_{\hspace{1.5mm}t}(t^\prime,x^\prime)T^t_{\hspace{1.5mm}x}(t,x)\sim\lim\limits_{\epsilon\rightarrow0^+} -i\left[\frac{F(t,x)}{{(\Delta\tilde{x}^\prime)}^4}+\frac{G(t,x)}{{(\Delta\tilde{x}^\prime)}^3}+\frac{(\bm{\xi}\cdot T^t_{\hspace{1.5mm}x})(t,x)}{{(\Delta\tilde{x}^\prime)}^2}+\frac{{\partial_{t}} T^t_{\hspace{1.5mm}x}(t,x)}{\Delta\tilde{x}^\prime}\right]\label{89}
\end{align}
with $F$, $G$ being two unspecified fields regular at $(t^\prime,x^\prime)=(t,x)$ and have scaling dimensions 0 and 1 respectively.

\medskip

Again, we expect the following bosonic exchange property to hold:
\begin{align*}
\mathcal{T}T^t_{\hspace{1.5mm}x}({t^\prime,x^\prime})T^t_{\hspace{1.5mm}t}({t,x})=\mathcal{T}T^t_{\hspace{1.5mm}t}({t,x})T^t_{\hspace{1.5mm}x}({t^\prime,x^\prime})
\end{align*}
When applied to \eqref{88}, this implies that:
\begin{align}
&T^t_{\hspace{1.5mm}t}(t^\prime,x^\prime)T^t_{\hspace{1.5mm}x}(t,x)\nonumber\\
\sim&\lim\limits_{\epsilon\rightarrow0^+}-i\left[\frac{D(t^\prime,x^\prime)}{{(\Delta\tilde{x}^\prime)}^4}-\frac{E(t^\prime,x^\prime)}{{(\Delta\tilde{x}^\prime)}^3}+\frac{{{2}}T^t_{\hspace{1.5mm}t}(t^\prime,x^\prime)}{{(\Delta\tilde{x}^\prime)}^2}-\frac{{\partial_{x^\prime}}T^t_{\hspace{1.5mm}t}(t^\prime,x^\prime)}{\Delta\tilde{x}^\prime}-({t^\prime}-{t})\frac{4C_3}{{(\Delta\tilde{x}^\prime)}^5}\right]\label{90}
\end{align}
while from \eqref{89}, it leads to:
\begin{align}
T^t_{\hspace{1.5mm}x}({t^\prime,x^\prime})T^t_{\hspace{1.5mm}t}(t,x)\sim\lim\limits_{\epsilon\rightarrow0^+}-i\left[\frac{F(t^\prime,x^\prime)}{{(\Delta\tilde{x}^\prime)}^4}-\frac{G(t^\prime,x^\prime)}{{(\Delta\tilde{x}^\prime)}^3}+\frac{(\bm{\xi}\cdot T^t_{\hspace{1.5mm}x})(t^\prime,x^\prime)}{{(\Delta\tilde{x}^\prime)}^2}-\frac{{\partial_{t^\prime}} T^t_{\hspace{1.5mm}x}(t^\prime,x^\prime)}{\Delta\tilde{x}^\prime}\right]\label{91}
\end{align}

\medskip

We need to keep in mind that \eqref{90} and \eqref{91} are not OPEs yet because an OPE has the following general form:
\begin{align*}
{\Phi}_1(t_1,x_1){\Phi}_2(t_2,x_2)=\sum_kC^k_{12}(t_{12},x_{12})\text{ }{\Phi}_k(t_2,x_2)
\end{align*} 
Thus, to turn them into OPEs, we need to Taylor-expand the numerators in \eqref{90} and \eqref{91} around $(t^\prime,x^\prime)=(t,x)$ upto (maximally) the third order. 

\medskip

Performing the required Taylor-expansions and comparing \eqref{90} with \eqref{89} and \eqref{91} with \eqref{88} and finally, using the conservation equation:
\begin{align*}
\partial_{x} T^t_{\hspace{1.5mm}t}=\partial_{t} T^t_{\hspace{1.5mm}x}
\end{align*}
we obtain the following consistency relations, in line with the required scaling dimensions of the fields $D$, $E$, $F$ and $G$:
\begin{align}
C_3=0\hspace{2.5mm};\hspace{2.5mm}E&=0=G\hspace{2.5mm};\hspace{2.5mm}D=F=\frac{C_2}{2}\text{ (constant)}\nonumber\\
&\bm{\xi}\cdot T^t_{\hspace{1.5mm}x}=2T^t_{\hspace{1.5mm}t}
\end{align}
This condition confirms that:
\begin{align*}
\text{the EM tensor components form a rank-$\frac{1}{2}$ multiplet with pCB charge $\xi=2$ .}
\end{align*}
Thus, we are led to the following three $TT$ OPEs:
\begin{align}
&T^t_{\hspace{1.5mm}t}({t^\prime,x^\prime})T^t_{\hspace{1.5mm}t}(t,x)\sim \text{ regular}\label{92}\\
&T^t_{\hspace{1.5mm}x}({t^\prime,x^\prime})T^t_{\hspace{1.5mm}t}(t,x)\sim\lim\limits_{\epsilon\rightarrow0^+}-i\left[\frac{\frac{C_2}{2}}{{(\Delta\tilde{x}^\prime)}^4}+\frac{2T^t_{\hspace{1.5mm}t}(t,x)}{{(\Delta\tilde{x}^\prime)}^2}+\frac{{\partial_{x}} T^t_{\hspace{1.5mm}t}(t,x)}{\Delta\tilde{x}^\prime}\right]\label{93}\\
&T^t_{\hspace{1.5mm}t}({t^\prime,x^\prime})T^t_{\hspace{1.5mm}x}(t,x)\sim\lim\limits_{\epsilon\rightarrow0^+}-i\left[\frac{\frac{C_2}{2}}{{(\Delta\tilde{x}^\prime)}^4}+\frac{2 T^t_{\hspace{1.5mm}t}(t,x)}{{(\Delta\tilde{x}^\prime)}^2}+\frac{{\partial_{t}} T^t_{\hspace{1.5mm}x}(t,x)}{\Delta\tilde{x}^\prime}\right]\label{94}
\end{align}

\medskip

To obtain the remaining one, we again appeal to the general form \eqref{42} to write:
\begin{align*}
&T^t_{\hspace{1.5mm}x}({t^\prime,x^\prime})T^t_{\hspace{1.5mm}x}(t,x)\sim\lim\limits_{\epsilon\rightarrow0^+} -i\left[\frac{U(t,x)}{{(\Delta\tilde{x}^\prime)}^4}+\frac{V(t,x)}{{(\Delta\tilde{x}^\prime)}^3}+\frac{2T^t_{\hspace{1.5mm}x}(t,x)}{{(\Delta\tilde{x}^\prime)}^2}+\frac{{\partial_{x}}T^t_{\hspace{1.5mm}x}(t,x)}{\Delta\tilde{x}^\prime}\right.\nonumber\\
&\left.\hspace{55mm}-({t^\prime}-{t})\left(\frac{2C_2}{{(\Delta\tilde{x}^\prime)}^5}+\frac{4 T^t_{\hspace{1.5mm}t}(t,x)}{{(\Delta\tilde{x}^\prime)}^3}+\frac{{\partial_{t}} T^t_{\hspace{1.5mm}x}(t,x)}{{(\Delta\tilde{x}^\prime)}^2}\right)\right]
\end{align*}
where the fields $U$ and $V$ are regular at $(t^\prime,x^\prime)=(t,x)$ and have scaling dimensions 0 and 1 respectively.

\medskip

Again, the following bosonic exchange property must be satisfied:
\begin{align*}
\mathcal{T}T^t_{\hspace{1.5mm}x}({t^\prime,x^\prime})T^t_{\hspace{1.5mm}x}({t,x})=\mathcal{T}T^t_{\hspace{1.5mm}x}({t,x})T^t_{\hspace{1.5mm}x}({t^\prime,x^\prime})
\end{align*}
Repeating then the arguments elaborated above, our goal is reached:
\begin{align}
&T^t_{\hspace{1.5mm}x}({t^\prime,x^\prime})T^t_{\hspace{1.5mm}x}(t,x)\sim\lim\limits_{\epsilon\rightarrow0^+} -i\left[\frac{-i\frac{C_1}{2}}{{(\Delta\tilde{x}^\prime)}^4}+\frac{2T^t_{\hspace{1.5mm}x}(t,x)}{{(\Delta\tilde{x}^\prime)}^2}+\frac{{\partial_{x}}T^t_{\hspace{1.5mm}x}(t,x)}{\Delta\tilde{x}^\prime}\right.\nonumber\\
&\left.\hspace{55mm}-({t^\prime}-{t})\left(\frac{2C_2}{{(\Delta\tilde{x}^\prime)}^5}+\frac{4 T^t_{\hspace{1.5mm}t}(t,x)}{{(\Delta\tilde{x}^\prime)}^3}+\frac{{\partial_{t}} T^t_{\hspace{1.5mm}x}(t,x)}{{(\Delta\tilde{x}^\prime)}^2}\right)\right]\label{95}
\end{align}
with $C_1$ being a constant\footnote{The reason to choose dissimilar definitions for the constants $C_1$ and $C_2$ will become clear in the section \ref{131} .}.

\medskip

Noticing the absence of the third order poles at appropriate places in the four $TT$ OPEs and comparing those to \eqref{81}, one readily concludes that:
\begin{align*}
\text{the EM tensor components transform as $1+1$D CC quasi-primary multiplet fields.}
\end{align*}

\medskip

Thus, we have shown that the assumption of the non-existence of negative scaling dimension along with the bosonic exchange property is sufficient to completely determine all the poles of order $\geq2$ in the $TT$ OPEs. As a by-product, our method `proves' that the $T^t_{\hspace{1.5mm}x}$ and $T^t_{\hspace{1.5mm}t}$ form a quasi-primary multiplet of rank-$\frac{1}{2}$ and pCB charge $\xi=2$ .

\medskip

We can easily find the $\langle TT\rangle$ correlators from the $TT$ OPEs: we just need to put the correlator symbol $\langle\ldots\rangle$ in both sides of \eqref{92}-\eqref{95}. Since VEV of the EM tensor components on the global CCA$_{1+1}$ invariant vacuum must vanish, we obtain the 2-point correlators as: 
\begin{align}
&\langle T^t_{\hspace{1.5mm}t}({t_1,x_1})T^t_{\hspace{1.5mm}t}(t_2,x_2)\rangle=0\nonumber\\
&\langle T^t_{\hspace{1.5mm}t}({t_1,x_1})T^t_{\hspace{1.5mm}x}(t_2,x_2)\rangle=\langle T^t_{\hspace{1.5mm}x}({t_1,x_1})T^t_{\hspace{1.5mm}t}(t_2,x_2)\rangle=\lim\limits_{\epsilon\rightarrow0^+}-i\frac{\frac{C_2}{2}}{\tilde{x}_{12}^4}\\
&\langle T^t_{\hspace{1.5mm}x}({t_1,x_1})T^t_{\hspace{1.5mm}x}(t_2,x_2)\rangle=\lim\limits_{\epsilon\rightarrow0^+}\frac{1}{\tilde{x}_{12}^{4}}\left(-\frac{C_1}{2}+2iC_2\frac{t_{12}}{\tilde{x}_{12}}\right)\nonumber
\end{align}   
As expected, these are in the form of the general 2-point quasi-primary correlators \eqref{45} with $\lambda=-i\frac{C_2}{4}$ and $\mu=-\frac{C_1}{2}$. This is in agreement with the result found in \cite{Bagchi:2010vw}.

\medskip

Equipped with the $TT$ OPEs, we are now in a position to derive the algebra of the EM tensor modes.

\medskip

\subsubsection*{The quantum CCA$_{1+1}$}
Finally, we show that the EM tensor modes generate the centrally extended (quantum) version of the CCA$_{1+1}$.

\medskip

We first deduce the infinitesimal $1+1$D CC transformation properties of the EM tensor components by applying the prescription \eqref{52} to the $TT$ OPEs. Following is a sample calculation:
\begin{align*}
[M_n\text{ },\text{ }T^t_{\hspace{1.5mm}x}(t,x)]&=\frac{1}{2\pi i}\oint\limits_{C_u} dx^\prime\text{ }{x^\prime}^{n+1}T^t_{\hspace{1.5mm}t}(t^+,x^\prime)T^t_{\hspace{1.5mm}x}(t,x)\text{\hspace{10mm}(prescription \eqref{52})}\\
&=\frac{-i}{2\pi i}\oint\limits_{C_u} dx^\prime\text{ }{x^\prime}^{n+1}\lim\limits_{\epsilon\rightarrow0^+}\left[\frac{\frac{C_2}{2}}{{(\Delta\tilde{x}^\prime)}^4}+\frac{2 T^t_{\hspace{1.5mm}t}(t,x)}{{(\Delta\tilde{x}^\prime)}^2}+\frac{{\partial_{t}} T^t_{\hspace{1.5mm}x}(t,x)}{\Delta\tilde{x}^\prime}\right]\text{\hspace{5mm}(OPE \eqref{94})}\\
&=\frac{-i}{2\pi i}\oint\limits_{x} dx^\prime\text{ }{x^\prime}^{n+1}\text{ }\left[\frac{\frac{C_2}{2}}{{(\Delta{x}^\prime)}^4}+\frac{2 T^t_{\hspace{1.5mm}t}(t,x)}{{(\Delta{x}^\prime)}^2}+\frac{{\partial_{t}} T^t_{\hspace{1.5mm}x}(t,x)}{\Delta{x}^\prime}\right]\text{\hspace{3mm}(contour deformation)}\\
&=-i\left[\frac{C_2}{12}(n^3-n)x^{n-2}+(n+1)x^n\cdot2T^t_{\hspace{1.5mm}t}(t,x)+x^{n+1}\partial_{t} T^t_{\hspace{1.5mm}x}(t,x)\right]
\end{align*}
All the transformation properties resulting from similar calculations are collected below:
\begin{align}
&[M_n\text{ },\text{ }T^t_{\hspace{1.5mm}t}(t,x)]=0\nonumber\\
&[L_n\text{ },\text{ }T^t_{\hspace{1.5mm}t}(t,x)]=-i\left[(n^3-n)x^{n-2}\frac{C_2}{12}+(n+1)x^n\cdot2T^t_{\hspace{1.5mm}t}+x^{n+1}\partial_{x} T^t_{\hspace{1.5mm}t}\right](t,x)\nonumber\\
&[M_n\text{ },\text{ }T^t_{\hspace{1.5mm}x}(t,x)]=-i\left[(n^3-n)x^{n-2}\frac{C_2}{12}+(n+1)x^n\cdot2T^t_{\hspace{1.5mm}t}+x^{n+1}\partial_{t} T^t_{\hspace{1.5mm}x}\right](t,x)\label{96}\\
&[L_n\text{ },\text{ }T^t_{\hspace{1.5mm}x}(t,x)]=-i\left[-(n^3-n)x^{n-2}\frac{iC_1}{12}+(n^3-n)(n-2)tx^{n-3}\frac{C_2}{12}+2(n+1)x^nT^t_{\hspace{1.5mm}x}\right.\nonumber\\
&\left.\hspace{28mm}+x^{n+1}\partial_xT^t_{\hspace{1.5mm}x}+n(n+1)tx^{n-1}\cdot2T^t_{\hspace{1.5mm}t}+(n+1)tx^{n}\partial_tT^t_{\hspace{1.5mm}x}\right](t,x)\nonumber
\end{align}

\medskip

We note that for $n\in\{0,\pm1\}$, the infinitesimal transformation rules \eqref{96} of the EM tensor components resemble those of the primary fields given in \eqref{53}. This observation reassures that the EM tensor transforms as a quasi-primary field.

\medskip

We now employ the EM tensor mode expansions \eqref{eq:37} in the both sides of \eqref{96}. Comparing the coefficients of the powers of $x$ in both sides, we get what is recognized to be the centrally extended $1+1$D Carrollian conformal or the BMS$_3$ algebra:
\begin{align}
&i\left[M_n\hspace{1mm},\hspace{1mm}M_m\right]=0\nonumber\\
&i\left[L_n\hspace{1mm},\hspace{1mm}M_m\right]=(n-m)M_{n+m}+\frac{C_2}{12}(n^3-n)\delta_{n+m,0}\label{130}\\
&i\left[L_n\hspace{1mm},\hspace{1mm}L_m\right]=(n-m)L_{n+m}-i\frac{C_1}{12}(n^3-n)\delta_{n+m,0}\nonumber
\end{align} 
where $n,m\in\mathbb{Z}$. Clearly, all $L_n$ and $M_n$ commute with $C_1$ and $C_2$. The constants $C_1$ and $C_2$ are hence called the central charges of this algebra.

\medskip

Thus, we have shown that those are indeed the EM tensor modes that generate the quantum CCA$_{1+1}$.

\medskip

\subsection{Hermitian conjugation}\label{131}
We shall now introduce a hermitian conjugation relation for the quantum Carrollian conformal fields on $\mathbb{R}\times S^1$.

\medskip

First, we note that the space coordinate $x\in\mathbb{R}\cup\{\infty\}$ on plane can be thought of as being the stereographic projection of the (periodic) coordinate $\theta\in[0,2\pi)$ with $\theta\sim\theta+2\pi$ on the Riemann-circle $S^1$. The 1D stereographic map is explicitly given by\footnote{More precisely, it is actually the stereographic projection from the `north-pole' $(X,Y)=(0,1)$ of the Riemann circle $(X,Y)=(-\sin\theta,\cos\theta)$ onto the $y=0$ line, given by $x=\frac{X}{1-Y}$ .}:
\begin{align}
x=-\cot\frac{\theta}{2}
\end{align}
so that $x$ monotonically increases ranging over the whole of the real line $\mathbb{R}$; moreover, the identification $0\sim2\pi$ for $\theta$ results into the one point (at $\infty$) compactification for the range of $x$.

\medskip

We can consider the above stereographic map as a part of a $1+1$D CC transformation \eqref{55} from the $(\tau,\theta)$ coordinates to the $(t,x)$ coordinates, that has the following form:
\begin{align}
\theta\rightarrow x=-\cot\frac{\theta}{2}\hspace{2.5mm};\hspace{2.5mm}\tau\rightarrow t=\frac{\tau}{2}\csc^2\frac{\theta}{2}\label{83}
\end{align}
We shall name this CC map the `stereographic map'. 

\medskip

Next, to define the hermitian conjugation relations for fields in a $1+1$D CCFT, we draw inspiration from the relativistic CFTs where the hermitian conjugation relation for a field on the cylinder is nothing but the transformation property of that field under the space-time (spherical-)inversion transformation \cite{Os}. We shall define the $1+1$D CC hermitian conjugate fields in almost the similar way. For this purpose, we note down the $1+1$D CC inversion transformation:
\begin{align}
x\rightarrow x^\prime=-\frac{1}{x}\text{ { },{ } }t\rightarrow t^\prime=-\frac{t}{x^2}\label{128}
\end{align}
Clearly, the inversion is not an element of the $1+1$D CC group $\cong ISO(1,2)$ since it is not connected to the identity; it is a discrete transformation. This fact is readily inferred from the negative definiteness of the determinant of the Jacobian of the inversion. Rather, it belongs to the group $IO(1,2)$. 

\medskip

Substituting the stereographic map \eqref{83} into \eqref{128}, we obtain the following effects of inversion on the $(\tau,\theta)$ coordinates:
\begin{align}
\theta\rightarrow \theta^\prime=\theta+\pi\text{ { },{ } }\tau\rightarrow \tau^\prime=-\tau
\end{align}  
i.e. the inversion maps a point $\theta$ on the Riemann circle into its anti-podal point $(\theta+\pi)$ while the $\tau$ coordinate undergoes a reversal. Thus, the spherical inversion in $(t,x)$ space-time simply corresponds to a spatio-temporal reflection in the $(\tau,\theta)$ space-time.

\medskip

Now, we propose the following the hermitian conjugation relation for a $1+1$D bosonic quasi-primary multiplet with pCB rank $l$, charge $\xi$ and scaling dimension $\Delta$:
\begin{align}
\left[\Phi_{(l)}^m(t,x)\right]^\dagger:=(-)^{l+m}x^{-2\Delta}{\left[e^{2\xi \frac{t}{x}\mathbf{J}_{(l)}^{\mathbf{-}}}\right]}^m_{\hspace{2mm} m^\prime}{\Phi}_{(l)}^{m^\prime}(-\frac{t}{x^2},-\frac{1}{x})\label{129}
\end{align} 
We recognize that barring the phase-factor, the rest of the R.H.S. is just the $1+1$D global CC transformation rule \eqref{eq:3} applied to inversion, for bosonic quasi-primary multiplets.

\medskip

Below we list the VEVs $\left\langle\left[\Phi^{s_1}_{\Delta_1,\xi_1}(t,x)\right]^\dagger\Phi^{s_2}_{\Delta_2,\xi_2}(t,x)\right\rangle$ where $\Phi^{s_i}_{\Delta_i,\xi_i}(t,x)$ are (the components of) bosonic vector quasi-primary multiplets. This operator product inside the correlator is automatically time-ordered if $t<0$ or anti-time-ordered if $t>0$, as is seen from the inversion map \eqref{128}. When $t>0$, one can make this product time-ordered by virtue of the assumed bosonic exchange property between the fields involved. So, these VEVs can be interpreted as (time-ordered) correlators without any problem. Thus, applying the above hermitian conjugation relation and using the two-point bosonic vector quasi-primary correlators \eqref{45}, we obtain the following results:
\begin{align}
\left\langle\left[\Phi^{-}_{\Delta_1,\xi_1}(t,x)\right]^\dagger\Phi^{-}_{\Delta_2,\xi_2}(t,x)\right\rangle&=0\nonumber\\
\left\langle\left[\Phi^{+}_{\Delta_1,\xi_1}(t,x)\right]^\dagger\Phi^{-}_{\Delta_2,\xi_2}(t,x)\right\rangle&=-\lim\limits_{\epsilon\rightarrow0^+}\frac{\lambda\xi_1\delta_{\Delta_1,\Delta_2}}{{\left(x^\prime\right)}^{2\Delta_1}}\nonumber\\
\left\langle\left[\Phi^{-}_{\Delta_1,\xi_1}(t,x)\right]^\dagger\Phi^{+}_{\Delta_2,\xi_2}(t,x)\right\rangle&=\lim\limits_{\epsilon\rightarrow0^+}\frac{\lambda\xi_2\delta_{\Delta_1,\Delta_2}}{{\left(x^\prime\right)}^{2\Delta_1}}\text{\hspace{7.5mm}(with $x^\prime:=x^2+1-i\epsilon t(x+\frac{1}{x})$ )}\nonumber\\
\left\langle\left[\Phi^{+}_{\Delta_1,\xi_1}(t,x)\right]^\dagger\Phi^{+}_{\Delta_2,\xi_2}(t,x)\right\rangle&
=-\lim\limits_{\epsilon\rightarrow0^+}\frac{\mu\delta_{\Delta_1,\Delta_2}}{{\left(x^\prime\right)}^{2\Delta_1}}
\end{align}
The VEVs of the oppositely ordered products are exactly the same as the above. 

\medskip

The most important property of the above VEVs is that all of them are independent of $t$ (after explicitly taking the limit $\epsilon\rightarrow0^+$). This $t$-independence holds for any VEV (of the type considered above) involving two quasi-primary multiplets of arbitrary (and unequal) ranks. It can be verified by applying the hermitian conjugation relation \eqref{129} and then making use of the general 2-point bosonic quasi-primary correlator \eqref{49}.

\medskip

In view of the $t$-independence (hence, also $\tau$-independence) of these VEVs and the $1+1$D CC inversion transformation being the space-time reflection in $(\tau,\theta)$ coordinates, it will be interesting to explore the implications of these facts on the $1+2$D flat holography.

\medskip

We conclude this work by pointing out the hermitian conjugation properties of the EM tensor modes. Using the relation \eqref{129} in the form suitable for the EM tensor (a rank-$\frac{1}{2}$ multiplet) and comparing the modes on both sides after using the EM tensor mode-expansion \eqref{eq:37}, one reaches the following mode-conjugation properties\footnote{$t$ and $x$ remain unchanged under hermitian conjugation.}:
\begin{align}
L_n^\dagger=(-)^{n+1}L_{-n}\hspace{4.5mm}\text{ and }\hspace{4.5mm}M_n^\dagger=(-)^{n}M_{-n}
\end{align}
Thus, the action of $M_0$ in the space of the quantum fields is hermitian while that of $L_0$ is anti-hermitian. Furthermore, since they mutually commute, $L_0$ and $M_0$ can be simultaneously diagonalized in the Hilbert space. This is to be contrasted with the classical action of the pCB generator on the space-time, that is non-diagonalizable. 

\medskip

Keeping this difference in mind, one needs to explicitly check if a unique quantum field (operator) corresponds to a (and only that one) state in the Hilbert space of the theory. We leave this exercise for future work. Such a state-operator correspondence in the `radial-quantization' scheme \cite{Hao:2021urq} was assumed in the literature \cite{Bagchi:2009pe,Chen:2020vvn}, in direct analogy with the relativistic CFTs. But, as said in section \ref{84}, in this work we have not performed any radial-quantization. In this context, it is worth noting that, recently in \cite{Agia:2022srj}, it was shown that an alternative quantization scheme named the `angular quantization' of 2D CFT does not give rise to any state-operator correspondence. 

\medskip

Finally, taking the hermitian conjugate on both sides of the centrally-extended (quantum) CCA$_{1+1}$ commutation relations \eqref{130}, we discover that both of the central-charges $C_1$ and $C_2$ are real. This is why these constants were defined the way they were.
\acknowledgments
We are deeply indebted to Arjun Bagchi, Aritra Banerjee, Shamik Banerjee, Rudranil Basu, Ritankar Chatterjee, Diptarka Das, Sudipta Dutta, Rishabh Kaushik, Nilay Kundu, Alok Laddha, Saikat Mondal and Debmalya Sarkar for useful discussions at various stages of this work. We would like to specially thank Arjun Bagchi for his valuable comments on the initial version of the draft. A.S. is financially supported by the PMRF fellowship, MHRD, India.

\end{document}